\UseRawInputEncoding
\documentclass[11pt]{article}
\usepackage{geometry}
\usepackage{amsmath, amsthm, amssymb, dsfont, mathrsfs, BOONDOX-cal, upgreek, makecell}
\usepackage{graphicx, caption, float, multirow}
\usepackage{chngcntr, etoolbox}

\usepackage{fancyhdr}
\usepackage{hyperref, xcolor}
\hypersetup{colorlinks=true, linkcolor=blue, citecolor=blue}

\counterwithin*{equation}{section}
\counterwithin*{equation}{subsection}
\renewcommand\theequation{\ifnumgreater{\value{subsection}}{0}{\thesubsection.}{\thesection.}\arabic{equation}}

\allowdisplaybreaks

\newtheoremstyle{theorem}
  {10pt}
  {10pt}
  {\sl}
  {\parindent}
  {\bf}
  {. }
  { }
  {}
\theoremstyle{theorem}
\newtheorem{assumption}{Assumption}
\newtheorem{definition}{Definition}
\newtheorem{lemma}{Lemma}
\newtheorem{theorem}{Theorem}
\newtheorem{proposition}{Proposition}
\newtheorem{corollary}{Corollary}

\begin{document}

\title{\textbf{Optimal Consumption--Investment Problems under Time-Varying Incomplete Preferences\footnote{This paper is based in part on the author's doctoral thesis for the Degree of PhD in Mathematical Finance at Boston University. The author is greatly indebted to his principal advisor, Andrew Lyasoff, and discussants Jin Ma, Scott Robertson, Jiang Wang, Hao Xing, and Jianfeng Zhang for their interest in this work and for providing constructive feedback and helpful comments.}}}
\author{Weixuan Xia\thanks{Boston University Questrom School of Business, Boston, MA, USA. Email: gabxia@bu.edu}}
\date{2023}
\maketitle

\begin{abstract}
  The main objective of this paper is to develop a martingale-type solution to optimal consumption--investment choice problems (\cite[Merton, 1969]{M3} and \cite[Merton, 1971]{M4}) under time-varying incomplete preferences driven by externalities such as patience, socialization effects, and market volatility. The market is composed of multiple risky assets and multiple consumption goods, while in addition there are multiple fluctuating preference parameters with inexact values connected to imprecise tastes. Utility maximization is a multi-criteria problem with possibly function-valued criteria. To come up with a complete characterization of the solutions, first we motivate and introduce a set-valued stochastic process for the dynamics of multi-utility indices and formulate the optimization problem in a topological vector space. Then, we modify a classical scalarization method allowing for infiniteness and randomness in dimensions and prove results of equivalence to the original problem. Illustrative examples are given to demonstrate practical interests and method applicability progressively. The link between the original problem and a dual problem is also discussed, relatively briefly. Finally, using Malliavin calculus with stochastic geometry, we find optimal investment policies to be generally set-valued, each of whose selectors admits a four-way decomposition involving an additional indecisiveness risk-hedging portfolio. Our results touch on new directions for optimal consumption--investment choices in the presence of incomparability and time variation, also signaling potentially testable assumptions on the variability of asset prices. Simulation techniques for set-valued processes are studied for how solved optimal policies can be computed in practice. \vspace{0.1in} \vspace{0.1in} \vspace{0.1in}
\end{abstract}

\newcommand{\dd}{{\rm d}}
\newcommand{\pd}{\partial}
\newcommand{\PP}{\mathbb{P}}
\newcommand{\E}{\mathbb{E}}
\newcommand{\0}{\mathbf{0}}
\newcommand{\1}{\mathbf{1}}
\newcommand{\Int}{\mathrm{int}}
\newcommand{\dom}{\mathrm{dom}}
\newcommand{\im}{\mathrm{im}}
\newcommand{\cl}{\mathrm{cl}}
\newcommand{\co}{\mathrm{co}}
\renewcommand{\Pi}{\varPi}
\renewcommand{\Gamma}{\varGamma}
\renewcommand{\Upsilon}{\varUpsilon}
\renewcommand{\Psi}{\varPsi}

\medskip

\section{Introduction}\label{s:1}

In the present paper we study optimal consumption--investment choices in a diffusion market where an investor (he) has incomplete preferences due to imprecise tastes over risky prospects. Incomplete preferences are represented by a collection of multi-attribute utility functions with parameters that also fluctuate, leading to time variation in preferences. The investor's objective is, broadly speaking, to optimize, striking a balance, a possibly infinite number of conflicting criteria embodying his tastes for distinct attributes, characteristics, and functionalities embedded in the consumption goods, which hinder full comparability and perfect substitution. The multiplicity of the chosen criteria is directly linked to inexact values of preference parameters. By simultaneously maximizing all of these utility functions the investor is looking for a set of optimal consumption policies which are equally efficient in the sense of optimizing the mutual tradeoff among tastes. If the investor relies on consumption financing by contemporaneous investment, the optimal investment policies will form a set as well, whose elements are expected to line with obtained consumption policies. The full characterization of both types of these policies is the main question we are trying to address in this study.

There is a simple illustration of having many optimal actions and the importance of finding them all by way of the classical problem in portfolio management to maximize expected return and (simultaneously) minimize variance (see the summary by \cite[Liu, 2004]{L1}). The inherently conflicting nature of these two criteria hints at a compromise between return and risk and the set of efficient frontier portfolios are viewed as the optimal investment policies for an investor who has mean--variance preferences but is unaware of his precise risk aversion degree. It can be equivalently taken as the result of maximizing a collection of utility functions of the mean--variance type with risk aversion degrees taking all (positive) values, where maximization is in the sense that a solution cannot be further improved without having to either decrease the expected return or increase variance. In more detail, if $\mu$, $\sigma$, and $\Pi$ are the mean vector of risky returns ($R$), the volatility matrix (with $\sigma\sigma^{\intercal}$ being the covariance matrix) of risky returns, and the corresponding vector of portfolio weights, respectively, then the problem can be stated in the following form:
\begin{equation}\label{1.1}
  \sup_{\Pi^{\intercal}\1=1}\bigg\{\Pi^{\intercal}\mu,-\frac{1}{2}\|\sigma^{\intercal}\Pi\|^{2}_{2}\bigg\} =\sup_{\Pi^{\intercal}\1=1}\bigg\{\Pi^{\intercal}\mu-\frac{p}{2}\|\sigma^{\intercal}\Pi\|^{2}_{2}:\;p>0\bigg\},
\end{equation}
with solutions forming the set of efficient frontier portfolios
\begin{equation}\label{1.2}
  \Pi^{\ast}=\bigg\{\frac{(\sigma\sigma^{\intercal})^{-1}\1}{\|\sigma^{-1}\1\|^{2}_{2}} +\frac{(\sigma\sigma^{\intercal})^{-1}}{p}\bigg(\mu-\frac{\1\1^{\intercal}(\sigma\sigma^{\intercal})\mu}{\|\sigma^{-1}\1\|^{2}_{2}}\bigg):\; p\in(0,\infty]\bigg\}.
\end{equation}
The so-called ``efficient portfolio frontier,'' which is the image of the frontier portfolios under the two criteria, or $\{(\|\sigma^{\intercal}\Pi\|^{2}_{2},\Pi^{\intercal}\mu)^{\intercal}:\Pi\in\Pi^{\ast}\}$, gives us full access to asset allocations that optimize the risk--return tradeoff.

In principle, the number of conflicting criteria or different utility functions can increase without bound (\cite[Evren, 2014, \text{Sect.} 3]{E4}). For example, (\ref{1.2}) can be immediately generalized to four dimensions if the investor considers return asymmetry and tail risk, which leads to mean--variance preferences adjusted for higher-order moments (see, e.g., \cite[P\'{e}zier and White, 2008]{PW2} for related concepts). More generally, the infinite dimensionality feature is determined by the structure of utility functions in use, which property will be thoroughly investigated. The significance of characterizing the full set of solutions to a utility maximization problem with incomplete preferences is in understanding the reasonable range within which an investor may vary his optimal actions, including consumption and investment policies. Although he can only implement a single action at a time, disclosing all equally optimal actions gives him opportunities to temporally alter his investment policies, driven by a preference for randomization -- even for the above mean--variance investor, he is not bound to the same portfolio variance weight in allocating wealth.

It is worth noting that characterizing the full solution set of a maximization problem involving many utility functions is not equivalent to solving a usual utility maximization problem with indefinite parameter values.\footnote{In this regard, (\ref{1.1}) happens to be a very special instance, because the imprecise risk aversion serves as a scaling factor of the portfolio variance, or a scale parameter of the underlying criteria.} While there are plenty of solutions that cannot be reconstructed from those to the latter problem, we will seek a unified approach only requiring one-time computation via a multi-valued characterization. This approach is particularly appealing to situations where the investor's preferences are also allowed to fluctuate over time, in ways suggested by empirical evidence. A canonical example is time-varying risk aversion, especially risk aversion driven by market fear (see, e.g., \cite[Guiso et al., 2018, \text{Sect.} 4 and 5]{GSZ}). In the context of the above mean--variance illustration, suppose that returns have stochastic volatility, i.e., $\sigma$ is instead the conditional volatility on a future (relative to the time of estimation) date, and suppose that the first risky asset's returns, $R_{1}$, exhibit strong volatility leverage effect so that the first element, $\sigma_{1,1}$, can be accepted as a measure of fear. Then, his risk aversion degree starts varying with time as, e.g., $p\sigma_{1,1}$, with which the problem (\ref{1.1}) evolves into\footnote{For similar formulations regarding volatility state-contingent risk aversion specifically, see also the recent developments in \cite[Li et al., 2022]{LWZ} and \cite[He and Hong, 2022]{HH}.}
\begin{equation}\label{1.3}
  \sup_{\Pi^{\intercal}\1=1}\bigg\{\Pi^{\intercal}\mu-\frac{p}{2}\E\big[\sigma_{1,1}((\Pi^{\intercal}R)^{2} -(\Pi^{\intercal}\mu)^{2})\big]:\;p>0\bigg\}.
\end{equation}
This no longer corresponds to mean--variance preferences because of the volatility dependence of risk aversion. It is implied in such a situation that the investor is aware that his risk aversion should change with $\sigma_{1,1}$ but not of its exact rate of change.

Time variation will be a key difference between the problem to be studied and classical multi-criteria optimization problems. For similar works concerning multivariate utility functions we highlight \cite[Campi and Owen, 2011]{CO1}, \cite[Benedetti and Campi, 2012]{BC}, and especially \cite[Hamel and Wang, 2017]{HW} and \cite[Rudloff and Ulus, 2021]{RU}, which studied portfolio optimization in a frictional currency market under either multivariate single-valued or univariate multi-valued utility functions. For consumption, reasonable consideration of utility functions with values in general function spaces will lead to an extension of their framework into infinite stochastic dimensions, and the multivariate feature will be particularly important in that substitution at the commodity level generates interchangeable utility across outcome dimensions (the same with currency exchanges). With fluctuating preferences, it turns out that the computational benefits from adopting a multi-valued perspective can be well retained by modifying the set of criteria into a random structure.

In an equilibrium framework, finding the full solution set underlies the exploration of the temporal variability of asset prices, when economic agents deliberately randomize over different consumption--investment policies over time.\footnote{Typical documented reasons for preferential randomization mostly include hedging motives and a desire for diversification across bundles; see, e.g., \cite[Agranov and Ortoleva, 2017]{AO}.} Since psychological preferences are not directly observable from revealed choices, realized variability in asset prices may be ascribed to agents' preferences being fluctuating or incomplete, or to a mutual effect of both. In particular, our approach will allow for (and hence reconciles) both externally driven randomness in preferences, based on strong empirical support from the market, including socialization effects and volatility risk, and static incompleteness in preferences due to involvement of ambiguous prospects both in consumption goods and in the driving externalities,\footnote{We refer to \cite[Agranov and Ortoleva, 2017]{AO} and \cite[Cettolin and Riedl, 2019]{CR1} for a detailed analysis of these two explanations and their relative importance to revealed choice randomness.} which are expected to lead to testable assumptions related to dynamic incompleteness in psychological preferences. Comprehensibly, this task could not be accomplished without taking the first step to deeply understand the individual investor's optimization problem.

In relatively technical terms, the present paper will focus on the following four main contributions:

\begin{itemize}
  \item{Construct a set-valued stochastic process that encompasses arbitrary patterns of parametric changes in incomplete preferences.}
  \item{Provide a formulation for multi-utility maximization involving consumption and investment in continuous time, which gives rise to a new multi-stochastic criteria optimization problem.}
  \item{Refine scalarization techniques to account for both infiniteness and randomness in dimensions for a complete characterization of optimal consumption policies and propose a novel stochastic geometry-based method to identify corresponding optimal investment policies.}
  \item{Characterize the composition of optimal consumption--investment policies according to evidenced psychological effects.}
\end{itemize}

\subsection{Literature review}\label{ss:1.1}

Characterizations of incomplete preferences to extend classical Debreu's utility representation theorem have led to representations by multifunctions, with the appellation ``multi-utility,'' as formalized in \cite[Ok, 2002]{O} and \cite[Evren and Ok, 2011]{EO}. In expected utility theory, the first attempt to do away with the completeness axiom was made by \cite[Aumann, 1962]{A1}, while the very notion of expected multi-utility representation was first given in \cite[Dubra et al., 2004]{DMO}, and in \cite[Evren, 2014]{E4} an expected multi-utility representation theorem was proposed to characterize the class of reflexive and transitive preference relations in terms of a compact set of continuous expected utility functions.

Noteworthily, these (expected) multi-utility representations are expressly proposed for incomplete preferences arising from imprecise tastes, which are to be distinguished from beliefs. Notably, preference incompleteness can also be realized by imprecise beliefs (see, e.g., \cite[Rigotti and Shannon, 2005]{RS}, \cite[Kelsey and Yalcin, 2007]{KY}, and also \cite[Galaabaatar and Karni, 2012]{GK}), due to the inability to assess outcome possibilities with precision. This second source of preference incompleteness, corresponding to beliefs being imprecise, is typically captured by a Knightian uncertainty model as pioneered in \cite[Bewley, 2002]{B1}. One may refer to \cite[Nau, 2006]{N1} and \cite[Ok et al., 2012]{OOR} for a thorough discussion of the similarities and differences between these two sources of preference incompleteness. Plainly speaking, Knightian uncertainty speaks to a shortage of quantifiable knowledge about certain aspects of the market, such as those embedded into model parameters, while imprecise tastes are deep-rooted in the investor's preferences and can be fundamentally quantified, e.g., by specified multi-utility.

The formulations of expected utility subject to these two sources also turn out to be organically different -- e.g., if beliefs were to be imprecise the set representing multiple utility functions would have to be replaced by a pool of equivalent probability measures alongside a single exogenously specified utility function. In this paper, our focus will be on the first source, imprecise tastes, for the following reason.

From a comparative viewpoint, under imprecise beliefs (namely Knightian uncertainty), consumption--investment choice problems have been studied extensively so far (albeit in Markovian settings) via the so-called ``robust utility maximization,'' assuming the investor to be ambiguous about certain aspects of a presumed market model. We highlight, among others, \cite[Fouque et al., 2016]{FPW}, \cite[Biagini and P{\i}nar, 2017]{BP}, and \cite[Liang and Ma, 2020]{LM1}, which have considered, in terms of model complexity, up to jump--diffusion processes and ambiguity about L\'{e}vy characteristics. Via a procedure of robustification, these problems are usually solved in a ``worst-case'' scenario based on the assumption of ambiguity aversion stemming from pessimism towards model risks (\cite[Maccheroni et al., 2013]{MMR}), i.e., risks attached to the pool of probability measures. Relevant to incomplete preferences, robustification is built on the concept of outcome regret (\cite[Boutilier et al., 2006]{BPPS}) requiring that incompleteness come down to inadequate preference information.

Nevertheless, there seems to be no formal investigation of consumption--investment choices under the first source of preference incompleteness -- imprecise tastes, which deal with indecisive choice behavior. As pointed out in \cite[Ok et al., 2012]{OOR}, taste imprecision had rarely been studied in an environment with uncertainty in the literature, most likely due to challenges in quantifying patterns of indecisiveness and behavioral changes. Despite this, imprecise tastes have been proven to prevail in preferences, especially over multi-attribute alternatives, and there has been abundant experimental evidence to show preference incompleteness not coming from information shortage or outcome regret; we highlight \cite[Danan and Ziegelmeyer, 2006]{DP}, \cite[Deparis et al., 2012]{DMOPH}, \cite[Agranov and Ortoleva, 2017]{AO}, \cite[Sautua, 2017]{S1}, and \cite[Cettolin and Riedl, 2019]{CR1} relying on experimental design to test preference incompleteness from revealed choice behavior. Notably, aversion towards imprecision is out of question -- flexibility in preferences is anything but detrimental. On the contrary, the investor will have established grounds for deliberately randomizing over optimal actions in consumption (and investment), abiding by some (set) comparison rule.

The study of imprecise tastes is in no conflict with that of imprecise beliefs. Instead, it allows one to combine the two sources of preference incompleteness to cater for the general formulation of utility shapes in \cite[Nau, 2006]{N1} and contemplate incomplete preferences along with possibly misspecified market models. Some detailed directions will be outlined at the end of this paper.

\subsection{Time-varying incomplete preferences}\label{ss:1.2}

As already mentioned, in large part the novelty of our study comes from combining time variation with imprecise tastes. In a flow of uncertainty, we will inspect mechanisms by which the parametrization of a multivariate utility function permits random changes, as each parameter has its own temporal--material significance. First, regardless of the incompleteness, the investor's preferences can undergo temporal changes via subjective stochastic discounting (see, e.g., \cite[Roelofsma and Read, 2000]{RR3}), as linked to time-varying patience. Then, with incomplete preferences, the space of taste imprecision is also allowed to evolve in time, pointing to other mechanisms apart from patience by which incomplete preferences change.

By and large, introducing time variation renders the incomplete preferences being considered even temporally intransitive, relating them to the general concept of time preferences as well (studied in \cite[Ok and Masatlioglu, 2007]{OM} and \cite[Dubra, 2009]{D}), whereas it also allows for intransitivity on the material level (see \text{e.g.} the discussion in \cite[Mandler, 2005]{M2}). More specifically, it is likely that preferences become intransitive under the mutual influence of a loss of patience and a shift of tastes, when the time delay alone is insufficient to trigger intransitivity. Clearly, such joint effects cannot be adequately modeled through a single channel of fluctuations or incompleteness in preferences.

To elucidate the nature of the consumption--investment problems to be explored, let us take a sneak peek at three illustrative examples by which we will demonstrate the general method later on. To keep things simple, there are assumed to be only two goods in the market.

In \textbf{Example 1}, the investor has time-invariant incomplete preferences. Of particular interest are three cases. In Case (I), the investor thinks that the two goods are totally distinct and assesses their values independently. He wishes to benefit as much as possible from consuming the two goods simultaneously; hence, his utility maximization problem simply involves two criteria consisting of univariate utility functions imposed separately on the two goods.\footnote{If consumption levels are taken as positions held in assets valued in different currencies, this scenario could be compared to the settings in \cite[Hamel and Wang, 2017]{HW} and \cite[Rudloff and Ulus, 2021, \text{Ex.} 6.7]{RU}, up to a finite number of dimensions, where it is assumed that utility is noninterchangeable across assets.}  In Case (II), the investor is biased towards the second good, perceiving it to be universally more essential to him than the first; this situation oftentimes arises when the first good contains luxuries and the second contains necessities, or when the two goods are produced, respectively, overseas and domestically.\footnote{This phenomenon is documented as the ``consumption home bias,'' for which there is abundant regional evidence -- \cite[Coeurdacier and Rey, 2012]{CR2}; also see the recent survey \cite[Gaar et al., 2022]{GSS}.} Specifically, the second good's perceived importance is such that a bundle superior to another cannot give a smaller quantity of the second good, whereas the investor is always ready to forego some of the first good for a larger quantity of the second. The latter aspect can be well modeled by using a consumption-additive utility function showing a measurement of attention to the first good (see \cite[\c{C}anako\u{g}lu and \"{O}zekici, 2012]{CO2} and \cite[Wu et al., 2018]{WWZ}), which acts as a scaling factor of the first good's marginal rate of substitution relative to the second good.\footnote{With a domestic currency and a foreign currency, it would then imply that the latter carries such inescapable foreign exchange risks that the investor essentially assigns more weights to assets valued in the domestic currency. The idea is to allow interaction among utility across assets as each utility element can be a multivariate utility function (compare the settings in \cite[Campi and Owen, 2011]{CO1} and \cite[Rudloff and Ulus, 2021, \text{Sect.} 6.2.1]{RU} with \cite[Rudloff and Ulus, 2021, \text{Sect.} 6.2.2]{RU}).} As a result, the utility maximization problem involves a continuum of criteria each with a different attention degree, and it is exactly the range of attention that reflects the imprecision of the perceived importance. It is suspected that in such a case the investor can simply focus on the ``boundary'' criteria that govern all utility functions, which is to be justified afterwards. In Case (III), the investor knows that the two goods are adequate substitutes having common attributes, with a parsimonious assumption that the two goods be mutually utility-independent. However, the investor is not aware of the precise extent to which one good can be substituted by the other. Likewise, there is understandably a continuum of criteria in the resultant utility maximization problem, which can be reduced in dimensionality into only two ``boundary'' criteria. These simple settings show that the investor readily faces a multi-criteria optimization problem whenever he is unable to perfectly compare consumption bundles and has imprecise evaluation of various aspects. The essence is to understand the available optimal actions, including consumption and investment policies, that he may implement in a self-financing portfolio.

\textbf{Example 2} and \textbf{Example 3} progress to inject time variation into the established incomplete preferences. Based on Case (II) of Example 1, in Example 2 the investor further expects increasing attention degrees to the first good as a result of socialization effects, classifiable as external habit formation. In the case of a luxury good, driving forces include status consumption, security risks, and technological innovations (see \cite[Jansen and Jager, 2001]{JJ} and \cite[Mrad et al., 2020]{MMCE} for support from experimental studies and market research), while in the case of a foreign good, it may be directly attributed to foreign exchange rate increases or optimism about foreign production technologies; see \cite[Levy and Levy, 2014]{LL1} for empirical evidence concerning the \text{U.S.} market.\footnote{Such a mechanism is also connected to multi-currency investments for one who overweighs assets of multinational corporations due to estimated appreciation of a foreign currency; empirical analysis of the time-varying importance of exchange rate exposure has been done in \cite[Doidge et al., 2006]{DGW}.} As the perceived relative importance of the first good shifts, time variation develops in preferences and utility maximization will involve a random continuum of criteria, which can still be verified to be reducible to only two criteria. In Example 3, the investor also has imprecise degrees of risk preferences directly related to the market volatility, in catering for fear-driven risk aversion with strong empirical evidence (\cite[Loomes and Pogrebna, 2014]{LP} and \cite[Guiso et al., 2018]{GSZ}). The added layers of complexity reflect the investor's sophistication in preferences, and he will be maximizing a random, irreducible continuum of criteria, with both time-varying attention and time-varying risk aversion. For an investor with these dynamic preference shapes, many equally optimal consumption--investment policies are expected, and the decomposition of the investment policies should also reveal a demand for hedging additional risks emerging from his indecisiveness.

As we have by now noticed, the formation of the set of multiple criteria to be optimized is fundamentally connected to the ranges of preference parameters, as soon as utility functions can be used. These parametric ranges are generally uncountable, with a continuum cardinality, for infiniteness is the natural, and yet a realistic, situation if only ranges or bounds of preference parameters are precise (see, again, \cite[Evren, 2014, \text{Sect.} 3]{E4}).

When there is time variation, the investor's perceptions about how those preference parameters change with external factors appear in the set of criteria, which will dynamically confine these parameters to certain value ranges. From his perspective, such value ranges at a future date cannot be established beforehand but are outcomes of preference fluctuations, i.e., the investor does not choose certain dynamics of preference parameters in order to match a value range; contrarily, parametric dynamics are specified based on perceived relationships with the external factors, whose consequence is a time-varying value range to describe the representing multi-utility family for incomplete preferences. Clearly, parametric dynamics back-engineered from value ranges are not expected to be unique. With the external factors involving risks, a suitable model for the representing multi-utility family at a future date is naturally a random variable with set values, or a set-valued stochastic process with the passage of time. This set-valued stochastic process should take values in the collection of closed convex subsets of some known global parameter space and its selectors, namely those belonging to the process in all states of nature, correspond to the arbitrary patterns of random changes in preferences. We will give a detailed description of the preference formation process in Section \ref{s:2}.

With the foregoing aspects in mind, we invoke set-valued stochastic calculus that has been elaborated by \cite[Zhang et al., 2009]{ZLMO}, \cite[Li et al., 2010]{LLL}, \cite[Kisielewicz, 2012]{K2}, and more comprehensively in \cite[Kisielewicz, 2020]{K3} from a seemingly separate field, in order to formulate a possibly infinite-dimensional, non-separable multi-utility maximization problem under time-varying incomplete preferences, also allowing for interactions at the consumption good level. This comes in stark contrast to the quoted works specifically studying multi-utility maximization problems, which have focused on vector-valued optimization involving a finite number of fixed criteria.

On the implementation side, our methodology gives rise to results that are compatible with classical scalarization techniques, which serve to turn multi-criteria optimization problems into single-criterion ones, albeit subject to significant modifications to allow for infinite stochastic dimensions (or numbers of criteria). To briefly explain, the investor can employ an indefinite floating totaling rule, known as a ``scalarization functional,'' on the multiple criteria that arise from imprecise tastes and, by varying this rule in its own to-be-identified value range, exhaust the set of optimal actions. For example, in the mean--variance frontier illustration, the totaling rule can simply be a vector containing two (possibly different) risk aversion degrees, i.e., (\ref{1.1}) is transformable into the following collection of problems:
\begin{equation*}
  \sup_{\Pi^{\intercal}\1=1}\big(w_{1}\Pi^{\intercal}\mu-w_{2}\|\sigma^{\intercal}\Pi\|^{2}_{2}\big),\quad w\equiv(w_{1},w_{2})^{\intercal}\in\mathds{R}^{2}_{+},
\end{equation*}
by solving which the exact same efficient frontier is recovered. Moreover, a Fenchel-type duality approach is available along the lines of \cite[Hamel et al., 2015]{HHLRS}, with which method the investor transforms the given criteria into lattice structures via set relations. More insight will be provided in Sections \ref{s:3} through Section \ref{s:5}.

\subsection{Structure of paper}\label{ss:1.3}

The remainder of this paper is structured as follows. In Section \ref{s:2} we set up the market and write down a formal multi-utility maximization problem ((\ref{2.3.2}) and (\ref{2.3.3})) involving consumption and bequest, assuming indecisiveness in tastes. Some classical results on multi-utility representations are reviewed while we lay out the skeleton of time-varying preferences. In Section \ref{s:3} we discuss solution by way of scalarization. By convention, we prove equivalence (Theorem \ref{thm:2}) between the original problem and the scalarized problem under suitable conditions and discuss how to apply them sparingly. The three examples mentioned above are analyzed in detail to demonstrate the applicability of these methods under various psychological effects. Section \ref{s:4} considers duality results (Theorem \ref{thm:3}) for the same problem. We subsequently derive in Section \ref{s:5} a general formula (Theorem \ref{thm:4}) for the corresponding optimal investment policy, based on what scalarization has found as optimal consumption, which boils down to a simpler formula (Corollary \ref{cor:1}) upon consumption-additivity assumptions. To numerically solve the general problem, we advance several extant results on the simulation of set-valued stochastic processes in Section \ref{s:6}. Conclusions are drawn in Section \ref{s:7}, including rules for implementing the optimal policies in practice, along with future research directions. All mathematical proofs are presented in \ref{A}, and \ref{B} briefly reviews the definitions and key properties of set-valued stochastic processes. For readers' convenience, \ref{C} also provides a list of advanced mathematical symbols.

\vspace{0.2in}

\section{Problem formulation}\label{s:2}

We begin by considering a complete filtered probability space $(\Omega,\mathcal{F},\PP;\mathbb{F}\equiv\{\mathscr{F}_{t}\}_{t\in[0,T]})$ over a finite time horizon, on which an $m$-dimensional Brownian motion $W$ is defined. We take $\mathbb{F}$ to be the augmented natural filtration of $W$ and all stochastic processes to be $\mathbb{F}$-non-anticipating, unless otherwise specified, with the understanding that $\mathcal{F}=\mathscr{F}_{T}$.

\subsection{Market setup}\label{ss:2.1}

The probability space supports a complete financial market consisting of one risk-free asset with a bounded short rate process $r\equiv(r_{t})_{t\in[0,T]}>0$ and $m\geq1$ risky assets that generate a vector-valued price process solving the stochastic differential equation (SDE)
\begin{equation}\label{2.1.1}
  S_{t}+\int^{t}_{0}D_{s}\dd s=S_{0}+\int^{t}_{0}\mathrm{diag}(S_{s})\mu_{s}\dd s+\int^{t}_{0}\mathrm{diag}(S_{s})\sigma_{s}\dd W_{s},\quad t\in[0,T],
\end{equation}
with given initial price vector $S_{0}\in\mathds{R}^{m}_{++}$, an integrable $m$-vector-valued dividend process $D\equiv(D_{t})_{t\in[0,T]}$ with all nonnegative components, an integrable $m$-vector-valued drift coefficient process $\mu\equiv(\mu_{t})_{t\in[0,T]}$, and a square-integrable invertible ($m\otimes m$)-matrix-valued volatility coefficient process $\sigma\equiv(\sigma_{t})_{t\in[0,T]}$, and which, despite a general diffusion, need not be a Markov process. Besides, the market offers a total of $n\geq2$ distinct consumption goods.

The choice of the number $n$ of consumption goods is fairly flexible with classifications. It can be fundamentally determined based on domesticity (as explained in Subsection \ref{ss:1.2}), durability, tangibility, or individual buying patterns. With $n=2$ a rough classification would lead to domestic versus foreign goods, durable versus nondurable goods, or tangible versus intangible goods; with $n=2,3$, it would also include, on the habit level, convenience goods, shopping goods, and specialty goods. All of these categories can be further subdivided, and for a much more detailed classification of consumption goods, we refer to the classical paper \cite[Bucklin, 1963]{B3}, and in the meantime wish to be as general as possible in the problem formulation. Let $P\equiv(P_{t})_{t\in[0,T]}$ denote the vector of the $n$ commodity prices, which evolve according to the SDE
\begin{equation}\label{2.1.2}
  P_{t}=P_{0}+\int^{t}_{0}\mathrm{diag}(P_{s})\mu_{P,s}\dd s+\int^{t}_{0}\mathrm{diag}(P_{s})\sigma_{P,s}\dd W_{s},\quad t\in[0,T],
\end{equation}
where, similarly, $P_{0}\in\mathds{R}^{n}_{++}$ is the initial price vector and $\mu_{P}\equiv(\mu_{P,t})_{t\in[0,T]}$ and $\sigma_{P}\equiv(\sigma_{P,t})_{t\in[0,T]}$ are drift coefficient and volatility coefficient processes valued in $\mathds{R}^{n}$ and $\mathds{R}^{n\otimes m}$, respectively.

\subsection{Construction of time-varying incomplete preferences}\label{ss:2.2}

The investor has imprecise tastes among certain bundles (or combinations) of available consumption goods. The incomplete preference relation $\succeq$ embodies that bundles cannot always be ranked. We present the concept of a multi-utility function that represents such a preference relation thanks to \cite[Evren and Ok, 2011]{EO}.

\begin{definition}\label{def:1}
Let $\mathcal{I}$ be a nonempty closed convex subset of the Euclidean space $\mathds{R}^{d}$ with $d\in\mathds{N}_{++}$ and define a set of utility elements $\{u_{i}:i\in\mathcal{I}\}$. Suppose that $c\succeq c'$ if and only if the utility elements $u_{i}(c)\geq u_{i}(c')$ for every $i\in\mathcal{I}$; then $\mathcal{I}$ is referred to as a (multi-utility representation) index set for the preference relation $\succeq$.
\end{definition}

The index set $\mathcal{I}$ serves the purpose of easy labeling of the utility elements by $d$ different parameters. It has been chosen to have cardinality at most that of a continuum $(\mathfrak{c})$ to keep with the space of real continuous functions.\footnote{This means that defining a multi-utility function by a set of parameter values is the same as doing so by a set of utility elements (compare the original definition in \cite[Evren and Ok, 2011, \text{Sect.} 2]{EO}) and sets with strictly greater cardinalities, such as (general subsets of) the function spaces $2^{\mathds{R}}$ and $\mathds{R}^{\mathds{R}}$, are avoidable for practical purposes.} It generalizes any interval in which a single parameter is valued into higher dimensions for multiple parameters. To formalize its structure, let us denote the $d$ parameters by $i_{k}$, $k\in\mathds{N}\cap[1,d]$, with respective global ranges $R_{k}$'s which are all closed convex subsets of $\mathds{R}$. Then, the index set $\mathcal{I}$ should have values in the space $\mathrm{Cl}\big(\prod^{d}_{k=1}R_{k}\big)$ of nonempty closed convex subsets of the (Cartesian) parametric product space.

As noted earlier, with time-varying incomplete preferences, the investor's multi-utility can evolve in a material fashion in addition to entirely temporal influence.\footnote{The modeling idea can be directly compared to the changing ``confidence sets'' of model parameters proposed in \cite[Liang and Ma, 2020]{LM1} when constructing deterministically varying imprecise beliefs. In this paper, such variation can be stochastic, whose evolution is also made explicit.} This makes the index set $\mathcal{I}\equiv(\mathcal{I}_{t})_{t\in[0,T]}$ ideally a stochastic process, whose design is supposed to encompass flexible connections between the evolution of the market with (\ref{2.1.1}) and (\ref{2.1.2}) and that of a $d$-parameter multi-attribute utility function.\footnote{As an example, an $n$-variate CRRA-type utility function can be characterized by up to $d=2n$ parameters, including attention degrees. When market equilibria are considered, a useful implication from a stochastic index set is that imprecise tastes can be established over empirically evidenced confidence intervals of these (utility) parameters as well (e.g., \cite[Eisenhauer and Ventura, 2003]{EV}), subject to certain mechanisms of temporal evolution.} At first, since these connections are all exogenous, they can be roughly classified into two types -- without sophistication and with sophistication, the latter being in the sense that certain monotonicity constraints are imposed by the investor on how his indecisiveness in tastes varies with time, in favor of existing knowledge.

On the one hand, there are apparent linkages between changes in preferences and market characteristics, which are not globally monotone. A famous example is risk preference degrees changing with market volatility, including volatility in the commodity prices. The idea is that when facing higher market volatility the investor has a tendency to become more averse to market risk. While such variation is immediately related to dynamic volatility risk premia (e.g., \cite[Bollerslev et al., 2011]{BGZ}) or market fear (again, see \cite[Guiso et al., 2018]{GSZ} for recent evidence), psychologically it is rooted in a desire for secureness (refer to \cite[Frost and Shows, 1993]{FS} and \cite[Patalano and Wengrowitz, 2007]{PW1}). Consequently, risk aversion coefficients can fluctuate as certain functionals of the volatility matrix processes $\sigma$ and $\sigma_{P}$\footnote{To our knowledge, \cite[Bekaert et al., 2009]{BEX} was the first to consider exogenously fluctuating risk aversion, in a discrete-time setting, which assumed that risk aversion coefficients vary due to a ``consumption surplus ratio.''} which are also likely to differ across consumption goods depending on their functionalities.

On the other hand, with sophistication, preference changes are limited to monotone evolution, in which situation the investor's indecisiveness degree either increases or decreases. There are a variety of psychological effects supporting this type of perception. For instance, stochastic patience is constantly increasing or decreasing over time, both of which patterns are equally conspicuous in reality (\cite[Read and Roelofsma, 2003, Experiments]{RR1}). Another intriguing aspect is when the investor is aware beforehand that consumption of certain (e.g., luxury-linked) goods, subject to socialization effects (\cite[Elias, 1982]{E3}, \cite[Janssen and Jager, 2001]{JJ}, and \cite[Mrad et al., 2020]{MMCE}) affecting perceived valuation, tends to increase their (state-dependent) degrees of attention (mentioning \cite[\c{C}anako\u{g}lu and \"{O}zekici, 2012]{CO2} and \cite[Wu et al., 2018]{WWZ} again), triggering temporal increases in his indecisiveness degree; meanwhile, the indecisiveness degree is expected to decrease among other obsolete goods that are inversely related to these effects. Much the same can be said about the consumption of foreign (relative to domestic) goods amid reduced investment costs and attenuating informational barriers in expectation (\cite[Levy and Levy, 2014]{LL1}). Differently, when attention degrees change with respect to the commodity prices ($P$), they will generate non-set-monotone preference changes that do not reflect sophistication in the stated sense.

As brought up in Section \ref{s:1}, to describe the evolution of time-varying incomplete preferences, it is necessary to consider three aspects: (i) static incompleteness, (ii) time variation in preferences, and (iii) their interactions. First, if preferences are initially incomplete but do not fluctuate, the multi-utility index set $\mathcal{I}$ is just a constant set of parameter values. With this choice there are simply no preference changes with external factors. Second, if incomplete preferences are driven by a single external factor, then $\mathcal{I}$ becomes a set with fixed shape and capacity being repositioned by fluctuations in that external factor. Third, if incomplete preferences can vary with numerous factors, then it can either be that dependence on any factor is universal in time, i.e., only on one factor through the course of investment, regardless of market conditions, or that factor dependence can also vary with time and market conditions. In both these situations, the investor is not fully aware of how preference changes depend on which factors (and in what states), thus having time-varying incomplete preferences, for which set-valued stochastic processes must be used. In the second, clearly more general situation, the index set $\mathcal{I}$ is a constant parameter set subject to movements driven by chosen market factors, their linear combinations, and all state-contingent combinations.

To expand the above descriptions, let us recall the mean--variance frontier illustration. With no volatility dependence, the imprecise risk aversion degree $p$ is valued as some fixed interval $I_{0}\subseteq\mathds{R}_{+}$. If risk aversion should fluctuate with the instantaneous volatility (e.g., $\sigma_{1,1}\equiv(\sigma_{1,1,t})$) of a risky asset, then its value is the sum (in the obvious sense) of a constant set and a single-valued stochastic process, i.e., $I_{0}+\{p\sigma_{1,1}\}$. If the risk aversion degree can depend on any weighted average of the instantaneous volatility and the scaled CBOE Volatility Index (a standard market fear gauge), the risk aversion degree becomes a set-valued stochastic process $I$ with values in $\mathrm{Cl}(\mathds{R})$, which permits state-contingent combinations of the two factors if and only if the decomposability property is satisfied (see \ref{B} for its detailed definition); a simple example is risk aversion depending on the maximum of the instantaneous volatility and the scaled Volatility Index, which is a nonlinear decomposable combination -- namely $\sigma_{1,1}\vee\text{VIX}=\sigma_{1,1}\mathds{1}_{[\text{VIX},\infty)}(\sigma_{1,1})+\text{VIX}\mathds{1}_{(0,\text{VIX}]}(\sigma_{1,1})$. The last situation hints at a lack of a precise measure of fear (also see the discussion in \cite[Guiso et al, 2018, \text{Sect.} 5]{GSS}), in the sense that indecisiveness is also reflected in the choice of external factors across different states and times.

With market characteristics conforming to a variety of stylized facts, there is no uniform yardstick for the patterns of preference changes with external factors for the choice of models. For example, pertinent to fear-driven risk aversion, mean reversion is most likely required for the volatility process $\sigma$ under normal circumstances (e.g., \cite[Fouque et al., 2000]{FPS}); for patience, subjective discount rates typically evolve as a uniformly bounded process (e.g., \cite[Luttmer and Mariotti, 2003, Sect. II.B]{LM2}), and attention degrees across consumption goods can well have an unconditionally positive growth rate (e.g., \cite[\c{C}anako\u{g}lu and \"{O}zekici, 2012, \text{Sect.} 4]{CO2}). With the Brownian motion $W$ being the sole source of randomness, it is standard to model the dynamics of external factors using classical It\^{o} processes, which become set-valued It\^{o} processes when modeling joint dynamics under incomplete preferences, according to the foregoing descriptions. Furthermore, if sophistication implies monotone preference changes, the simplest relevant operations are to take the intersection or union in the time variable. Doing so signifies that temporally decreasing or increasing indecisiveness is determined by the temporal ranges of external factors, and with this additional layer of flexibility, monotonicity in preference changes does not imply monotonicity in the driving external factors -- instead of parallel co-movements, they may exhibit delayed reaction or stickiness. For example, an increasing attention degree (or correspondingly increasing indecisiveness) can result from a not strictly monotone but overall upward-trending price of the reference good, direct investment costs (see again \cite[Levy and Levy, 2014]{LL1}), or many other macro-finance state variables.

As a means to aggregate effects from different channels of indecisiveness changes, we propose to perform (in $\mathds{R}^{d}$) the Minkowski addition. Beside most simple properties of (single-valued) addition such as commutativity, associativity, and distributivity, and preservation of set convexity, it is always possible to isolate channels of indecisiveness changes from one another given their Minkowski sum, because the $d$ preference parameters have been collocated into $d$ ordered positions of the parameter vector (belonging to $\mathcal{I}$) and the Minkowski addition represents element-wise vector addition in the sets being operated on.

Lastly, it is crucial to ensure that the index set $\mathcal{I}$ is concentrated on meaningful parameter values, for which the utility elements are well-defined. The precise idea is then to treat the index set as an aggregation of transformed set-valued It\^{o} process restricted to the global product space of the $d$ parameters, namely $\mathcal{R}:=\prod^{d}_{k=1}R_{k}$.

\begin{figure}[H]
  \centering
  \includegraphics[scale=0.21]{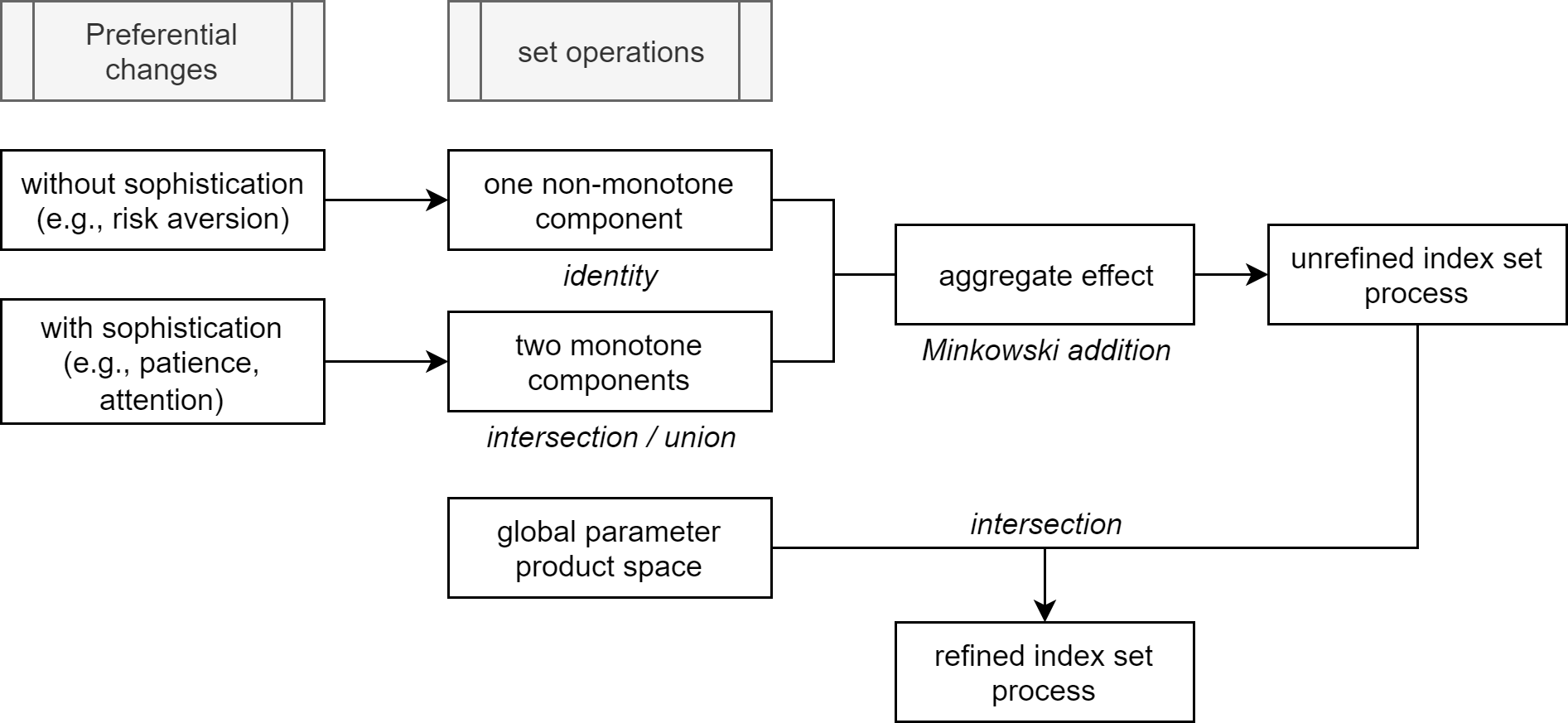}\\
  \caption{Construction of multi-utility index set process}
  \label{fig:1}
\end{figure}

The above procedures to construct a general multi-utility index set process are summarized by the diagram in Figure \ref{fig:1}. In mathematical language, it translates into the following dynamics:\footnote{In risk management, (\ref{2.2.1}) could also be contrasted with the propagation of set-valued risk preferences, which were first modeled by \cite[Ararat and Feinstein, 2021]{AF} using (backward) set-valued stochastic difference equations in discrete time (see also \cite[Ararat et al., 2023]{AMW} for a continuous-time setting). Differently, in our setup the dynamics is imposed on the number $d$ of (characterizing) parameters rather than those ($m$ \text{resp.} $n$) of risky assets and consumption goods, being very rich when it comes to utility design.}
\begin{equation}\label{2.2.1}
  \mathcal{I}_{t}(\omega):=\mathcal{R}\cap\cl_{\mathds{R}^{d}}\Bigg(I_{1,t}(\omega) +\bigcap_{s\in[0,t\wedge\mathcal{t}(\omega)]}I_{2,s}(\omega)+\overline{\co}_{\mathds{R}^{d}}\bigcup_{s\in[0,t]}I_{3,s}(\omega)\Bigg), \quad(t,\omega)\in[0,T]\times\Omega,
\end{equation}
where $\mathcal{R}$ is recalled to be the global product space of the $d$ parameters and
\begin{equation}\label{2.2.2}
  \mathcal{t}(\omega):=\inf\Bigg\{t\in[0,T]:\mathrm{card}\bigcap_{s\in[0,t]}I_{2,s}(\omega)=1\Bigg\},\quad\omega\in\Omega
\end{equation}
is the first time at which the intersection yields a singleton. The intersection with $\mathcal{R}$ is controlled to be nonempty. For each $q\in\{1,2,3\}$, $I_{q}$ is a general set-valued It\^{o} process that can be written as the (Minkowski) sum of an Aumann stochastic integral\footnote{Set-valued integrals of this type were first studied by \cite[Aumann, 1965]{A2}, hence the name.} and a set-valued It\^{o} integral,
\begin{equation}\label{2.2.3}
  I_{q,t}=\cl_{\mathds{L}^{1}}\bigg(I_{q,0}+\int^{t}_{0}f_{q,s}\dd s+\int^{t}_{0}\overline{\co}_{\mathds{L}^{2}}G_{q,s}\dd W_{s}\bigg),\quad t\in[0,T].
\end{equation}
Here, $I_{q,0}$ is an $\mathscr{F}_{0}$-measurable nonempty compact convex subset of $\mathds{R}^{d}$, $f_{q}:[0,T]\times\Omega\mapsto\mathrm{Cl}(\mathds{R}^{d})$ is a compact convex set-valued stochastic process, and $G_{q}:=\{(g_{q,k}:[0,T]\times\Omega\mapsto\mathds{R}^{d\otimes m}):k\in\mathds{N}_{++}\}$ is a family of continuous ($d\otimes m$)-dimensional processes, satisfying that
\begin{align}\label{2.2.4}
  &\sup_{t\in[0,T]}\E[\mathcal{d}_{\rm H}(f_{1,t},\{\0\})]+\E\Big[\sup_{t\in[0,T]}\mathcal{d}_{\rm H}(f_{3,t},\{\0\})\Big]<\infty, \nonumber\\
  &\sum^{\infty}_{k=1}\sup_{t\in[0,T]}\E\big[\|g_{1,k,t}\|^{2}_{\mathrm{F}}\big]+\sum^{\infty}_{k=1}\E\Big[\sup_{t\in[0,T]} \|g_{3,k,t}\|^{2}_{\mathrm{F}}\Big]<\infty,
\end{align}
where $\mathcal{d}_{\rm H}$ stands for the Hausdorff distance (measured in $\mathds{R}^{d}$) and $\|\;\|_{\mathrm{F}}$ the Frobenius norm. As indicated, in (\ref{2.2.3}), the inner convex closure is understood in $\mathds{L}^{2}_{\mathbb{F}}([0,T]\times\Omega;\mathds{R}^{d\otimes m})$ (the spaces of all $\mathbb{F}$-non-anticipating square-integrable processes) while the outer closure and the Minkowski sums are taken in $\mathds{L}^{1}_{\mathscr{F}_{t}}(\Omega;\mathds{R}^{d})$ and $\mathds{L}^{1}_{\mathbb{F}}([0,T]\times\Omega;\mathds{R}^{d})$, respectively. Some important properties of set-valued It\^{o} processes along with their economic intuition are summarized in \ref{B}.

Regarding the structure of (\ref{2.2.1}), the random time $\mathcal{t}$ and the operator $\overline{\co}$ (taken in $\mathds{R}^{d}$) are rather functional -- the former precludes emptiness of the index set after intersecting, and the latter ensures closed-ness and convexity after taking union. In a broad sense, these properties are technically beneficial and standard for multi-criteria optimization.

Observably, there are three subcomponents in each set-valued It\^{o} process $I_{q}$. The a-priori known set $I_{q,0}$ marks the initial range of imprecise parameters, namely the investor's space of imprecise tastes before preferences start to change. The Aumann integral and the set-valued It\^{o} integral, which play the same role as a Lebesgue stochastic integral and a usual It\^{o} integral, respectively, in a single-valued It\^{o} process,\footnote{The reason to impose a closed convex hull (in $\mathds{L}^{2}$) on the It\^{o} integrand as a countable collection of integrable processes is rather technical: In the plainest of words, It\^{o} integrals built from a truly set-valued process are generally not integrably bounded, thus obstructing applications; for details we refer to \cite[Michta, 2015]{M5} and also \ref{B}.} capture long-term persistence and short-term noises in his preference changes over time through the $d$ parameters. As discussed before, with preference changes linked to the market coefficients (\ref{2.1.1}) and (\ref{2.1.2}) taking the form of generic It\^{o} processes, none of these subcomponents is superfluous. The general set-valued nature of the Aumann and It\^{o} integrands $f_{q}$'s and $G_{q}$'s gives the possibility of having varying indecisiveness degrees, beside fundamental changes in any external factor, and allows to achieve time-varying spans in imprecise tastes as well. In its original form, the structure (\ref{2.2.1}) gives the desired flexibility to analyze time-varying incomplete preferences comprising the aforementioned three aspects including static incompleteness, time variation, and their interactions. In the special case where $f_{q}\equiv\{\0\}$ and $G_{q}\equiv\{\0\}$ for all $q$ the standard setting (as adopted in \cite[Hamel and Wang, 2017]{HW} and \cite[Rudloff and Ulus, 2021]{RU}) of time-invariant incomplete preferences is restored.

The structure (\ref{2.2.1}) has some limitations despite the stated generality. First, it does not bring in additional subjective randomness originating from the investor, i.e., preference changes are solely driven by the Brownian motion $W$. However, this presumption is in keeping with individual rationality and there is no substantial benefit in including independent randomness only for preferences, as $W$ may well be unobserved. Second, (\ref{2.2.1}) does not permit preference changes driven by endogenous habits (controllable by the investor). Clearly, establishment of endogenous habits would require a different mechanism of indecisiveness change -- the mere-exposure effect (\cite[Bornstein, 1989]{B2}, also in \cite[Janssen and Jager, 2001]{JJ}), which will have to presuppose how the index set $\mathcal{I}$ evolves as a multifunction of cumulative consumption and result in another type of optimization problem. This subject is definitely worth a further investigation and will be addressed in a separate paper.

The following proposition is about some technical properties of the index set process $\mathcal{I}$.

\begin{proposition}\label{pro:1}
The set-valued process $\mathcal{I}$ is $\mathbb{F}$-non-anticipating, integrably bounded, and continuous $\PP$-a.s.
\end{proposition}

Going forward we shall adopt a formal representation of the investor's incomplete preference relation by a collection of multivariate time-varying utility functions under the index set process $\mathcal{I}$, i.e.,
\begin{equation}\label{2.2.5}
  u(t,c)\equiv u(t,c|\mathcal{I}_{t})=
  \begin{cases}
    \{u_{i}(t,c):i\in\mathcal{I}_{t}\},&\quad\text{if }c\in\mathds{R}^{n}_{++},\\
    -\infty,&\quad\text{o.w.},
  \end{cases}\quad t\in[0,T],
\end{equation}
where $u_{i}:[0,T]\times\mathds{R}^{n}_{++}\mapsto\mathds{R}$, $i\in\mathcal{I}_{t}$, are utility elements that are c\`{a}dl\`{a}g in the first argument (time) and continuous in the second argument (vector of consumption levels). At any time $t\in[0,T]$, the multi-utility $u(t,\cdot)\equiv u(t,\cdot|\mathcal{I}_{t})$ as a whole can be thought of as an $\mathcal{I}_{t}$-parameterized multifunction whose range is $\mathcal{P}(\mathds{R})$, $\mathcal{P}$ denoting the family of nonempty subsets. The index function $\mathds{R}^{d}\ni i\mapsto u_{i}\in\mathds{R}$ is also a continuous and bounded. Continuity in consumption levels is a rather standard rule for utility functions, while parametric continuity ensures that each utility element can be tracked down without uncertainty, once the index set $\mathcal{I}_{t}$ gets revealed. A direct consequence from such continuity is that the multi-utility value range $u(t,\mathds{R}^{n}_{+})$ for every $t\in[0,T]$ forms a closed subset of $\mathcal{C}_{\rm b}(\mathds{R}^{d};\mathds{R})$, the space of bounded continuous functions. It is helpful to make sure that none of the utility elements is redundant in the sense that no element can be reconstructed as a linear combination of the other elements, as was shown by the mean--variance frontier illustration in Section \ref{s:1}. More generally, if there exists a finite subset $J\subsetneq\mathcal{I}$ that is $\mathbb{F}$-non-anticipating and such that for every $t\in[0,T]$,
\begin{equation}\label{2.2.6}
  u(t,c|\mathcal{I}_{t})=\overline{\co}_{\mathcal{C}_{\rm b}}u(t,c|J_{t}),\quad\PP\text{-a.s.},
\end{equation}
where $\overline{\co}$ is taken in $\mathcal{C}_{\rm b}(\mathds{R}^{d};\mathds{R})$, then we can reduce $\mathcal{I}$ into the effective (multi-utility) index set $J$, and consider alternatively the simpler multi-utility $u(t,c|J_{t})$, and it is reasonable to assume that $\mathrm{card}J$ is fixed over time, despite that $J$ can still be time-varying.

Of course, the investor can choose not to use up all of his wealth for consumption and hence generate utility from his terminal bequest, measured by the continuous real-valued univariate function $U(x)$ if $x>0$ and $-\infty$ otherwise.\footnote{Still, one may desire for dependency of the bequest utility on the terminal multi-utility index set $\mathcal{I}_{T}$, through some measurable selector (e.g., a support function), so that the investor's valuation of terminal wealth also changes with time. This is indeed possible following our analysis. However, since $U$, unlike $u$, has values only focused on one time point, it is much more convenient to encode such uncertainty into parameters exclusive to $U$, which we decide to adopt throughout.} To rank his aggregate multi-utility combining $u$ and $U$ at time $t\in[0,T]$, the investor resorts to a general closed convex cone
\begin{equation*}
  \mathcal{K}_{t}\subseteq\bigcup_{s\in[0,t]}\prod_{i\in\mathcal{I}_{s}}\im(u_{i}(s,\cdot)+U)\ni\0, \quad t\in[0,T].
\end{equation*}
For every $t\in[0,T]$, the right side of the above relation specifies a ``collocated'' value range in which ranking of the multi-utility can be reasonably done; the product takes into account the images of the consumption utility elements across all index dimensions while the union serves to enlarge this space due to time variation. The cone $\mathcal{K}_{t}$ is understood to take values in $\mathrm{Cl}(\mathcal{C}_{\rm b}(\mathds{R}^{d};\mathds{R}))$ and is assumed to always contain the zero element, so that it can be pointed (a property necessary for defining optimality), i.e., $\alpha_{1}k_{1}+\alpha_{2}k_{2}\in\mathcal{K}_{t}$ for any $k_{1},k_{2}\in\mathcal{K}_{t}$ and $\alpha_{1},\alpha_{2}\geq0$ and $\mathcal{K}_{t}\cap(-\mathcal{K}_{t})=\{\0\}$. Economically, $\mathcal{K}$ describes the time-dependent regions of comparability where the investor is able to evaluate utility differences over consumption--bequest levels. Whenever $\mathcal{I}$ can be reduced to the finite subset $J$, namely when (\ref{2.2.6}) holds, the cone becomes finitely generated, i.e.,
\begin{equation*}
  \mathcal{K}_{t}\subseteq\prod^{\mathrm{card}J_{t}}_{i=1}\im(u_{i}(t,\cdot)+U),\quad t\in[0,T],
\end{equation*}
where the cardinality is fixed by assumption. The investor's ability to contemporaneously rank any multi-utility levels requires measurability of the cone $\mathcal{K}$, as stated in the next proposition.

\begin{proposition}\label{pro:2}
The set-valued process $\mathcal{K}$ can be $\mathbb{F}$-non-anticipating.
\end{proposition}

There is a convenient spatial extension (see (\ref{A.1}) in \ref{A}) for the multi-utility function $u$ over consumption, which will be considered interchangeably with (\ref{2.2.5}) going forward. This understanding applies to the case where $\mathcal{I}$ is reducible to the finite subset $J$ (see (\ref{2.2.6})). To continue our analysis, some formal assumptions are due on the fundamental properties of the multi-utility.

\begin{assumption}\label{as:1}
The following are assumed to hold for every fixed $t\in[0,T]$. \vspace{0.1in}\\
(i) (Monotonicity): For any $c,c'\in\mathds{R}^{n}_{+}$ with $c-c'\in\mathds{R}^{n}_{+}$ and any $x\geq x'\geq0$,
\begin{equation*}
  u(t,c)-u(t,c')\in\mathcal{K}_{t}\quad\text{and}\quad U(x)-U(x')\geq0.
\end{equation*}
(ii) (Concavity): For any $\alpha\in[0,1]$, $c,c'\in\mathds{R}^{n}_{+}$, and $x,x'\geq0$,
\begin{equation*}
  u(t,\alpha c+(1-\alpha)c')\in\alpha u(t,c)+(1-\alpha)u(t,c')+\mathcal{C}_{\rm b}(\mathcal{I}_{t};\mathds{R}_{+})
\end{equation*}
and
\begin{equation*}
  U(\alpha x+(1-\alpha)x')\geq\alpha U(x)+(1-\alpha)U(x').
\end{equation*}
(iii) (Non-redundancy): $u_{i}\equiv0$ for every $i\in\mathcal{R}^{\complement}$.
\end{assumption}

While monotonicity and concavity are standard for utility functions under choice rationality and risk aversion, non-redundancy is imposed to rule out redundant utility elements associated with meaningless parameter values. We denote the space of all (extended) multi-utility over the space $\mathcal{I}$ that satisfies Assumption \ref{as:1} as $\mathfrak{U}_{\mathcal{I}}$ (regardless of (\ref{2.2.6})), and with the spatial extension (\ref{A.1}), indecisiveness can be equivalently thought of as being fixed in the universe $\mathfrak{U}_{\mathds{\mathds{R}}^{d}}$ of all possible types of tastes while the index set process $\mathcal{I}$ controls which types are effective from time to time.

The investor's incomplete preferences are directly induced from an established multi-utility, as the following proposition explains. Such induction is not only technically manageable but also freely connectible to a variety of practically interesting phenomena with empirical evidence (with details in Subsection \ref{ss:3.2}).\footnote{A similar approach can be found for vector-valued utility maximization in \cite[Hamel and Wang, 2017]{HW} and \cite[Rudloff and Ulus, 2021]{RU}, except that in the present setting the multi-utility can have infinite stochastic dimensions.}

\begin{proposition}\label{pro:3}
For any fixed $t\in[0,T]$ and a given multi-utility function $u\in\mathfrak{U}_{\mathcal{I}_{t}}$, define the $u(t,\cdot)$-induced preference relation on $\mathds{R}^{n}_{+}$ as the set
\begin{equation*}
  \succeq_{t}:=\{(c,c')\in\mathds{R}^{n}_{+}\times\mathds{R}^{n}_{+}:u(t,c)-u(t,c')\in\mathcal{K}_{t}\}.
\end{equation*}
Then $\succeq_{t}$ is reflexive and transitive but not necessarily complete.
\end{proposition}

We also remark that, despite transitivity at any fixed time, a preference relation induced in the fashion above can exhibit dynamic intransitivity, in the sense that $c\succeq_{t}c'$ and $c'\succeq_{t}c''$ for some $t\in[0,T]$ do not imply $c\succeq_{s}c''$ for all $s\in[0,T]$, due to the time-varying feature in (\ref{2.2.5}). Again, this feature is heavily tied to the notion of time preferences (\cite[Ok and Masatlioglu, 2007, \text{Sect.} 2]{OM}) and max-min multi-utility (\cite[Nishimura and Ok, 2016, \text{Sect.} 3.1]{NO}), in that every fixed time $t\in[0,T]$ gives rise to a set-valued random variable $\mathcal{I}_{t}$ that indexes the multi-utility.

The investor's preference relation on bequests is governed by $\geq$ which is clearly reflexive, transitive, complete, and time-invariant for the investment horizon. The following definition deals with the incomplete part induced from the representing multi-utility $u$, which pinpoints the regions where incomparability occurs for given consumption bundles.

\begin{definition}\label{def:2}
For any fixed $t\in[0,T]$, given a multi-utility function $u(t,\cdot)\in\mathfrak{U}_{\mathcal{I}_{t}}$, the incomplete part of the induced preference relation $\succeq_{t}$ is defined as
\begin{equation}\label{2.2.7}
  \circleddash_{t}:=\{(c,c')\in\mathds{R}^{n}_{+}\times\mathds{R}^{n}_{+}:u(t,c)-u(t,c')\in\pm\mathcal{K}_{t}\}^{\complement}.
\end{equation}
\end{definition}

Definition \ref{def:2} can be interpreted as the incomplete part being exactly the complement of the union of the upper and lower contour sets of $\mathds{R}^{n}_{+}$ as implied by every element of $u$. In the special case $n=2$ and $\mathcal{K}=\mathcal{C}_{\rm b}(\mathcal{I};\mathds{R}_{+})$, (\ref{2.2.7}) specializes into
\begin{equation}\label{2.2.8}
  \circleddash_{t}:=\bigg\{(c,c')\in\mathds{R}^{2}_{+}\times\mathds{R}^{2}_{+}:c'\in\bigcap_{i\in\mathcal{I}_{t}}\mathrm{epi}(K_{i}(c))^{\complement} \cap\bigcap_{i\in\mathcal{I}_{t}}\mathrm{hyp}(K_{i}(c))^{\complement}\bigg\},\quad t\in[0,T],
\end{equation}
where $K_{i}(c)$ denotes the contour line of the utility element $u_{i}$ at level $c$. This simple result is sufficient to show the general uncountable infiniteness of the index set $\mathcal{I}$, even if there are as few as two consumption goods.

\subsection{Multi-utility maximization}\label{ss:2.3}

After defining the multi-utility index set and its associated incomplete preferences, we are now ready to formulate the general multi-utility maximization problem. Given the market setup in (\ref{2.1.1}) and (\ref{2.1.2}), the investor's inter-temporal wealth follows the SDE
\begin{equation}\label{2.3.1}
  X_{t}\equiv X^{(c,\Pi)}_{t}=X_{0}+\int^{t}_{0}(r_{s}X_{s}-C_{s})\dd s+\int^{t}_{0}\langle\Pi_{s},(\mu_{s}-r_{s}\1)\dd s+\sigma_{s}\dd W_{s}\rangle_{m},\quad t\in[0,T],
\end{equation}
where $c$ stands for his consumption process of the $n$ goods (in quantities), $C:=\langle P,c\rangle_{n}$ the total consumption expenditure (in dollar amounts), with $P$ being the vector of commodity prices, and $\Pi$ his portfolio process (in dollar amounts) given the $m$ risky assets in the market. In this case, $T$ is thought of as the end of the investor's investment horizon. We introduce some classical assumptions on the consumption--investment (portfolio) policy $(c,\Pi)$.

\begin{assumption}\label{as:2}
The following are assumed to hold. \vspace{0.1in}\\
(i) $c_{t}\in\mathds{R}^{n}_{+}$, $\forall t\in[0,T]$, $\PP$-a.s., and $\E\big[\int^{T}_{0}C_{s}\dd s\big]<\infty$. \vspace{0.1in}\\
(ii) $\E\big[\int^{T}_{0}\|\Pi^{\intercal}_{s}\sigma_{s}\|^{2}_{2}\dd s\big]<\infty$ and $\E\big[\int^{T}_{0}|\langle\Pi_{s},\mu_{s}-r_{s}\1\rangle_{m}|\dd s\big]<\infty$.
\end{assumption}

We shall write $c\in\mathfrak{C}_{n}$ if (i) in Assumption \ref{as:2} is satisfied and $\Pi\in\mathfrak{P}_{m}$ if (ii) is satisfied.

\begin{definition}\label{def:3}
For an initial investment $X_{0}>0$, a consumption--investment policy $(c,\Pi)$ is said to be admissible if the wealth process in (\ref{2.3.1}) does not go negative over $[0,T]$. We define the augmented admissibility set as
\begin{equation*}
  \mathfrak{A}(X_{0}):=\{(c,\Pi)\in\mathfrak{C}_{n}\times\mathfrak{P}_{m}:X_{t}\geq0,\;t\in[0,T],\;\PP\text{-a.s.}\},\quad X_{0}>0.
\end{equation*}
\end{definition}

Over his investment horizon $[0,T]$, the investor solves the following dynamic multi-utility maximization problem:
\begin{equation}\label{2.3.2}
  \sup_{(c,\Pi)\in\mathfrak{A}(X_{0})}V(c,\Pi),
\end{equation}
with the objective function
\begin{equation}\label{2.3.3}
  V(c,\Pi):=\E\bigg[\int^{T}_{0}u(t,c_{t})\dd t+U(X_{T})\bigg].
\end{equation}
To assign meaning to the stated problem, the multi-utility function $u$, given any consumption level in $\mathds{R}^{n}_{+}$, is valued in the topological vector space $\mathcal{C}_{\rm b}(\bar{\mathcal{I}};\mathds{R})$, where $\bar{\mathcal{I}}$ is an ambient space taken as the closed convex hull of the set of all elements in the support of the set-valued random variable $\cl_{\mathds{R}^{d}}\bigcup_{t\in[0,T]}\mathcal{I}_{t}$; i.e., it is the smallest closed convex subset of $\mathds{R}^{d}$ measurable with respect to $\mathscr{F}_{0}$ such that $\mathcal{d}_{\rm H}\big(\cl_{\mathds{R}^{d}}\bigcup_{t\in[0,T]}\mathcal{I}_{t},\{\0\}\big)\leq\mathcal{d}_{\rm H}(\bar{\mathcal{I}},\{\0\})$, $\PP$-a.s. This construction is always possible with the spatial extension in (\ref{A.1}) employed. Thus, the (time) integral appearing in (\ref{2.3.3}) is a Bochner integral in the same vector space; it generalizes vector-valued integrals to infinite-dimensional spaces (under vector addition) and is not to be confused with Aumann integrals representing set operations (under Minkowski addition). Expectation is also taken pointwise in the sense of Bochner integration.\footnote{A parallel can be directly drawn with the vector-valued utility maximization problems considered in \cite[Hamel and Wang, 2017, \text{Sect.} 2]{HW} and \cite[Rudloff and Ulus, 2021, \text{Sect.} 4]{RU}. For a better understanding of the integration in stochastic dimensions, the consumption multi-utility can be thought of as an extended function $(u_{i}(t,c_{t})\mathds{1}_{\mathcal{I}_{t}}(i))(\omega)$ for $i\in\bar{\mathcal{I}}$ and $(t,\omega)\in[0,T]\times\Omega$; again, see the spatial extension in (\ref{A.1}).} If the positivity of consumption is violated, it is automatically agreed that $\E[-\infty]=-\infty$ by (\ref{2.2.5}). With Assumption \ref{as:1} and Assumption \ref{as:2}, the dominated convergence theorem implies that $V(c,\Pi)$ takes values in $\mathcal{C}_{\rm b}(\bar{\mathcal{I}};\mathds{R})$ as well. Then, maximization in (\ref{2.3.2}) can be defined with respect to some chosen $\mathscr{F}_{0}$-measurable pointed closed convex cone $\bar{\mathcal{K}}\supseteq\overline{\co}_{\mathcal{C}_{\rm b}}\bigcup_{t\in[0,T]}\mathcal{K}_{t}$ ($\PP$-a.s.) over the ambient space $\bar{\mathcal{I}}$, in the sense of Definition \ref{def:4} below. If $\mathcal{I}$ is reducible to the subset $J$, $V(c,\Pi)$ in (\ref{2.3.3}) can be effectively treated as being vector-valued in $\mathds{R}^{\mathrm{card}J}$; in such a case, we have particularly $\mathcal{K}_{t}=\bar{\mathcal{K}}$, $\forall t\in[0,T]$, with universal $\mathrm{card}J$ dimensions.

The following is a classical definition in multi-criteria optimization. It gives a precise description of the efficiency of admissible consumption--investment policies, with clear analogy to the mean--variance frontier mentioned in Section \ref{s:1}. For multi-utility maximization, the notion of weak $\bar{\mathcal{K}}$-maximality is of the essence, with which the investor optimizes a policy to the extent that his utility cannot be further increased simultaneously for all possible tastes.

\begin{definition}\label{def:4}
We say that $(c,\Pi)\in\mathfrak{A}(X_{0})$ is a $\bar{\mathcal{K}}$-maximal solution of the problem (\ref{2.3.1}) if $(V(c,\Pi)+\bar{\mathcal{K}})\cap V(\mathfrak{A}(X_{0}))=\{V(c,\Pi)\}$. On the other hand, it is said to be weakly $\bar{\mathcal{K}}$-maximal if $\Int\bar{\mathcal{K}}\neq\emptyset$ and $(V(c,\Pi)+\Int\bar{\mathcal{K}})\cap V(\mathfrak{A}(X_{0}))=\emptyset$.
\end{definition}

Since the financial market is complete, the state price density is uniquely identified as
\begin{equation}\label{2.3.4}
  \xi_{t}:=\exp\bigg(-\int^{t}_{0}\bigg(r_{s}+\frac{1}{2}\|\theta_{s}\|^{2}_{2}\bigg)\dd s-\int^{t}_{0}\langle\theta_{s},\dd W_{s}\rangle_{m}\bigg),\quad t\in[0,T],
\end{equation}
where $\theta:=\sigma^{-1}(\mu-r\1)$ defines the unique market price of risk, and it is natural to consider the corresponding static problem,
\begin{equation}\label{2.3.5}
  \sup_{(c,X_{T})\in\mathfrak{B}(X_{0})}V(c,X_{T}),
\end{equation}
within the budget set given an initial investment
\begin{equation}\label{2.3.6}
  \mathfrak{B}(X_{0}):=\bigg\{(c,X_{T})\in\mathfrak{C}_{n}\times\mathds{L}^{1}_{\mathcal{F}}(\Omega;\mathds{R}_{+}): \E\bigg[\int^{T}_{0}\xi_{s}C_{s}\dd s+\xi_{T}X_{T}\bigg]\leq X_{0}\bigg\},\quad X_{0}>0.
\end{equation}

The following theorem builds a useful connection between the dynamic problem for the consumption--investment policies $(c,\Pi)\in\mathfrak{A}(X_{0})$ and the static problem for the consumption--bequest policies $(c,X_{T})\in\mathfrak{B}(X_{0})$.

\begin{theorem}\label{thm:1}
Assume that $\Int\bar{\mathcal{K}}\neq\emptyset$. We have the following two assertions. \vspace{0.1in}\\
(i) If $(c,\Pi)\in\mathfrak{A}(X_{0})$, then $(c,X_{T})\in\mathfrak{B}(X_{0})$. \vspace{0.1in}\\
(ii) If $(c,X_{T})\in\mathfrak{B}(X_{0})$, then there exists $\Pi\in\mathfrak{P}_{m}$ such that $(c,\Pi)\in\mathfrak{A}(X_{0})$.
\end{theorem}

With Theorem \ref{thm:1} we can focus on the static optimization problem (\ref{2.3.5}) which is easier to handle in the absence of Markov properties. In particular, the very notions of (weak) $\bar{\mathcal{K}}$-maximality in Definition \ref{def:4} immediately apply with $\mathfrak{A}(X_{0})$ replaced by $\mathfrak{B}(X_{0})$ and with the value function taking the consumption--bequest policy as argument, namely $V\equiv V(c,X_{T})$ as in (\ref{2.3.5}). The possibility to recover the optimal investment policy from the optimal wealth is given by exploiting the martingale representation (see (\ref{A.3}) in \ref{A}) as in a one-good market, but a new method will be required to accommodate the stochastic set-valued nature of the problem, which we develop in Section \ref{s:5}.

\vspace{0.2in}

\section{Solution I: Scalarization}\label{s:3}

To solve the static problem (\ref{2.3.5}) in the sense of weak $\bar{\mathcal{K}}$-maximality as stated in Definition \ref{def:4}, we first consider the technique of scalarization. By transforming a multi-valued objective function into a single-valued one according to prescribed rules, scalarization is fairly amenable to numerical computations and will be the methodical focus of this paper. Due to the long strand of related literature, we only refer to two comprehensive surveys -- \cite[Ehrgott, 2005]{E1}, in which linear and nonlinear scalarization techniques are compared in depth, and \cite[Eichfelder, 2008]{E2}, which discusses parametric adaptive methods for numerical computing.

\subsection{A modified Gass--Saaty method}\label{ss:3.1}

Perhaps the most straightforward method to tackle a deterministic multi-criteria problem is by constructing a linear functional that collects all objective functions and projects them into $\mathds{R}$, producing a simple real-valued problem with no additional constraints. Also known as the weighted-sum method, it was initially proposed by \cite[Gass and Saaty, 1955]{GS} (see also \cite[Zadeh, 1963]{Z}).

We modify this method in our stochastic optimization setting by considering the following single-criterion problem:
\begin{equation}\label{3.1.1}
  \sup_{(c,X_{T})\in\mathfrak{B}(X_{0})}V(c,X_{T}|w),\quad w\in\mathcal{K}^{\dag},\;\sup_{t\in[0,T]}\|w(t)\|_{1}>0,\;\PP\text{-a.s.},
\end{equation}
where $w$ is a weight functional,\footnote{Taking values in the (topological) dual cone, $w$ depends on time only through dimensionality (see (\ref{A.5}) in \ref{A}), i.e., $w_{i}(s)=w_{i}(t)$ for any $i\in\mathcal{I}_{s\vee t}$ with $s\neq t$, and is not to be considered as a deterministic function of time. For each fixed $t$, $w(t)$ is nothing but a linear functional on $\mathcal{C}_{\rm b}(\mathcal{I}_{t};\mathds{R})$ (endowed with the topology of pointwise convergence). With a little abuse of notation, the angle brackets are reused to denote the duality pairing over the corresponding product space, with $w$ writable for a ``generalized density'' of the Radon measure thanks to Lebesgue's decomposition theorem.} $\|\;\|_{1}$ is the total variation norm (reducible to the Taxicab norm in finite dimensions if (\ref{2.2.6}) holds) on Radon measures of bounded variation, and the real-valued parameterized objective function is
\begin{equation}\label{3.1.2}
  V(c,X_{T}|w):=\E\bigg[\int^{T}_{0}\bigg\langle w(t),u(t,c_{t})+\frac{U(X_{T})}{T}\bigg\rangle_{\mathcal{I}_{t}}\dd t\bigg],
\end{equation}
$\mathcal{K}^{\dag}_{t}:=\big\{z\in(\mathcal{C}_{\rm b}(\mathcal{I}_{t};\mathds{R}))^{\dag}:\langle z,k\rangle_{\mathcal{I}_{t}}\geq0,\;\forall k\in\mathcal{K}_{t}\big\}$ being the (topological) dual cone of $\mathcal{K}_{t}$, for any $t\in[0,T]$. Clearly, $\mathcal{K}^{\dag}$ is an $\mathbb{F}$-non-anticipating closed convex-valued stochastic process. Unlike in the original objective function in (\ref{2.3.5}), the time integral and the expectation in (\ref{3.1.2}) can be understood in the usual sense for a real-valued stochastic process which the duality pairing has led to.

The economic meaning of the weight functional $w$ is a floating totaling rule applied inter-temporally to the multi-utility $u$ augmented by the time-scaled bequest utility $U/T$. Its indefiniteness is exactly what imprecise tastes entail. It is equivalent to say that $w$ has encoded all potential conflicts that the investor needs to account for as soon as incomparability occurs.

We will show the following equivalence result which explains how the solutions of the two problems (\ref{2.3.5}) and (\ref{3.1.1}), in stochastic dimensions, are fundamentally related to each another.

\begin{theorem}\label{thm:2}
Assume that $\Int\bar{\mathcal{K}}\neq\emptyset$. We have the following two assertions. \vspace{0.1in}\\
(i) If $(c^{\ast},X^{\ast}_{T})$ is a $\bar{\mathcal{K}}$-maximal solution of the multi-criteria problem (\ref{2.3.5}), then there exists $w(t)\in\mathcal{K}^{\dag}_{t}$ for every $t\in[0,T]$ with $\sup_{t\in[0,T]}\|w(t)\|_{1}>0$, $\PP$-a.s., such that $(c^{\ast},X^{\ast}_{T}|w)$ is a maximal solution of the single-criterion problem (\ref{3.1.1}). \vspace{0.1in}\\
(ii) If $(c^{\ast},X^{\ast}_{T}|w)$ is a maximal solution of (\ref{3.1.1}) then $(c^{\ast},X^{\ast}_{T}|w)$ is at least a weakly $\bar{\mathcal{K}}$-maximal solution of (\ref{2.3.5}).
\end{theorem}

What Theorem \ref{thm:2} has implied is that since the criterion space $V(\mathfrak{B}(X_{0}))$ is convex, the (modified) Gass--Saaty method is powerful enough to recover the entire set of (weakly) $\bar{\mathcal{K}}$-maximal solutions of (\ref{2.3.5}). We reiterate the significance of this result by the following proposition.

\begin{proposition}\label{pro:4}
Let $(c^{\ast},X^{\ast}_{T}|w)$ be a maximal solution of the single-criterion problem (\ref{3.1.1}) conditional on $w\in\mathcal{K}^{\dag}$. Then the set of weakly $\bar{\mathcal{K}}$-maximal solutions of the multi-criteria problem (\ref{2.3.5}) precisely equals
\begin{equation}\label{3.1.3}
  \mathcal{S}^{\ast}=\Big\{(c^{\ast},X^{\ast}_{T}|w):w\in\mathcal{K}^{\dag},\sup_{t\in[0,T]}\|w(t)\|_{1}=1,\;\PP\text{-a.s.}\Big\}.
\end{equation}
\end{proposition}

As the duality pairing is a continuous mapping, from Definition \ref{def:4} it follows from the closed-ness and convexity of the criterion space $V(\mathfrak{B}(X_{0}))$ that the solution set $\mathcal{S}^{\ast}$ in (\ref{3.1.3}) gives rise to a $\mathcal{B}([0,T])\otimes\mathcal{F}$-measurable $w$-parameterized augmented set-valued stochastic process $(c^{\ast},X^{\ast}_{T})$ with values in $\mathrm{Cl}\big(\mathfrak{C}_{n}\times\mathds{L}^{1}_{\mathcal{F}}(\Omega;\mathds{R}_{+})\big)$; that is, there is no need to close $\mathcal{S}^{\ast}$ for completeness. Notably, in this context the measurable selectors\footnote{For the optimal consumption $c^{\ast}$ and the optimal bequest $X^{\ast}_{T}$, the selectors are understood in the spaces $\mathfrak{C}_{n}$ and $\mathds{L}^{1}_{\mathcal{F}}(\Omega;\mathds{R}_{+})$, respectively.} of $(c^{\ast},X^{\ast}_{T})$ are exactly the parameterized augmented (single-valued) processes $(c^{\ast},X^{\ast}_{T}|w)$ with suitable weight functionals.

If the index set $\mathcal{I}$ is reduced to the subset $J$, with (\ref{2.2.6}), the preceding context can be understood as $\mathcal{K}^{\dag}\equiv\bar{\mathcal{K}}^{\dag}$ being a finite-dimensional dual cone and $w$ a dual vector of fixed $\mathrm{card}J$ dimensions. Consequently, the suprema in (\ref{3.1.1}) and (\ref{3.1.3}) can be dropped as time dependency is no longer required.

Generally speaking, the Gass--Saaty method is incommensurably simple in that scalarization is achieved without introducing additional constraints to the original problem, and yet it provides a complete characterization of the weakly $\bar{\mathcal{K}}$-maximal solutions of the multi-criteria problem (\ref{2.3.5}), or equivalently, the original problem (\ref{2.3.2}), based on convexity. For this reason, Proposition \ref{pro:4} is directly applicable to recover all the optimal consumption--bequest policies $(c,X_{T})$.

\subsection{Procedures and examples}\label{ss:3.2}

We outline the procedures of finding the solution set $\mathcal{S}^{\ast}$ in (\ref{3.1.3}), based on Proposition \ref{pro:4}. For computational convenience we focus on smooth utility functions by imposing some additional conditions.

\begin{assumption}\label{as:3}
For every $t\in[0,T]$ and $i\in\mathcal{I}_{t}$, either $u_{i}(t,\cdot)\in\mathcal{C}^{\infty}(\mathds{R}^{n}_{++};\mathds{R})$, which satisfies the Inada conditions that the first-order derivatives with respect to the $j$th consumption level, $c_{j}$, $\lim_{c_{j}\searrow0}u^{(j)}_{i}(t,c)=\infty$ and $\lim_{c_{j}\rightarrow\infty}u^{(j)}_{i}(t,c)=0$, for any $j\in\mathds{N}\cap[1,n]$, or $u_{i}\equiv-\infty$; similarly, $U\in\mathcal{C}^{\infty}(\mathds{R}_{++};\mathds{R})$, satisfying that $\lim_{x\searrow0}U(x)=\infty$ and $\lim_{x\rightarrow\infty}U(x)=0$.
\end{assumption}

As each multi-utility element $u_{i}$ constitutes a single-valued utility function, the above conditions are classical and they suffice for all the phenomena of practical interest mentioned earlier in Section \ref{s:1} and Section \ref{s:2}; still, they are relaxable to the case with non-smooth utility (e.g., as piecewise smooth functions) with reasonable effort. The Inada conditions are favorable because the consumption and wealth levels are restricted to nonnegativity (see (\ref{2.2.5}) and Assumption \ref{as:2}). We also recall the assumption of the c\`{a}dl\`{a}g property of each multi-utility element in time.

In the very first step, we construct the index set process $\mathcal{I}\subseteq\mathds{R}^{d}$ according to the recipe in (\ref{2.2.1}) and (\ref{2.2.3}) with appropriate coefficients $f_{q}$'s and $G_{q}$'s, for $q\in\{1,2,3\}$, compute its unconditional superset $\bar{\mathcal{I}}$, and then specify the multi-utility $u\in\mathfrak{U}_{\bar{\mathcal{I}}}$ as well as the bequest utility $U$. If the dimensionality reduction requirement (\ref{2.2.6}) is fulfilled, it is recommended to use the reduced finite subset $J$ instead of $\mathcal{I}$, with the reduced ordering cone $\mathcal{K}$ of fixed $\mathrm{card}J$ dimensions.

Then, when applying the Gass--Saaty method, the key step is to write down the Lagrangian
\begin{equation*}
  L_{w}(c,X_{T},\eta)=V(c,X_{T}|w)+\eta\bigg(X_{0}-\E\bigg[\int^{T}_{0}\xi_{s}C_{s}\dd s+\xi_{T}X_{T}\bigg]\bigg),\quad\eta\geq0,
\end{equation*}
with a weight functional $w\in\mathcal{K}^{\dag}$ satisfying $\sup_{t\in[0,T]}\|w(t)\|_{1}=1$ ($\PP$-a.s.) (clearly, $\|w(t)\|_{1}\equiv\|w\|_{1}=1$ if $J$ takes the place of $\mathcal{I}$) and $\eta$ being the multiplier to the budget set.

After employing a perturbation argument with $\pd L_{w}(c+\epsilon\Delta c,X_{T},\eta)/\pd\epsilon|_{\epsilon=0}=0$, $\forall\Delta c\in\mathfrak{C}_{n}$, $\pd L_{w}(c,X_{T}+\epsilon\Delta X_{T},\eta)/\pd\epsilon|_{\epsilon=0}=0$, $\forall\Delta X_{T}\in\mathds{L}^{1}_{\mathcal{F}}(\Omega;\mathds{R}_{+})$, and $\pd L_{w}(c,X_{T},\eta+\epsilon\Delta\eta)/\pd\epsilon|_{\epsilon=0}=0$, $\forall\Delta\eta\geq0$, the necessary optimality conditions can be cast as\footnote{For convenience we will present these general conditions in terms of the original index set $\mathcal{I}$; whenever it is reducible to $J$, duality pairings are automatically understood as taking place in the Euclidean space $\mathds{R}^{\mathrm{card}J}$ with no randomness dimensions.}
\begin{align}\label{3.2.1}
  \eta\xi P_{j}&=\langle w,u^{(j)}(\imath,c)\rangle_{\mathcal{I}},\quad j\in\mathds{N}\cap[1,n], \nonumber\\
  \eta\xi_{T}&=\frac{U'(X_{T})}{T}\int^{T}_{0}\langle w(t),\1\rangle_{\mathcal{I}_{t}}\dd t, \nonumber\\
  X_{0}&=\E\bigg[\int^{T}_{0}\xi_{t}C_{t}\dd t+\xi_{T}X_{T}\bigg].
\end{align}
The first condition is stated for every $t\in[0,T]$ and by $u^{(j)}$ we mean the parameterized multifunction $\big\{u^{(j)}_{i}(t,c):i\in\mathcal{I}_{t}\big\}$. Whenever $\|w(t)\|_{1}>0$ at time $t$, its attainability is guaranteed by the imposed Inada conditions (Assumption \ref{as:3}), in the same way as that of the second condition for the optimal bequest, while in case $\|w(t)\|_{1}=0$, it becomes a trivial condition and $c^{\ast}_{t}=\0$. The above conditions are also sufficient for ensuring (global) maximization of the single-criterion problem (\ref{3.1.1}). Indeed, suppose that the consumption--bequest policy $(c^{\ast},X^{\ast}_{T})$ satisfies (\ref{3.2.1}) and $(\hat{c},\hat{X}_{T})\in\mathfrak{B}(X_{0})$ is another policy, and then by the concavity property in Assumption \ref{as:1} we have $u_{i}(t,c^{\ast}_{t})\geq u_{i}(t,\hat{c}_{t})+\sum^{n}_{j=1}u^{(j)}_{i}(t,c^{\ast}_{t})(c^{\ast}_{j,t}-\hat{c}_{j,t})$, $\forall i\in\mathcal{I}_{t}$, $t\in[0,T]$, and $U(X^{\ast}_{T})\geq U(\hat{X}_{T})+U'(X^{\ast}_{T})(X^{\ast}_{T}-\hat{X}_{T})$; with the correspondence (\ref{A.5}) it follows that
\begin{align*}
  V(c^{\ast},X^{\ast}_{T}|w)&\geq\E\bigg[\int^{T}_{0}\bigg\langle w(t),u(t,\hat{c}_{t})+\frac{U(\hat{X}_{T})}{T}\bigg\rangle_{\mathcal{I}_{t}}\dd t\bigg]\\
  &\qquad+\E\Bigg[\int^{T}_{0}\Bigg\langle w(t),\sum^{n}_{j=1}u^{(j)}(t,c^{\ast}_{t})(c^{\ast}_{j,t}-\hat{c}_{j,t})+\frac{U'(X^{\ast}_{T})}{T}(X^{\ast}_{T}-\hat{X}_{T}) \Bigg\rangle_{\mathcal{I}_{t}}\dd t\Bigg]\\
  &=V(\hat{c},\hat{X}_{T}|w)+\eta\bigg(X_{0}-\E\bigg[\int^{T}_{0}\xi_{t}\hat{C}_{t}\dd t+\xi_{T}\hat{X}_{T}\bigg]\bigg),\quad\eta\geq0,
\end{align*}
from where it suffices to realize that the second term in the last equality is nonnegative by (\ref{2.3.6}).

\begin{figure}[H]
  \centering
  \includegraphics[scale=0.21]{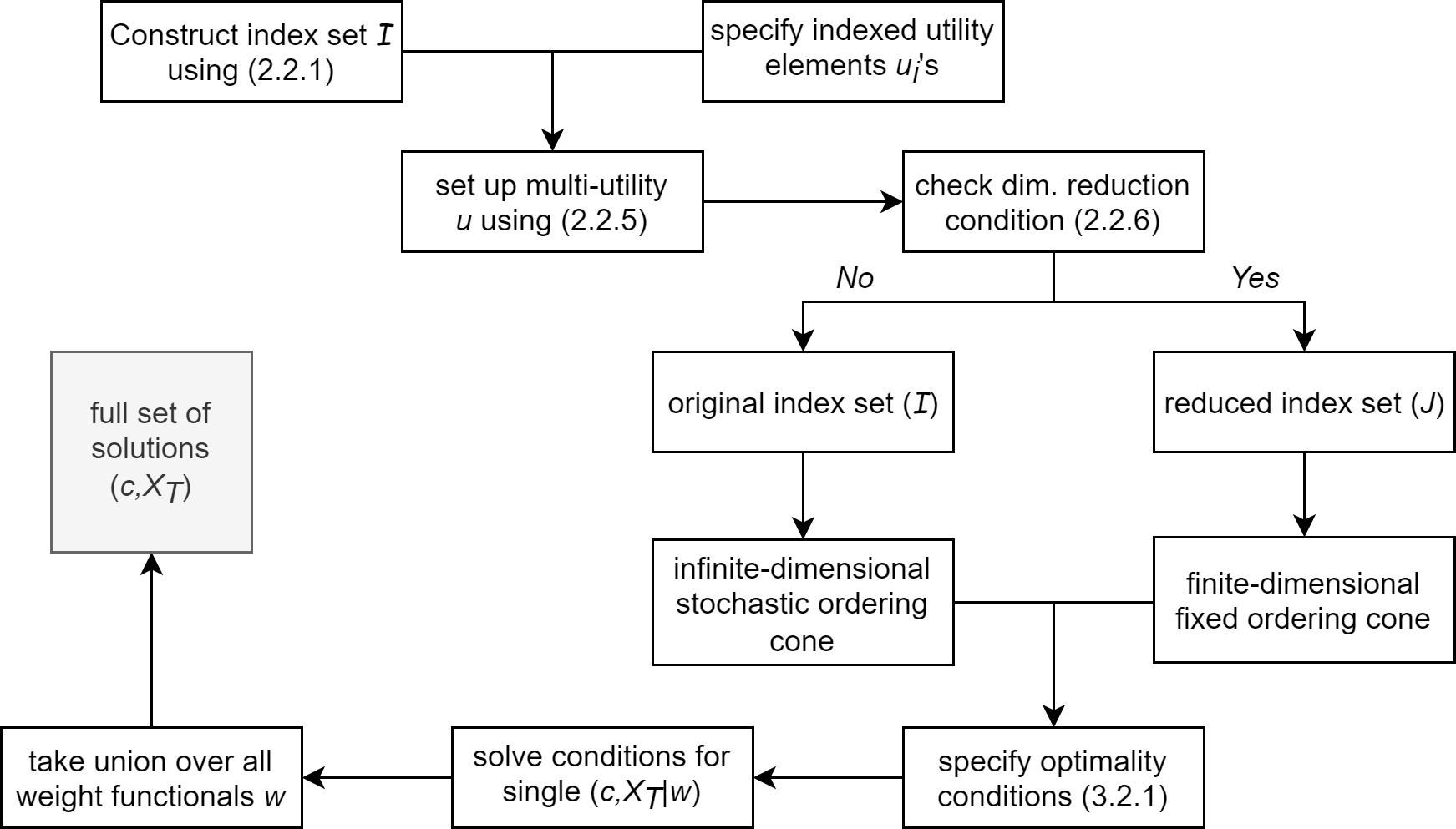}\\
  \caption{Solution procedures by way of Gass--Saaty method}
  \label{fig:2}
\end{figure}

On a second look, the conditions in (\ref{3.2.1}) form an $(n+2)$-dimensional nonlinear system connecting $(c,X_{T})$ and $\eta$, conditional on $w$. In particular, while consumption $c$, and hence the total expenditure $C$, as well as the terminal wealth $X_{T}$ can always be written as $w$-aggregated feedback functions of $\eta\xi$, or specifically,\footnote{According to the proof of Theorem \ref{thm:2} (see \ref{A}), we must have $\int^{T}_{0}\langle w(t),\1\rangle_{\mathcal{I}_{t}}\dd t\neq0$, $\PP$-a.s.} $c=\psi_{\mathcal{I}}(\eta\xi P|w)$ and $X_{T}=(U')^{-1}\big(\eta\xi_{T}T\big/\int^{T}_{0}\langle w(t),\1\rangle_{\mathcal{I}_{t}}\dd t\big)$, they are usually insufficient for a unique (namely $w$-free) determination unless $u$ is single-valued with $\mathrm{card}\bar{\mathcal{I}}=1$. Even so, it is possible to obtain a unique -- hence proper -- maximal solution $(c^{\ast},X^{\ast}_{T})$ on certain special occasions, e.g., when (individual) consumption elements in feedback form happen to be proportional to each other (see Example 1). In general, once we have the aforementioned feedback functions, by altering $w$ in its corresponding region in (\ref{3.2.1}) we can establish $\mathcal{S}^{\ast}$. To summarize all the necessary steps to take in implementing the Gass--Saaty method, we present a flowchart in Figure \ref{fig:2}.

Now we turn back to the illustrative examples outlined in Subsection \ref{ss:1.2}. For specificity we will only consider up to two index dimensions ($d\in\{1,2\}$)\footnote{These will be sufficient for studying CRRA-type utility and for showing why it is undesirable to stick to only one-dimensional indices.} and one risky asset ($m=1$), with exactly two consumption goods ($n=2$). It is an undemanding exercise to generalize the results to more risky assets and consumption goods. To isolate the effects of time-varying incomplete preferences, we let the market coefficients $r>0$ and $\mu\in\mathds{R}$ both be constant, with no dividend payments $D\equiv0$, and the commodity prices be uniformly one, namely $P\equiv\1$, upon setting $P_{0}=\1$, $\mu_{P}\equiv\0$, and $\sigma_{P}=\mathbf{O}$. The construction of the universe $\mathfrak{U}_{\mathds{R}^{d}}$ of multi-utility is based on regular choices in multi-attribute utility theory; for more details in how (multivariate) utility elements can be customized for the discussed features we refer to \cite[Keeney and Raiffa, 1993]{KR}. Natural ordering is adopted for all these examples for comparability. \vspace{0.1in}

\textbf{Example 1. (Invariant indecisiveness)}\quad The market volatility $\sigma$ is constant and the investor does not care about bequests ($U\equiv0$), consuming all of his wealth, so that $V(c,X_{T})\equiv V(c)$, and his preferences are time-invariant. Thus, in (\ref{2.2.3}), we have $I_{2,0}=I_{3,0}=\{0\}$, $f_{q}\equiv\{0\}$ and $G_{q}\equiv\{0\}$ for all $q\in\{1,2,3\}$, so that $\bar{\mathcal{I}}=\mathcal{I}$ are both constant, and let the multi-utility $u$ take the form of time-invariant power utility.

\begin{figure}[H]
  \centering
  \begin{minipage}{0.32\linewidth}
  \centering
  \includegraphics[scale=0.23]{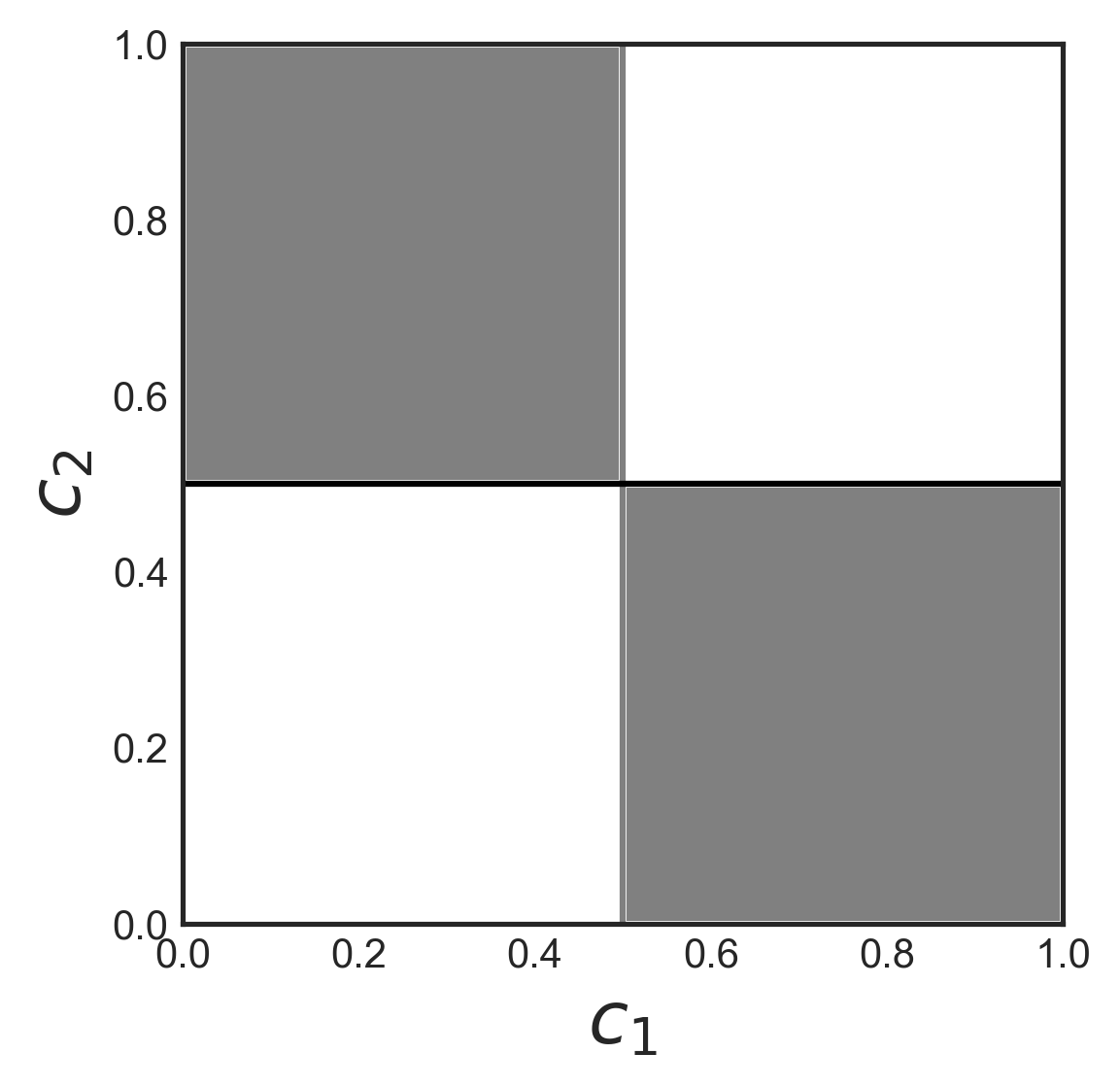}
  \caption*{Case (I)}
  \end{minipage}
  \begin{minipage}{0.32\linewidth}
  \centering
  \includegraphics[scale=0.23]{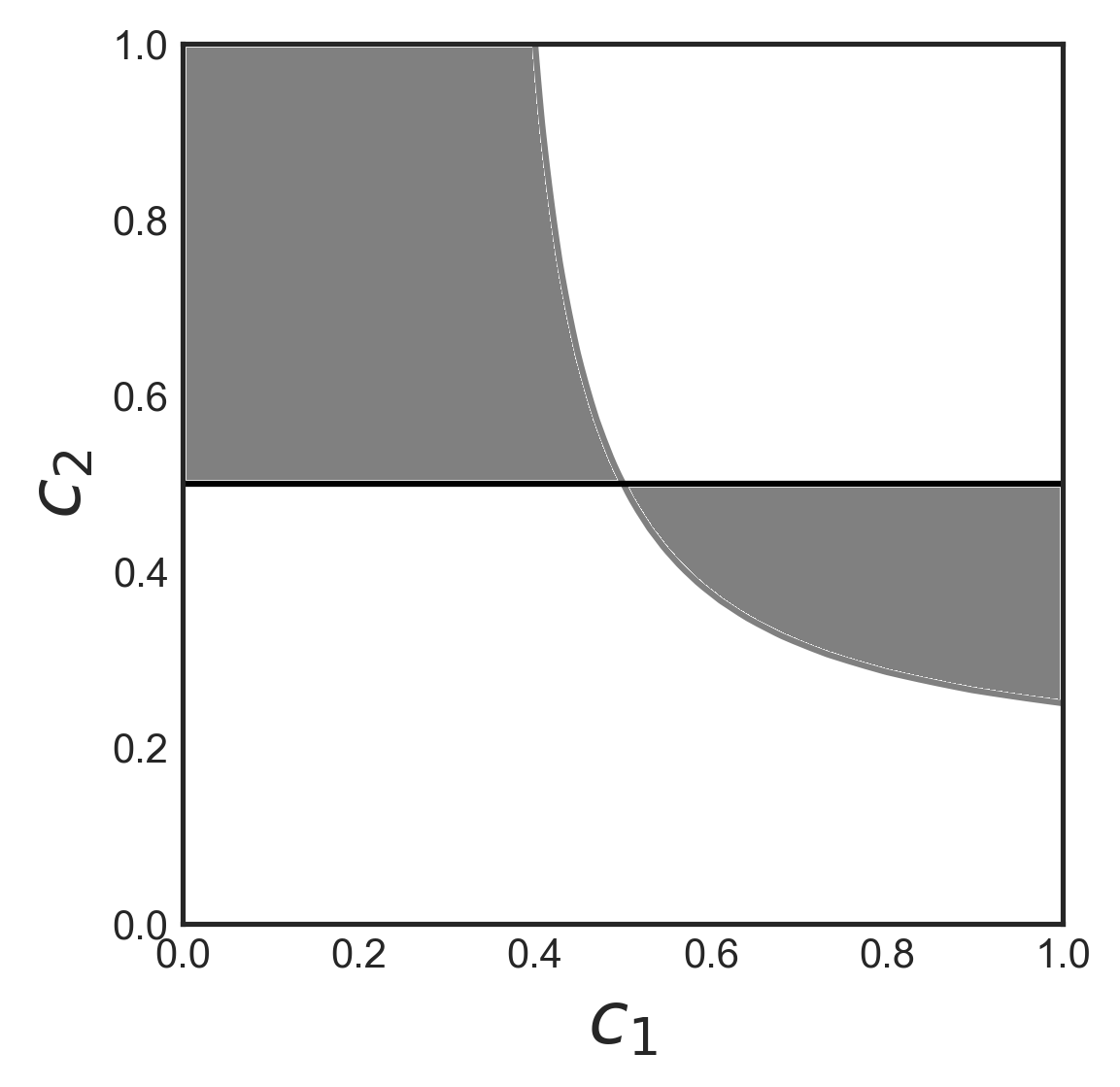}
  \caption*{Case (II)}
  \end{minipage}
  \begin{minipage}{0.32\linewidth}
  \centering
  \includegraphics[scale=0.23]{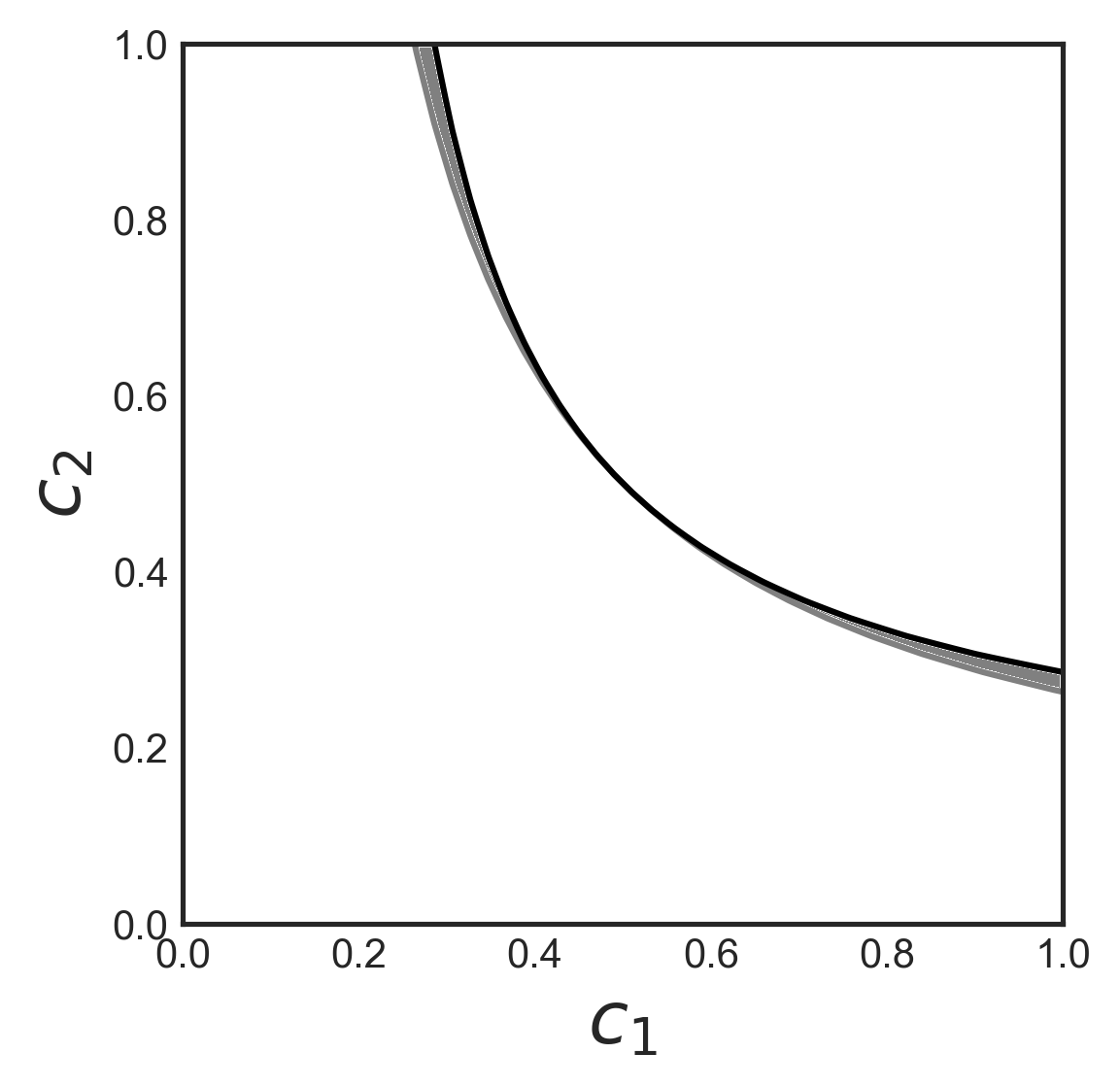}
  \caption*{Case (III)}
  \end{minipage}
  \caption{Incomplete part (gray area) of preference relation in Example 1}
  \label{fig:3}
\end{figure}

We will investigate the three cases mentioned in Subsection \ref{ss:1.2}. In Case (I), the two goods are totally incomparable, and according to the formula (\ref{2.2.8}), the preference relation $\succeq$ is time-invariant with exactly a cross-shaped incomplete part (see Figure \ref{fig:3}), $\circleddash=\{(c,c')\in\mathds{R}^{2}_{+}\times\mathds{R}^{2}_{+}:c-c'\notin\mathds{R}^{2}_{\pm}\}$. These preferences are representable by a simple multi-utility function each of whose elements depends on exactly one consumption good,
\begin{equation*}
  u_{i}(\imath,c)\equiv u_{i}(c)=
  \begin{cases}
    \displaystyle \frac{c^{1-p}_{i}-1}{1-p},\quad&\text{if }i\in\{1,2\},\\
    -\infty,\quad&\text{if }i\in(1,2),\\
    0,\quad&\text{o.w.},
  \end{cases} \tag{I}
\end{equation*}
where $p\in\mathds{R}_{++}\setminus\{1\}$ is the investor's universal risk aversion degree.

In Case (II), the second good is the unconditionally preferred good but the first generates imprecise attention. Such a static preference relation has an incomplete part of the form $\circleddash=\{(c,c')\in\mathds{R}^{2}_{+}\times\mathds{R}^{2}_{+}:c-c'\notin([h(c'_{2}),\infty)\times\mathds{R}_{+})\cup ((-\infty,h(c'_{2})]\times\mathds{R}_{-})\}$, where the function $h$ is strictly decreasing and convex with $h(0)=0$ (Figure \ref{fig:3}). This can be achieved through the following form of multi-utility:
\begin{equation*}
  u_{i}(\imath,c)\equiv u_{i}(c)=
  \begin{cases}
    \displaystyle \frac{i(c^{1-p}_{1}-1)+(c^{1-p}_{2}-1)}{1-p},\quad&\text{if }i\in[0,\chi],\\
    0,\quad&\text{o.w.},
  \end{cases} \tag{II}
\end{equation*}
where $i\geq0$ is the attention degree of the first good relative to the second and $\chi>0$ sets its upper bound. Such a structure also preserves preference independency from Case (I) (see \cite[Keeney and Raiffa, 1993, \text{Chap.} V]{KR}).

In Case (III), the two goods are utility-independent, and with restricted risk aversion $p>1$ the multi-utility can be set up as
\begin{equation*}
  u_{i}(\imath,c)\equiv u_{i}(c)=
  \begin{cases}
    \displaystyle \frac{c^{1-p}_{1}+c^{1-p}_{2}}{1-p}-\frac{i(c_{1}c_{2})^{1-p}}{(1-p)^{2}},\quad&\text{if }i\in[\varkappa_{1},\varkappa_{2}]\subsetneq\mathds{R}_{++},\\
    0,\quad&\text{o.w.},
  \end{cases} \tag{III}
\end{equation*}
where the parameter $i$ measures the degree of interaction between the two goods, which is only known to lie within the positive interval $[\varkappa_{1},\varkappa_{2}]$.\footnote{If the two goods were adequate complements, a similar construction would require an interval $[\varkappa_{1},\varkappa_{2}]\subsetneq\mathds{R}_{--}$ of negative values, with risk aversion restricted to $p\in(0,1)$.} Specifically, a larger value of $i$ implies a higher rate of substitution between more preferred quantities of different goods. The static preference relation, in this case, is generally of the form $\circleddash=\{(c,c')\in\mathds{R}^{2}_{+}\times\mathds{R}^{2}_{+}:c-c'\notin([h_{1}(c'_{2}),\infty)\times[h_{2}(c'_{1}),\infty))\cup ((-\infty,h_{1}(c'_{2})]\times(-\infty,h_{2}(c'_{1})]\}$, $h_{1}$ and $h_{2}$ being non-overlapping, strictly decreasing, and convex functions with $h_{1}(0)=h_{2}(0)=0$ (again, see Figure \ref{fig:3}).

Based on the above specifications it is easy to check that the dimensionality reduction condition (\ref{2.2.6}) indeed stands, since all parameters of interest are substantially scaling factors. The index set $\mathcal{I}$ is then reducible to the subset $J$ of cardinality two; in particular, we have $\mathcal{I}=I_{1,0}=[1,2],[0,\chi],[\varkappa_{1},\varkappa_{2}]$ and $J=\{1,2\},\{0,\chi\},\{\varkappa_{1},\varkappa_{2}\}$ in Case (I), Case (II), and Case (III), respectively, allowing us to work with the static cone $\mathcal{K}\equiv\mathds{R}^{2}_{+}\equiv\mathcal{K}^{\dag}$ in all three cases.

For Case (I), using the Gass--Saaty method, the optimality conditions in (\ref{3.2.1}) specialize into the three-dimensional system
\begin{equation}\label{3.2.2}
  \eta\xi=w_{1}c^{-p}_{1}=w_{2}c^{-p}_{2},\quad X_{0}=\E\bigg[\int^{T}_{0}\xi_{t}C_{t}\dd t\bigg].
\end{equation}
Consider the extremal case with $w_{1}=0$ and $w_{2}=1$. The second equation in (\ref{3.2.2}) then gives $\eta^{1/p}=(e^{\rho_{p}(r,\theta)T}-1)/(\rho_{p}(r,\theta)X_{0})$ and $C^{\ast}=c^{\ast}_{2}=\rho_{p}(r,\theta)X_{0}/(\xi^{1/p}(e^{\rho_{p}(r,\theta)T}-1))$,\footnote{Here, the optimal total consumption expenditure happens to be single-valued partly because of constant commodity prices. An direct observation from the second equation in (\ref{3.2.2}) is that when commodity prices are fluctuating (even in a deterministic manner), the expenditure is generally parameterized and hence admits multiple values at optimum.} where we have used (\ref{2.3.4}) for $\rho_{p}(r,\theta):=(1/p-1)r+(1-p)/(2p^{2})\theta^{2}$. The first equation in (\ref{3.2.2}) is redundant, which signifies that only the total consumption expenditure $C^{\ast}$ matters. Varying $w$ in $\mathds{R}^{2}_{+}$ is equivalent to having $c^{\ast}_{1}/c^{\ast}_{2}\in[0,\infty]$ with $C^{\ast}$ unchanged. Therefore, the set of $\mathds{R}^{2}_{+}$-maximal solutions sought by the investor is given by\footnote{In the limit as $p\rightarrow1$ one obtains the solution with log-type multi-utility, or $\lim_{p\rightarrow1}\mathcal{S}^{\ast}=\{C^{\ast}=X_{0}/(\xi T)\}$.}
\begin{equation}\label{3.2.3}
  \mathcal{S}^{\ast}=\bigg\{c\in\mathfrak{C}_{2}:C^{\ast}=\frac{\rho_{p}(r,\theta)X_{0}}{\xi^{1/p}(e^{\rho_{p}(r,\theta)T}-1)}\bigg\}.
\end{equation}
This implies that the investor only focuses on the total consumption expenditure at optimality and particular consumption bundles do not matter, with intuition given by the two consumption goods being ``equally important'' and assessed without interaction.

For Case (II), the optimality conditions in (\ref{3.2.1}) become
\begin{equation}\label{3.2.4}
  \eta\xi=w_{1}\chi c^{-p}_{1}=c^{-p}_{2},\quad X_{0}=\E\bigg[\int^{T}_{0}\xi_{t}C_{t}\dd t\bigg].
\end{equation}
Solving (\ref{3.2.4}) immediately gives $c_{1}=(\eta\xi/(w_{1}\chi))^{-1/p}$ and $c_{2}=(\eta\xi/(w_{1}+w_{2}))^{-1/p}$, from which the total consumption expenditure $C^{\ast}=\rho_{p}(r,\theta)X_{0}/(\xi^{1/p}(e^{\rho_{p}(r,\theta)T}-1))$ is still pinned down, but consumption elements have to satisfy the quotient bound $c^{\ast}_{1}/c^{\ast}_{2}\in[0,\chi^{1/p}]$. Therefore, the set of $\mathds{R}^{2}_{+}$-maximal solutions for the investor can be expressed as
\begin{align*}
  \mathcal{S}^{\ast}=\bigg\{c^{\ast}\in\mathfrak{C}_{2}:C^{\ast}=\frac{\rho_{p}(r,\theta)X_{0}}{\xi^{1/p}(e^{\rho_{p}(r,\theta)T}-1)};\; \frac{c^{\ast}_{1}}{c^{\ast}_{2}}\in[0,\chi^{1/p}]\bigg\}.
\end{align*}
Intuitively, while primarily focusing on his total consumption expenditure, the investor also limits the proportion of his consumption of the first good to no more than $\chi^{1/p}/(\chi^{1/p}+1)$, considering the second good to be more essential. This proportion is directly related to $\chi$ but inversely related to $p$. In other words, the limitation is severer when perceived importance of the first good declines or when the investor becomes more risk-averse; in the limit as $p\searrow0$ or $\chi\rightarrow\infty$ one recovers (\ref{3.2.3}).

Turning to Case (III), under the Gass--Saaty method, the optimality conditions in (\ref{3.2.1}) imply after some transformation
\begin{equation*}
  c_{1}=c_{2}=\psi(\eta\xi|w)\equiv\psi(\eta\xi|w_{1},w_{2}),
\end{equation*}
where $\psi(\eta\xi|w)$ is the unique solution to the transcendental equation
\begin{equation}\label{3.2.5}
  1-\frac{x^{1-p}(w_{1}\varkappa_{1}+w_{2}\varkappa_{2})}{1-p}=\eta\xi x^{p},\quad x\geq0.
\end{equation}
In this case, for a general $w\in\mathds{R}^{2}_{+}$, the total consumption expenditure is necessarily parameterized, while the budget constraint conditional on $w$ uniquely determines the multiplier $\eta$. Therefore, the set of $\mathds{R}^{2}_{+}$-maximal solutions for the investor can be written
\begin{equation*}
  \mathcal{S}^{\ast}=\bigcup_{w\in\mathds{R}^{2}_{+},\;\|w\|_{1}=1}\bigg\{c^{\ast}\in\mathfrak{C}_{2}:c^{\ast}_{1}=c^{\ast}_{2} =\psi(\eta\xi|w);\;\eta\rightsquigarrow2\int^{T}_{0}\E[\xi_{t}\psi(\eta\xi_{t}|w)]\dd t=X_{0}\bigg\}.
\end{equation*}
This result seems a bit convoluted, though, it is clear that at optimum the investor always consumes an equal amount of the two goods, owing to symmetric interaction. However, the optimal consumption expenditure paths can be chosen within a parameterized family of functionals given by $2\psi(\eta\xi|w)$. We observe that in (\ref{3.2.5}), with $\|w\|_{1}=1$, fixing $p$ and $\varkappa_{1}$, as $\varkappa_{2}$ increases the value range of solutions (in terms of various choices of $w_{1}$) increases, which is also observed as $p$ decreases with $\varkappa_{1,2}$ fixed. This points to increased flexibility of optimal consumption policies under high levels of ambivalence about the degree of interaction or low levels of risk aversion. \hfill{\scriptsize $\lozenge$}

\vspace{0.1in}

We summarize the economic significance of Example 1 as follows. First, when the two consumption goods are totally incomparable and assessed independently, the multi-utility is simply structured as a collection of univariate utility elements, and eventually only the total consumption expenditure is really optimized and actual combinations make no difference. Second, when one good is essentially preferred over the other, the multi-utility can be imposed as a collection of bivariate utility elements exactly one of which is univariate, and an additional scale parameter can be introduced to enlarge the rate of substitution; optimization is still centered around the consumption expenditure, but strict quotient bounds are placed on the less essential good despite a reasonable range of optimal combinations. Third, when the goods can be equally substituted for one another, the multi-utility can well be a reflection of a set of utility-independent preferences, with indecisiveness lying in the rate of substitution; in optimization, the quotient of consumption elements is largely determined by symmetry properties whereas their total expenditure is generally not determined. In other words, depending on the construction of multi-utility, the total consumption expenditure may or may not be set-valued at optimum, and understandably even the optimal consumption itself may be single-valued when indecisiveness is trifling enough. This further suggests that although the incomplete parts can adequately depict (or visualize) indecisiveness over consumption goods, they should not be relied upon for unraveling the optimal consumption patterns.

\vspace{0.1in}

\textbf{Example 2. (Socialization and increasing indecisiveness)}\quad The investor now has increasing indecisiveness for changing consumption attention. In particular, his initial attention of the first good is imprecise, but so is the speed of indecisiveness increase, based on the latent factor $W$. We retain the assumptions of constant market volatility and zero bequest utility. Then, in (\ref{2.2.3}), $\mathcal{R}=\mathds{R}_{+}$, $I_{1,0}=I_{2,0}=\{0\}$, $I_{3,0}=[0,1]$, $f_{q}\equiv\{0\}$ for all $q\in\{1,2,3\}$, $G_{1}\equiv G_{2}\equiv\{0\}$, and $G_{3}\equiv\{0,\lambda\}$ for a parameter $\lambda>0$. The conditions in (\ref{2.2.4}) are trivially satisfied and the index set process can be verified to be
\begin{equation*}
  \mathcal{I}_{t}=\mathcal{R}\cap\overline{\co}_{\mathds{R}}\bigcup_{s\in[0,t]}I_{3,s}=\big[0,\lambda W^{\uparrow}_{t}+1\big],\quad t\in[0,T],
\end{equation*}
where $W^{\uparrow}:=\sup_{s\in[0,\imath]}W_{s}$ denotes the running maximum of the Brownian motion $W$. Clearly, $\bar{\mathcal{I}}=\mathds{R}_{+}$. Also, similarly to Case (II) of Example 1 we take
\begin{equation*}
  u_{i}(t,c)=
  \begin{cases}
    \displaystyle \frac{\chi_{i}(c^{1-p}_{1}-1)+(c^{1-p}_{2}-1)}{e^{\beta t}(1-p)},\quad&\text{if }i\in\big[0,\lambda W^{\uparrow}_{t}+1\big],\\
    0,\quad&\text{o.w.}
  \end{cases}
\end{equation*}
In this design of multi-utility, $\beta>0$ is a universal subjective discount factor measuring the investor's patience as usual and his risk aversion degree across the two consumption goods stands at $p>0$, but the attention degree (or measurement of the first good's relative importance) $\chi:\mathds{R}_{+}\mapsto\mathds{R}_{+}$ is some customizable bounded, nondecreasing function of the index $i$. The additional parameter $\lambda$ is linked to the acceleration of indecisiveness increase, or the severity of addiction.\footnote{If indecisiveness were to decrease over time, one would need nontrivial specifications of the components in (\ref{2.2.3}) with $q=2$, which case is not illustrated in this paper for conciseness.}

\begin{figure}[H]
  \centering
  \includegraphics[scale=0.23]{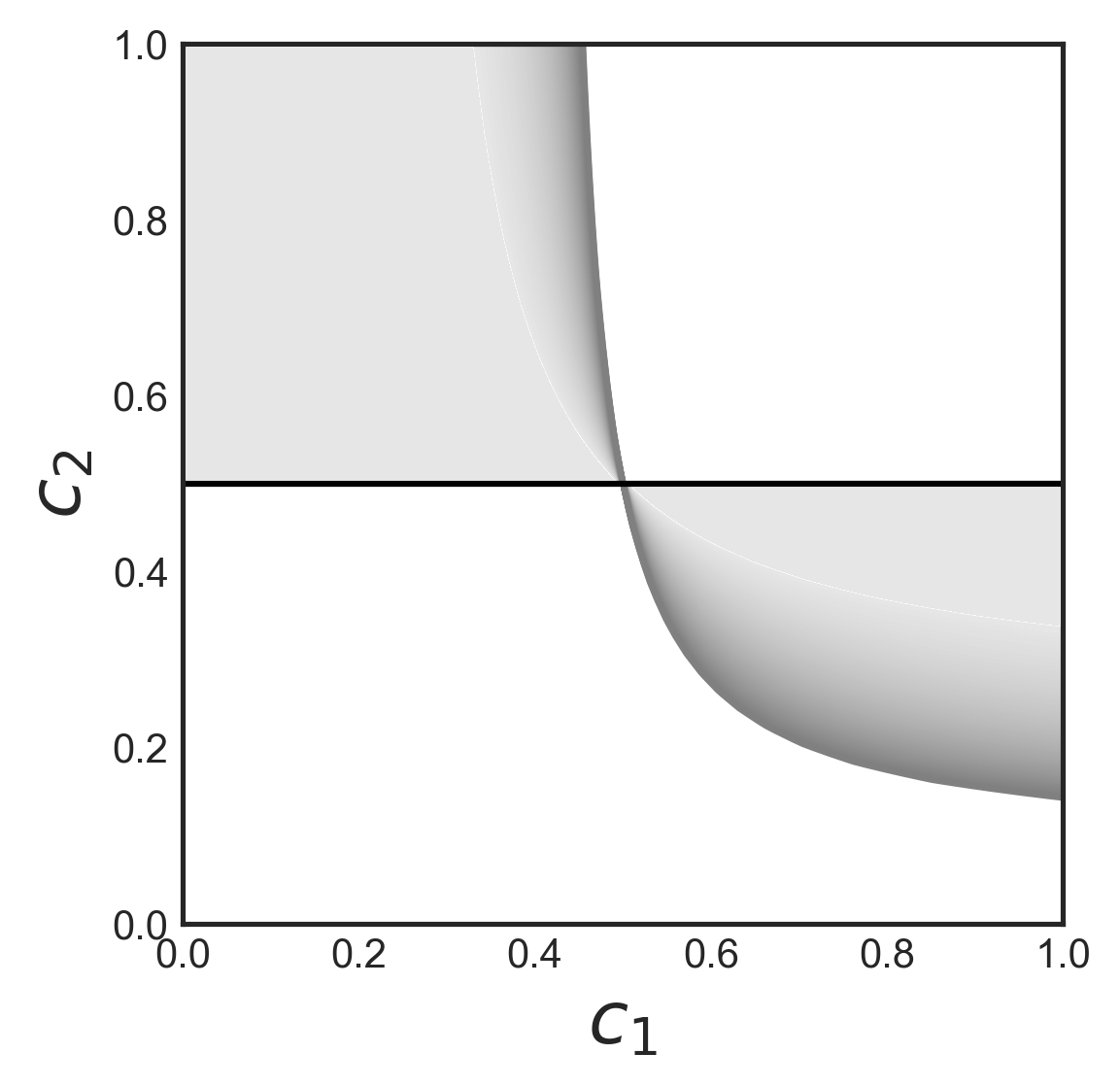}
  \caption{Incomplete part (gray area) of preference relation in Example 2}
  \label{fig:4}
\end{figure}

With the attention degree $\chi$ staying a scaling factor, the dimensionality reduction condition (\ref{2.2.6}) is satisfied, rendering the index set $\mathcal{I}$ reducible to the subset $J=\{0,\lambda W^{\uparrow}+1\}$ ($\mathrm{card}J=2$). Thus, the ordering cone is specifiable to $\mathcal{K}\equiv\mathds{R}^{2}_{+}\equiv\mathcal{K}^{\dag}$. Obviously, the investor's preference relation $\succeq$ is time-varying and addiction increases perceived importance of the first good and hence reduces its quantity that the investor is willing to forego to be better off or refuses to accept not to be worse off, as Figure \ref{fig:4} visualizes.

In applying the Gass--Saaty method, the optimality conditions (\ref{3.2.1}) imply that
\begin{equation*}
  c_{1}=\bigg(\frac{w_{1}\chi_{0}+w_{2}\chi_{\lambda W^{\uparrow}+1}}{\eta\xi e^{\beta\imath}}\bigg)^{1/p}\quad\text{and}\quad c_{2}=(\eta\xi e^{\beta\imath})^{-1/p},
\end{equation*}
with $w\in\mathds{R}^{2}_{+}$ and $\|w\|_{1}=1$, and
\begin{equation*}
  \eta^{1/p}=\frac{1}{X_{0}}\int^{T}_{0}\E\bigg[\frac{\xi^{1-1/p}_{t}}{e^{\beta t/p}}\big((w_{1}\chi_{0}+w_{2}\chi_{\lambda W^{\uparrow}+1})^{1/p}+1\big)\bigg]\dd t.
\end{equation*}
Combining the above conditions we can express the optimal consumption policy $c^{\ast}$ in terms of $w$ only. More specifically, following Case (II) of Example 1 it is not difficult to see the quotient relation $c^{\ast}_{1}/c^{\ast}_{2}\in\big[\chi^{1/p}_{0},\chi^{1/p}_{\lambda W^{\uparrow}+1}\big]$ with the admissible choice $c^{\ast}_{1}=0$. Putting everything together, the set of all $\mathds{R}^{2}_{+}$-maximal solutions sought by the investor has the semi-closed formula
\begin{align}\label{3.2.6}
  \mathcal{S}^{\ast}&=\bigcup_{w\in\mathds{R}^{2}_{+},\;\|w\|_{1}=1} \Bigg\{c^{\ast}\in\mathfrak{C}_{2}:C^{\ast}=\frac{(w_{1}\chi_{0}+w_{2}\chi_{\lambda W^{\uparrow}+1})^{1/p}+1}{(\eta\xi e^{\beta\imath})^{1/p}};\;\frac{c^{\ast}_{1}}{c^{\ast}_{2}}\in\big[\chi^{1/p}_{0},\chi^{1/p}_{\lambda W^{\uparrow}+1}\big]; \nonumber\\
  &\quad\eta^{1/p}=\frac{1}{X_{0}}\int^{T}_{0}\iint_{\mathds{R}_{+}\times(\infty,x_{1}]}\sqrt{\frac{2}{\pi t^{3}}}(2x_{1}-x_{2})\exp\bigg(\bigg(\frac{1}{p}-1\bigg)\bigg(r+\frac{\theta^{2}}{2}\bigg)t+\theta x_{2} \nonumber\\
  &\qquad-\frac{\beta t}{p}-\frac{(2x_{1}-x_{2})^{2}}{2t}\bigg)\big((w_{1}\chi_{0}+w_{2}\chi_{\lambda x_{1}+1})^{1/p}+1\big)\dd(x_{1},x_{2})\dd t\Bigg\},
\end{align}
where we have used the familiar joint density of the random vector $(W^{\uparrow}_{t},W_{t})$ (see, e.g., \cite[Lyasoff, 2017, \text{Sect.} 8.116]{L2}).

The result (\ref{3.2.6}) shows that when indecisiveness is time-varying uniqueness is not guaranteed even in the optimal total consumption expenditure, and the investor can flexibly specify optimal consumption paths under various choices of $w$. The optimal consumption quotient is a stochastic version of what has been obtained in Case (II) of Example 1. The larger the scale parameter $\lambda$, the more intensely the investor perceives the importance of the first consumption good amid socialization, leading to a larger permissible proportion of its consumption under optimality. \hfill{\scriptsize $\lozenge$}

\vspace{0.1in}

The setup of Example 2 has illustrated external influence on fluctuating imprecise tastes through a channel of importance perception of nonessential goods while maintaining necessity-based living standards. Viewing from outcomes, it is shown that when the investor has time-varying incomplete preferences over goods, his consumption policy is expected to adjust dynamically. Still, he places more emphasis on the second good by imposing a stochastic upper bound on the relative consumption level of the first. As the multi-utility index set $\mathcal{I}$ expands over time, he gains flexibility in altering his consumption policies. Example 2 also leaves room for exploring what velocity is the most appropriate for increasing indecisiveness, which is incorporated into the parametric function $\chi$, with the most conservative estimate being a linear one; in contrast, the additional parameter $\lambda>0$ controls the maximal acceleration of increases.

\vspace{0.1in}

\textbf{Example 3. (Socialization, market volatility, and changing indecisiveness)}\quad This is a hybrid situation where indecisiveness changes as a result of both a socialization effect and fluctuating market volatility. We keep the basic setup of consumption goods in Example 2 but suppose now that the market volatility is driven by an exponential Ornstein--Uhlenbeck process,
\begin{equation}\label{3.2.7}
  \sigma_{t}=\exp\bigg((\log\sigma_{0})e^{-\kappa t}+\varsigma\int^{t}_{0}e^{-\kappa(t-s)}\dd W_{s}\bigg),\quad t\in[0,T],
\end{equation}
with additional parameters $\sigma_{0}>0$, $\kappa>0$, and $\varsigma<0$, which stand for the initial volatility, mean reversion speed, and volatility of volatility with leverage effect, in proper order. In addition to attention degrees, the investor's risk aversion also changes with the market volatility, through another channel; this points especially to curvature increases of a utility function driven by fear (see, again, \cite[Guiso et al., 2018]{GSZ}). Establishing volatility-dependent indecisiveness is, in the matter of utility maximization, comparable to the investor with constant volatility-driven risk aversion viewing the volatility process as being set-valued, whereas there is no aversion to ambiguity as imprecision is in preferences.\footnote{For a formal treatment of ambiguous volatility we mention \cite[Epstein and Ji, 2013]{EJ}; along their lines is the proposed structure (\ref{2.2.1}) also an adequate generalization as the components $f_{q}$'s and $G_{q}$'s permit dependence on more sophisticated statistical measures such as the reversion speed and the volatility of volatility, apart from the persistence (drift) of the volatility of returns that is at least partially observable to the investor.} Given two mechanisms of changing preferences, we have $d=2$ and for simplicity assume that they are mutually independent, to construct the multi-utility index set process as a random rectangular area. In addition, the investor now values his bequests from investment.

In (\ref{2.2.3}), we have $\mathcal{R}=\mathds{R}_{+}\times[1,\infty)$, $I_{1,0}=\{(0,\sigma_{0})^{\intercal}\}$, $I_{2,0}=\{\0\}$, $I_{3,0}=[0,1]\times[1,2]$, $f_{1}=\{(0,\kappa(\varsigma^{2}/(2\kappa)-\log\sigma)\sigma)^{\intercal}\}$, $f_{2}=f_{3}=\{\0\}$, $G_{1}=\{(0,\varsigma\sigma)^{\intercal}\}$, $G_{2}\equiv\{\0\}$ and $G_{3}=\{\0,(\lambda,0)^{\intercal}\}$. It is clear that the conditions in (\ref{2.2.4}) are satisfied and a straightforward application of It\^{o}'s formula to $\sigma$ (see, e.g., \cite[Lyasoff, 2017, \text{Sect.} 11.60]{L2}) yields with (\ref{3.2.7})
\begin{equation*}
  \mathcal{I}_{t}=[0,\lambda W^{\uparrow}_{t}+1]\times[\sigma_{t}+1,\sigma_{t}+2],\quad t\in[0,T],
\end{equation*}
and $\bar{\mathcal{I}}=\mathds{R}_{+}\times[1,\infty)$.

We design the multi-utility in a similar fashion as
\begin{equation*}
  u_{i}(t,c)=
  \begin{cases}
    \displaystyle \frac{\chi_{i_{1}}(c^{1-p_{i_{2}}}_{1}-1)+(c^{1-p_{i_{2}}}_{2}-1)}{e^{\beta t}(1-p_{i_{2}})},\quad&\text{if }i\in\mathcal{I}_{t},\\
    0,\quad&\text{o.w.},
  \end{cases}
\end{equation*}
where all elements are as specified in Example 2 except that the risk aversion parameter $p:[1,\infty)\mapsto\mathds{R}_{++}\setminus\{1\}$ is some bounded nondecreasing function. Risk aversion is still assumed to be identical across consumption goods. The bequest utility function is taken to be $U(x)=e^{-\beta T}(x^{1-p_{\circ}}-1)/(1-p_{\circ})$, for $x>0$, under the same subjective discount factor and a terminal wealth-specific risk aversion coefficient $p_{\circ}\in\mathds{R}_{++}\setminus\{1\}$.

\begin{figure}[H]
  \centering
  \includegraphics[scale=0.23]{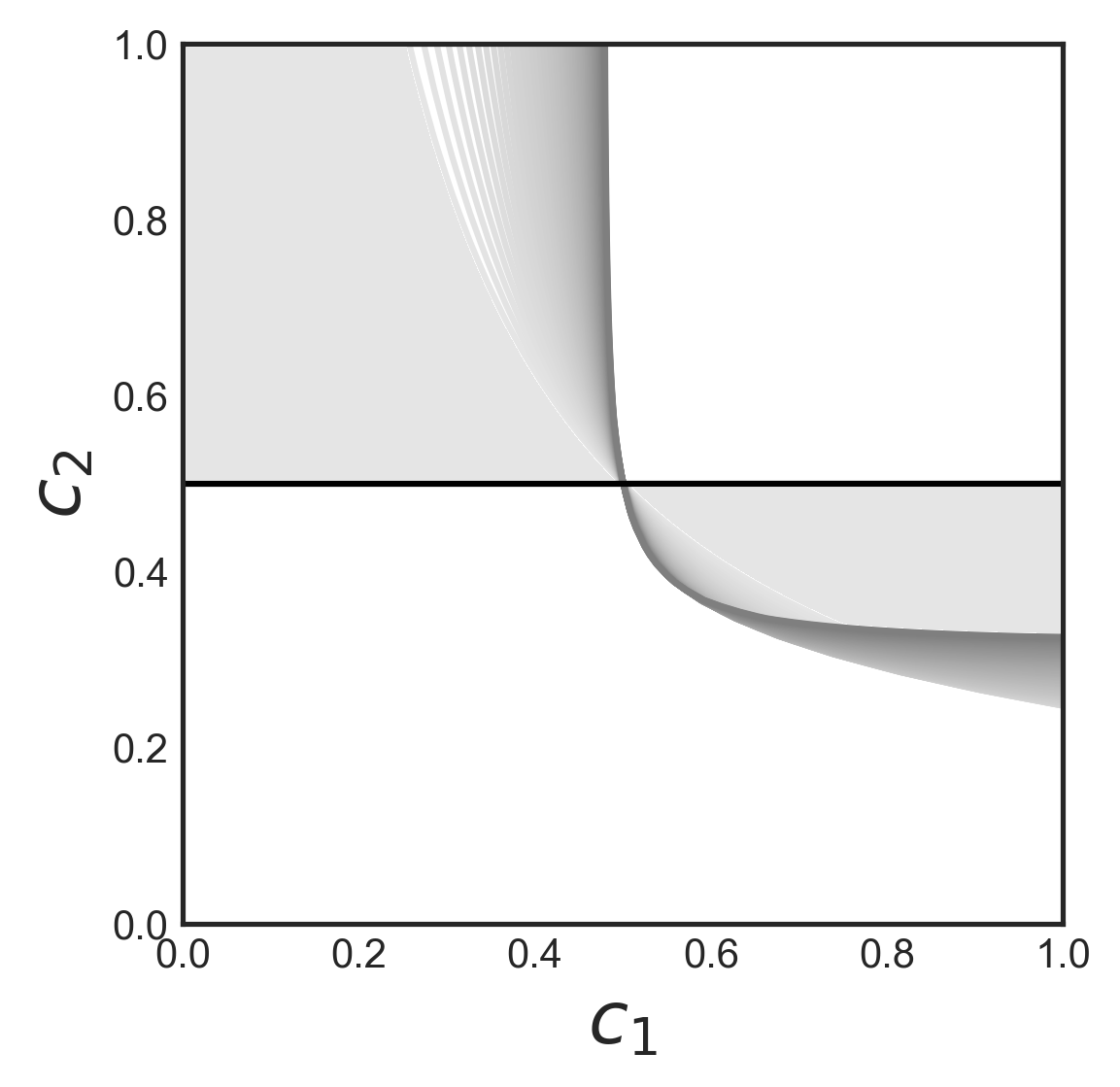}
  \caption{Incomplete part (gray area) of preference relation in Example 3}
  \label{fig:5}
\end{figure}

Since the risk aversion $p$ is a shape parameter with no scaling-invariance properties, the dimensionality reduction condition (\ref{2.2.6}) fails and as a consequence the corresponding ordering cones are necessarily time-varying (consult Figure \ref{fig:2}): $\mathcal{K}=\mathcal{C}_{\rm b}(\mathcal{I};\mathds{R}_{+})$ and $\bar{\mathcal{K}}=\mathcal{C}_{\rm b}(\mathds{R}_{+}\times[1,\infty);\mathds{R}_{+})$, which have nonempty interiors. The preference relation $\succeq$ is time-varying as well, whose incomplete part moves according to rotation ($\chi$-channel) and curvature ($p$-channel). Intuitively, while addiction and rising market volatility both reduce the quantity of the first good the investor benefits from foregoing, rising market volatility contrarily increases the quantity he is certainly worse off by accepting, due to increased risk aversion. This is visualized in Figure \ref{fig:5}.

For a linear functional $w_{2}\in(\mathcal{C}_{\rm b}([\sigma+1,\sigma+2];\mathds{R}_{+}))^{\dag}$ we define the following risk-linked function:
\begin{equation}\label{3.2.8}
  \mathds{R}_{+}\ni x\mapsto\vartheta_{p,w_{2};\sigma}(x):=\int^{\sigma+2}_{\sigma+1}w_{2,i_{2}}x^{-p_{i_{2}}}\dd i_{2}\in[0,\infty],
\end{equation}
which is strictly decreasing if $\|w_{2}\|_{1}>0$. Then the optimality conditions in (\ref{3.2.1}) specialize to
\begin{equation*}
  c_{1}=\vartheta^{-1}_{p,w_{2};\sigma}\bigg(\frac{\eta\xi e^{\beta\imath}}{\langle w_{1},\chi\rangle_{[0,\lambda W^{\uparrow}+1]}}\bigg),\quad c_{2}=\vartheta^{-1}_{p,w_{2};\sigma}\bigg(\frac{\eta\xi e^{\beta\imath}}{\langle w_{1},\1\rangle_{[0,\lambda W^{\uparrow}+1]}}\bigg)
\end{equation*}
and
\begin{equation*}
  X_{T}=\bigg(\frac{\int^{T}_{0}\|w(t)\|_{1}\dd t}{\eta\xi_{T}Te^{\beta T}}\bigg)^{1/p_{\circ}},
\end{equation*}
with the understanding that $w=w_{1}\otimes w_{2}$. Similarly to Case (III) of Example 1, given $w$, the budget set uniquely determines the multiplier $\eta$, which in turn makes the total consumption expenditure $C^{\ast}$ undeterminable. Note also that in this setting the market price of risk $\theta=(\mu-r)/\sigma$ is stochastic, with the state price density $\xi=\exp(-\int^{\imath}_{0}(r+\theta^{2}_{s}/2)\dd s+\int^{\imath}_{0}\theta_{s}\dd W_{s})$. After some simplifications the set of all $\bar{\mathcal{K}}$-maximal solutions can be expressed as follows:
\begin{align}\label{3.2.9}
  \mathcal{S}^{\ast}=&\bigcup_{\substack{\bar{w}\in(\mathcal{C}_{\rm b}(\mathds{R}_{+}\times[1,\infty);\mathds{R}_{+}))^{\dag}, \\ \|\bar{w}\|_{1}=1}}\Bigg\{(c^{\ast},X^{\ast}_{T})\in\mathfrak{C}_{2}\times\mathds{L}^{1}_{\mathcal{F}}(\Omega;\mathds{R}_{+}): c^{\ast}_{1}=\vartheta^{-1}_{p,\bar{w}_{2};\sigma}\bigg(\frac{\eta\xi e^{\beta\imath}}{\int^{\lambda W^{\uparrow}+1}_{0}\bar{w}_{1,i_{1}}\chi_{i_{1}}\dd i_{1}}\bigg);\nonumber\\
  &\quad c^{\ast}_{2}=\vartheta^{-1}_{p,\bar{w}_{2};\sigma}\bigg(\frac{\eta\xi e^{\beta\imath}}{\int^{\lambda W^{\uparrow}+1}_{0}\bar{w}_{1,i_{1}}\dd i_{1}}\bigg);\; X^{\ast}_{T}=\bigg(\frac{\int^{T}_{0}\int_{[0,\lambda W^{\uparrow}_{t}+1]\times[\sigma_{t}+1,\sigma_{t}+2]}\bar{w}_{i}\dd i\dd t}{\eta\xi_{T}Te^{\beta T}}\bigg)^{1/p_{\circ}}; \nonumber\\ &\quad\eta\rightsquigarrow\int^{T}_{0}\E\bigg[\xi_{t}\bigg(\vartheta^{-1}_{p,\bar{w}_{2};\sigma_{t}}\bigg(\frac{\eta\xi_{t}e^{\beta t}}{\int^{\lambda W^{\uparrow}_{t}+1}_{0}\bar{w}_{1,i_{1}}\chi_{i_{1}}\dd i_{1}}\bigg)+\vartheta^{-1}_{p,\bar{w}_{2};\sigma_{t}}\bigg(\frac{\eta\xi_{t}e^{\beta t}}{\int^{\lambda W^{\uparrow}_{t}+2}_{1}\bar{w}_{1,i_{1}}\dd i_{1}}\bigg)\bigg)\bigg]\dd t \nonumber\\
  &\qquad+\E\Bigg[\xi_{T}^{1-1/p_{\circ}}\bigg(\frac{\int^{T}_{0}\int_{[0,\lambda W^{\uparrow}_{t}+1]\times[\sigma_{t}+1,\sigma_{t}+2]}\bar{w}_{i}\dd i\dd t}{\eta Te^{\beta T}}\bigg)^{1/p_{\circ}}\Bigg]=X_{0}\Bigg\},
\end{align}
with $\bar{w}=\bar{w}_{1}\otimes\bar{w}_{2}$ and $\vartheta^{-1}_{p,\bar{w}_{2};\sigma}\equiv\vartheta^{-1}_{p,\bar{w}_{2}\upharpoonright_{[\sigma+1,\sigma+2]};\sigma}$ and where $\eta$ can only be computed using simulation (see Section \ref{s:6}) based on (\ref{3.2.7}) and (\ref{3.2.8}). Since in the current setting the major difference from Example 2 is the inclusion of changing risk aversion, let us focus on the $\bar{w}_{2}$-dimension. For example, if we restrict our attention to extended weights of the degenerate form $\bar{w}_{2}=\{\delta_{\{\varepsilon+2\}}(i_{2}):i_{2}\geq1\}$ for some small $\varepsilon>0$ indicating a perceived minimal volatility, in which case $\vartheta^{-1}_{p,\bar{w}_{2};\sigma}(x)=x^{-1/p_{\varepsilon+2}}$ for $\sigma\in[\varepsilon,\varepsilon+1]$, then a subset of (\ref{3.2.9}) reads
\begin{align}\label{3.2.10}
  \tilde{\mathcal{S}}^{\ast}=&\bigcup_{\substack{\bar{w}_{2}\in(\mathcal{C}_{\rm b}([1,\infty);\mathds{R}_{+}))^{\dag},\\ \|\bar{w}_{1}\|_{1}=1,
  \varepsilon>0}}\Bigg\{(c^{\ast},X^{\ast}_{T})\in\mathfrak{C}_{2}\times\mathds{L}^{1}_{\mathcal{F}}(\Omega;\mathds{R}_{+}):\; c^{\ast}_{1}=\bigg(\frac{\int^{\lambda W^{\uparrow}+1}_{0}\bar{w}_{1,i_{1}}\chi_{i_{1}}\dd i_{1}}{\eta\xi e^{\beta\imath}}\bigg)^{1/p_{\varepsilon+2}}; \nonumber\\
  &\quad c^{\ast}_{2}=\bigg(\frac{\int^{\lambda W^{\uparrow}+1}_{0}\bar{w}_{1,i_{1}}\dd i_{1}}{\eta\xi e^{\beta\imath}}\bigg)^{1/p_{\varepsilon+2}};\;X^{\ast}_{T}=\bigg(\frac{\int^{T}_{0}\int^{\lambda W^{\uparrow}_{t}+1}_{0}\bar{w}_{i_{1}}\dd i_{1}\dd t}{\eta\xi_{T}Te^{\beta T}}\bigg)^{1/p_{\circ}}; \nonumber\\ &\;\eta\rightsquigarrow\int^{T}_{0}\E\Bigg[\frac{\xi^{1-1/p_{\varepsilon+2}}_{t}}{e^{\beta t/p_{\varepsilon+2}}}\Bigg(\bigg(\int^{\lambda W^{\uparrow}_{t}+1}_{0}\bar{w}_{1,i_{1}}\chi_{i_{1}}\dd i_{1}\bigg)^{1/p_{\varepsilon+2}}+\bigg(\int^{\lambda W^{\uparrow}_{t}+1}_{0}\bar{w}_{1,i_{1}}\dd i_{1}\bigg)^{1/p_{\varepsilon+2}}\Bigg)\Bigg]\dd t \nonumber\\
  &\qquad\times\eta^{-1/p_{\varepsilon+2}}+\E\Bigg[\frac{\xi_{T}^{1-1/p_{\circ}}}{(Te^{\beta T})^{1/p_{\circ}}}\bigg(\int^{T}_{0}\int^{\lambda W^{\uparrow}_{t}+1}_{0}\bar{w}_{1,i_{1}}\dd i_{1}\dd t\bigg)^{1/p_{\circ}}\Bigg]\eta^{-1/p_{\circ}}=X_{0}\Bigg\},
\end{align}
provided $\sigma_{t}\in[\varepsilon,\varepsilon+1]$, $\forall t\in[0,T]$. The equation of $\eta$ is of the same type as (\ref{3.2.5}), which can only be solved explicitly when $p_{\varepsilon+2}=p_{\circ}$. Besides, (\ref{3.2.10}) shows that the consumption--bequest policies are optimal up to a variety of risk aversion coefficients $p_{\varepsilon+2}$ restricted by the value of $\sigma$; as $\sigma$ increases, so do the permissible values of $\varepsilon$. Since the index span of volatility-driven indecisiveness is constant ($\mathrm{Leb}_{\mathds{R}}([\sigma+1,\sigma+2])\equiv1$), the overall flexibility in adjusting such policies is expected to stay unchanged over time, whereas the actual optimal policies will become more conservative or condensed as the power $1/p_{\varepsilon+2}$ decreases, which implies decreasing indecisiveness as duly explicable by increased risk aversion. On the other hand, the lower the mean reversion speed $\kappa$ or the higher the volatility of volatility $\varsigma$ in (\ref{3.2.7}), the more prone $\sigma$ is to rising to a higher level. \hfill{\scriptsize $\lozenge$}

\vspace{0.1in}

To summarize, Example 3 establishes a tractable framework for modeling situations where imprecise tastes over consumption goods undergo both socialization effects and influence from market volatility. As in Example 2, the former is realized through perception of the importance of nonessential goods while the latter is linked to volatility-dependent risk aversion. Under independence assumptions, different channels of changing indecisiveness can be modeled by a Cartesian product of random sets which allows the impact on optimal policies to be interpreted separately. From the outcome perspective, even if increased market volatility does not shrink the index span of indecisiveness, indecisiveness can well decrease in terms of an ``effective'' diversity of optimal policies once the investor has become more averse to market risks.

Altogether, the three examples have shown that while there is sufficient space for designing the multi-utility and analyzing the incomplete part of its induced preference relation in solving a specific problem of interest, it is somewhat exorbitant to hope for a simple expression for the full set of solutions, especially when incomplete preferences also fluctuate.

\vspace{0.2in}

\section{Solution II: Duality}\label{s:4}

In this section we present Fenchel-type duality results for the multi-utility maximization problem (\ref{2.3.2}), or its static equivalent (\ref{2.3.5}). The duality method provides yet another, arguably more intuitive way of thinking about the problem than direct scalarization. We employ the convex duality theory developed in \cite[Hamel, 2009]{H} and \cite[Hamel and L\"{o}hne, 2014]{HL} (see also the overview in \cite[Hamel et al., 2015, \text{Chap.} II]{HHLRS}) for multifunctions and attempt to incorporate it into our stochastic optimization setting. It is worth highlighting that, instead of analyzing the boundaries of the criterion space, this inherently set-valued way of solving the problem is based on establishing a lattice structure for ranking sets (of adjusted shapes) and is yet capable of fully recovering solutions that are (weakly) maximal in the sense of Definition \ref{def:4}, regardless of problem convexity.

It is enough to consider the general indexing with $\mathcal{I}$, ignoring the dimensionality reduction condition (\ref{2.2.6}). Let us be reminded that $\bar{\mathcal{I}}$ is defined as the smallest closed subset of $\mathds{R}^{d}$ covering $\sup_{t\in[0,T]}\mathcal{I}_{t}$ $\PP$-a.s. To give a meaningful rank for sets, a modified version of the problem (\ref{2.3.5}) is introduced first,
\begin{equation}\label{4.1}
  \sup_{(c,X_{T})\in\mathfrak{B}(X_{0})}(V(c,X_{T})-\bar{\mathcal{K}}),
\end{equation}
where $\bar{\mathcal{K}}$ is recalled to be the select closed convex cone over $\bar{\mathcal{I}}$ and which will be referred to as the primal problem in this section.

According to \cite[Hamel et al., 2015, \text{Chap.} II]{HHLRS}, we can define a complete lattice on the space $\mathcal{C}_{\rm b}(\bar{\mathcal{I}};\mathds{R})$ and the cone $\bar{\mathcal{K}}$ by constructing
\begin{equation*}
  \mathcal{L}(\mathcal{C}_{\rm b}(\bar{\mathcal{I}};\mathds{R}),\bar{\mathcal{K}}):= \big\{A\in\mathcal{P}(\mathcal{C}_{\rm b}(\bar{\mathcal{I}};\mathds{R})):A=\overline{\co}_{\mathcal{C}_{\rm b}}(A-\bar{\mathcal{K}})\big\},
\end{equation*}
along with the closed Minkowski addition $\oplus$ and multiplication by a nonnegative number $\alpha\geq0$; for any $A,B\in\mathcal{L}(\mathcal{C}_{\rm b}(\bar{\mathcal{I}};\mathds{R}),\bar{\mathcal{K}})$, we write
\begin{equation*}
  A\oplus B=
  \begin{cases}
    \cl_{\mathcal{C}_{\rm b}}\{a+b:a\in A,\;b\in B\},\quad&\text{if }A,B\neq\emptyset,\\
    \emptyset,\quad&\text{o.w.}
  \end{cases}
\end{equation*}
and
\begin{equation*}
  \alpha A=
  \begin{cases}
    \{\alpha a:a\in A\}\quad&\text{if }\alpha>0,A\neq\emptyset,\\
    \emptyset\quad&\text{if }\alpha>0,A=\emptyset,\\
    -\bar{\mathcal{K}},\quad&\text{if }\alpha=0.
  \end{cases}
\end{equation*}
It is seen that the above construction makes it possible to regard the set inclusion $\subseteq$ as a partial order on $\mathcal{L}(\mathcal{C}_{\rm b}(\bar{\mathcal{I}};\mathds{R}),\bar{\mathcal{K}})$ so that optimality of the modified problem (\ref{4.1}) can be understood in the complete lattice $(\mathcal{L}(\mathcal{C}_{\rm b}(\bar{\mathcal{I}};\mathds{R}),\bar{\mathcal{K}}),\subseteq)$ as
\begin{equation}\label{4.2}
  \sup\tilde{V}(c,X_{T})=\overline{\co}_{\mathcal{C}_{\rm b}}\bigcup_{\varpi\in\tilde{V}(c,X_{T})}\varpi,\quad(c,X_{T})\in\mathfrak{B}(X_{0})
\end{equation}
with $\tilde{V}(c,X_{T}):=V(c,X_{T})-\bar{\mathcal{K}}$ and the agreement that $\tilde{V}\equiv\emptyset$ whenever $V\equiv-\infty$. Similarly, for every $t\in[0,T]$ we can construct a time-dependent (complete) lattice on $(\mathcal{C}_{\rm b}(\mathcal{I}_{t};\mathds{R}),\mathcal{K}_{t})$,
\begin{equation*}
  \mathcal{L}(\mathcal{C}_{\rm b}(\mathcal{I}_{t};\mathds{R}),\mathcal{K}_{t}):=\big\{A\in\mathcal{P}(\mathcal{C}_{\rm b}(\mathcal{I}_{t};\mathds{R})): A=\overline{\co}_{\mathcal{C}_{\rm b}}(A-\mathcal{K}_{t})\big\},
\end{equation*}
and the (modified) Minkowski and multiplication operations are understood in the same way. We remark the requirement that $\bar{\mathcal{K}}\supseteq\overline{\co}_{\mathcal{C}_{\rm b}}\bigcup_{t\in[0,T]}\mathcal{K}_{t}$ $\PP$-a.s.

The next definition is about a modified version for the multi-utility $u$ which helps introduce the notion of (set-valued) Lagrangian duality.

\begin{definition}\label{def:5}
For every $t\in[0,T]$ and the multi-utility given in (\ref{2.2.5}), define
\begin{equation}\label{4.3}
  \tilde{u}(t,c):=u(t,c)-\mathcal{K}_{t}.
\end{equation}
The (set-valued) Fenchel--Legendre conjugate of $\tilde{u}$ is then defined as the function
\begin{equation}\label{4.4}
  (-\tilde{u})^{\dag}(t,y,\lambda):=\sup_{c\in\mathds{R}^{n}_{+}}\{\tilde{u}(t,c)\oplus\{z\in\mathcal{C}_{\rm b}(\mathcal{I}_{t};\mathds{R}):\langle y,c\rangle_{n}-\langle\lambda,z\rangle_{\mathcal{I}_{t}}\geq0\}\},\quad(y,\lambda)\in\mathds{R}^{n}_{+}\times\mathcal{K}^{\dag}_{t}
\end{equation}
and its (set-valued) bi-conjugate as
\begin{align*}
  \tilde{u}^{\ddag}(t,c)&:=\inf_{(y,\lambda)\in\mathds{R}^{n}_{+}\times\mathcal{K}^{\dag}_{t}} \{(-\tilde{u})^{\dag}(t,y,\lambda)\oplus\{z\in\mathcal{C}_{\rm b}(\mathcal{I}_{t};\mathds{R}):\langle y,c\rangle_{n}+\langle\lambda,z\rangle_{\mathcal{I}_{t}}\leq0\}\}\\
  &=\bigcap_{(y,\lambda)\in\mathds{R}^{n}_{+}\times\mathcal{K}^{\dag}_{t}} \{(-\tilde{u})^{\dag}(t,y,\lambda)\oplus\{z\in\mathcal{C}_{\rm b}(\mathcal{I}_{t};\mathds{R}):\langle y,c\rangle_{n}+\langle\lambda,z\rangle_{\mathcal{I}_{t}}\leq0\}\}
\end{align*}
\end{definition}

Note that in the above definition all of $\tilde{u}(t,\cdot)$, $\tilde{u}^{\dag}(t,\cdot,\cdot)$ and $\tilde{u}^{\ddag}(t,\cdot)$ are valued in the lattice $\mathcal{L}(\mathcal{C}_{\rm b}(\mathcal{I}_{t};\mathds{R}),\mathcal{K}_{t})$, at time $t\in[0,T]$. In particular, based on Assumption \ref{as:1}, they preserve two desirable properties from the single-valued setting.

\begin{theorem}\label{thm:3}
For every $t\in[0,T]$, let $u(t,\cdot)\in\mathfrak{U}_{\mathcal{I}_{t}}$ and $\tilde{u}(t,\cdot)$ be its modified version in (\ref{4.3}). Then the following assertions hold. \vspace{0.1in}\\
(i) $\tilde{u}^{\ddag}(t,\cdot)=\tilde{u}(t,\cdot)$. \vspace{0.1in}\\
(ii) $\tilde{u}(t,c)\oplus\{z\in\mathcal{C}_{\rm b}(\mathcal{I}_{t};\mathds{R}):\langle y,c\rangle_{n}-\langle\lambda,z\rangle_{\mathcal{I}_{t}}\geq0\}\subseteq(-\tilde{u})^{\dag}(t,y,\lambda)$, for any $y\in\mathds{R}^{n}_{+}$ and $\lambda\in\mathcal{K}^{\dag}_{t}$.
\end{theorem}

As a remark, assertion (ii) is nothing but a set-valued version of the well-known Fenchel--Young inequality put into the complete lattice. Likewise, for the bequest utility function, its Fenchel--Legendre conjugate is simply $(-U)^{\dag}(y)=\sup_{x\geq0}(yx-U(x))$ and it is familiar that $U^{\ddag}=U$ and $U(x)-yx\leq(-U)^{\dag}(-y)$ for $y\geq0$.

Now let us return to the primal (modified) problem in (\ref{4.1}), but before the results of Theorem \ref{thm:3} can be reasonably applied, a notion of set-valued Lagrangian is needed.

\begin{definition}\label{def:6}
For a dual variable $y\in\big(\mathfrak{C}_{n}\times\mathds{L}^{1}_{\mathcal{F}}(\Omega;\mathds{R}_{+})\big)^{\dag}$ and a multiplier $\bar{\lambda}\in\bar{\mathcal{K}}^{\dag}$, the Lagrangian for the primal problem (\ref{4.1}) is given by the $\mathcal{L}(\mathcal{C}_{\rm b}(\bar{\mathcal{I}};\mathds{R}),\bar{\mathcal{K}})$-valued function
\begin{equation}\label{4.5}
  L((c,X_{T}),y,\bar{\lambda}):=\tilde{V}(c,X_{T})\oplus\bigcup_{(c,X_{T})\in\mathfrak{B}(X_{0})}\big\{z\in\mathcal{C}_{\rm b}(\bar{\mathcal{I}};\mathds{R}): y((c,X_{T}))-\langle\bar{\lambda},z\rangle_{\bar{\mathcal{I}}}\geq0\big\}.
\end{equation}
\end{definition}

Just as in single-criterion optimization problems, by forming the Lagrangian we want to transform the constrained primal problem into an unconstrained dual problem. In light of the previous definition, the objective function of such a dual problem is taken to be the supremum (in the sense of (\ref{4.2})) of the Lagrangian,
\begin{equation*}
  \mathcal{H}(y,\bar{\lambda}):=\sup_{(c,X_{T})\in\mathfrak{C}_{n}\times\mathds{L}^{1}_{\mathcal{F}}(\Omega;\mathds{R}_{+})} L((c,X_{T}),y,\bar{\lambda}),
\end{equation*}
giving rise to the following dual problem:
\begin{equation}\label{4.6}
  \inf_{\substack{y\in(\mathfrak{C}_{n}\times\mathds{L}^{1}_{\mathcal{F}}(\Omega;\mathds{R}_{+}))^{\dag},\\ \bar{\lambda}\in\bar{\mathcal{K}}^{\dag}}}\mathcal{H}(y,\bar{\lambda}) =\bigcap_{\substack{y\in(\mathfrak{C}_{n}\times\mathds{L}^{1}_{\mathcal{F}}(\Omega;\mathds{R}_{+}))^{\dag},\\ \bar{\lambda}\in\bar{\mathcal{K}}^{\dag}}}\mathcal{H}(y,\bar{\lambda}).
\end{equation}

As the next proposition shows, it suffices to consider an extremal version of the dual problem (\ref{4.6}), which facilitates identification of maximal solutions in the sense of (\ref{4.2}) to some extent.

\begin{proposition}\label{pro:5}
The set of maximal solutions for the (modified) primal problem (\ref{4.1}) is given by
\begin{align}\label{4.7}
  \tilde{\mathcal{S}}^{\ast}&=\bigcap_{\eta\geq0}\big\{(c^{\ast},X^{\ast}_{T})\in\mathfrak{C}_{n} \times\mathds{L}^{1}_{\mathcal{F}}(\Omega;\mathds{R}_{+}):\; \tilde{u}(t,c^{\ast}_{t})-\eta\xi_{t}C^{\ast}_{t}=(-\tilde{u})^{\dag}(t,-\eta\xi_{t}\1,\0); \nonumber\\
  &\qquad U(X^{\ast}_{T})-\eta\xi_{T}X^{\ast}_{T}=(-U)^{\dag}(-\eta\xi_{T})\big\},
\end{align}
where $(-\tilde{u})^{\dag}$ is as given in (\ref{4.4}).
\end{proposition}

Proposition \ref{pro:5} provides a different perspective towards characterizing the solution set of the multi-utility maximization problem.
Specifically, present the additional conditions in Assumption \ref{as:3}, then to ensure attainment of the equality in (\ref{4.7}) one immediately obtains $X^{\ast}_{T}=(U')^{-1}(-\eta\xi_{T})$ (subject to nonnegativity), with $\eta>0$, while resorting to scalarization techniques on the modified multi-utility $\tilde{u}(\imath,c)$ one also has
\begin{equation*}
  \langle\tilde{w},u^{(j)}(\imath,c)\rangle_{\mathcal{I}}\leq\eta\xi,\quad j\in\mathds{N}\cap[1,n],
\end{equation*}
for some linear functional $\tilde{w}\in\bar{\mathcal{K}}^{\dag}$ with $\|\tilde{w}\|_{1}>0$ (the same as $\bar{w}$ that appears in (\ref{A.5})). These inferred results are equivalent to the optimality conditions (\ref{3.2.1}), up to scaling the multiplier, as $\int^{T}_{0}\langle w(t),\1\rangle_{\mathcal{I}_{t}}\dd t\neq0$, $\PP$-a.s. Speaking of outcomes, the duality method essentially leads to the same characterization of solutions of the utility maximization problem as with the direct Gass--Saaty scalarization method.

\vspace{0.2in}

\section{Optimal investment}\label{s:5}

After finding the optimal consumption--bequest policy $(c^{\ast},X^{\ast}_{T})$ for the equivalent static problem (\ref{2.3.5}), the next step is to determine the optimal investment policy $\Pi^{\ast}$ that guarantees attainment of the optimal bequest in connection with the original dynamic problem (\ref{2.3.2}). We notice that the optimal portfolio only depends on consumption through the total expenditure, $C^{\ast}$, apart from the optimal wealth $X^{\ast}$.

\subsection{Portfolio structure}\label{ss:5.1}

As we have seen from the previous three examples, even if optimal consumption levels are multi-valued, the total consumption expenditure can still be single-valued, which may occur when $\mathcal{I}$ is a constant set and no bequest utility is present (compare Cases (I) and (II) with Case (III) of Example 1). However, fluctuations in the commodity prices $P$ aside, when the investor's imprecise tastes are time-varying (even deterministically), or if he also values his terminal wealth, the optimal tuple $(C^{\ast},X^{\ast}_{T})$ is generally an $\mathbb{F}$-non-anticipating set-valued process augmented by a measurable set-valued random variable (consult Example 2 and Example 3). With the Gass--Saaty method, this set-valued tuple is parameterized by the weight $w$, and hence finding the set of optimal portfolios consists in computing $\Pi^{\ast}$ given each $\mathbb{F}$-non-anticipating selector of $C^{\ast}$ as a set-valued process.

Let us consider the solution set with total consumption expenditure,
\begin{equation*}
  \bar{\mathcal{S}}^{\ast}:=\{(C^{\ast},X^{\ast}_{T}):(c^{\ast},X^{\ast}_{T})\in\mathcal{S}^{\ast}\},
\end{equation*}
where $\mathcal{S}^{\ast}$ is as in (\ref{3.1.3}). Then, for every $(C^{\ast},X^{\ast}_{T})\in\bar{\mathcal{S}}^{\ast}$, the optimal wealth process can be written as the present value of future cash flows, namely
\begin{equation*}
  X^{\ast}_{t}=\xi^{-1}_{t}\E\bigg[\int^{T}_{t}\xi_{s}C^{\ast}_{s}\dd s+\xi_{T}X^{\ast}_{T}\bigg|\mathscr{F}_{t}\bigg],\quad t\in[0,T].
\end{equation*}

For the following results we inherit the smoothness assumptions of the utility (Assumption \ref{as:3}) and focus on the Gass--Saaty method. Consulting the optimality conditions (\ref{3.2.1}) together with the supporting remarks we can write
\begin{equation*}
  C^{\ast}=\Psi_{\mathcal{I}}(\eta\xi P|w):=\langle P,\psi_{\mathcal{I}}(\eta\xi P|w)\rangle_{n}\quad\text{and}\quad X^{\ast}_{T}=(U')^{-1}\bigg(\frac{\eta\xi_{T}T}{\int^{T}_{0}\langle w(t),\1\rangle_{\mathcal{I}_{t}}\dd t}\bigg).
\end{equation*}
The following main theorem of this section gives an implementable characterization of the entire set of optimal investment policies, which comes from the art of Malliavin calculus in a set-valued setting.

\begin{theorem}\label{thm:4}
The optimal investment policy is given by the set-valued process
\begin{align}\label{5.1.1}
  \Pi^{\ast}_{t}&=\cl_{\mathds{L}^{1}}\bigg\{\xi^{-1}_{t}\E\bigg[\int^{T}_{t}\xi_{s}P^{\intercal}_{s}\gamma_{\mathcal{I}_{s}}(\eta\xi_{s}P_{s}|w(s))\dd s+\xi_{T}\Gamma(\eta\xi_{T}|w)\bigg|\mathscr{F}_{t}\bigg](\sigma^{\intercal}_{t})^{-1}\theta_{t} \nonumber\\
  &\quad-\xi^{-1}_{t}(\sigma^{\intercal}_{t})^{-1}\E\bigg[\int^{T}_{t}\xi_{s}\big((\Psi_{\mathcal{I}_{s}}(\eta\xi_{s}P_{s}|w(s)) -P^{\intercal}_{s}\gamma_{\mathcal{I}_{s}}(\eta\xi_{s}P_{s}|w(s)))H_{\xi,t,s} \nonumber\\
  &\quad-\big(H_{P,t,s}+\sigma^{\intercal}_{P,t}\big)\mathrm{diag}(P_{s})(\psi_{\mathcal{I}_{s}}(\eta\xi_{s}P_{s}|w(s)) -\gamma_{\mathcal{I}_{s}}(\eta\xi_{s}P_{s}|w(s)))+\upsilon(t,s|w(s))\big)\dd s \nonumber\\
  &\quad+\xi_{T}\bigg(\bigg((U')^{-1}\bigg(\frac{\eta\xi_{T}T}{\int^{T}_{0}\langle w(s),\1\rangle_{\mathcal{I}_{s}}\dd s}\bigg)-\Gamma(\eta\xi_{T}|w)\bigg)H_{\xi,t,T}+\Upsilon(t,T|w)\bigg)\bigg|\mathscr{F}_{t}\bigg]: \nonumber\\
  &\quad w(s)\in\mathcal{K}^{\dag}_{s},\;\forall s\in[t,T];\;\sup_{s\in[0,T]}\|w(s)\|_{1}>0,\;\PP\text{-a.s.}\bigg\},\quad t\in[0,T],
\end{align}
where
\begin{equation*}
  \gamma_{\mathcal{I}}(\eta\xi P|w):=-\eta\xi\big(\langle w,(u^{(j)})^{(j')}(\imath,\psi_{\mathcal{I}}(\eta\xi P|w))\rangle\big)^{-1}_{j,j'\in\mathds{N}\cap[1,n]}P
\end{equation*}
and
\begin{equation*}
  \Gamma(\eta\xi_{T}|w):=-\frac{\eta\xi_{T}T}{\int^{T}_{0}\langle w(s),\1\rangle_{\mathcal{I}_{s}}\dd s}((U')^{-1})'\bigg(\frac{\eta\xi_{T}T}{\int^{T}_{0}\langle w(s),\1\rangle_{\mathcal{I}_{s}}\dd s}\bigg)
\end{equation*}
are parameterized risk tolerance functions associated with the consumption multi-utility and the bequest utility, respectively, $H_{\xi,t,}$ and $H_{P,t,}$ are some yet-to-be-determined $\mathbb{F}$-non-anticipating process associated with $(r,\theta)$ and $(\mu_{P},\sigma_{P})$, respectively (see (\ref{A.8}) and (\ref{A.13}) in \ref{A}), and
\begin{align*}
  \upsilon(t,\imath|w)&:=\Bigg(\big(\langle w,(u^{(j)})^{(j')}(\imath,\psi_{\mathcal{I}}(\eta\xi P|w))\rangle\big)^{-1}_{j,j'\in\mathds{N}\cap[1,n]}\\
  &\qquad\times\bigg(\int_{\pd\mathcal{I}}\mathbf{v}(i,W)\lrcorner\big(w_{i}u^{(j)}_{i}(\imath,\psi_{\mathcal{I}}(\eta\xi P|w))\dd i\big)\mathds{1}_{(0,\infty)}(\|w\|_{1})\bigg)_{j\in\mathds{N}\cap[1,n]}\Bigg)^{\intercal}P
\end{align*}
and
\begin{equation*}
  \Upsilon(t,T|w):=\eta\xi_{T}T((U')^{-1})'\bigg(\frac{\eta\xi_{T}T}{\int^{T}_{0}\langle w(s),\1\rangle_{\mathcal{I}_{s}}\dd s}\bigg)\frac{\int^{T}_{t}\int_{\pd\mathcal{I}_{s}}\mathbf{v}(i,W_{s})\lrcorner(w_{i}(s)\dd i)\dd s}{\big(\int^{T}_{0}\langle w(s),\1\rangle_{\mathcal{I}_{s}}\dd s\big)^{2}}
\end{equation*}
are parameterized psychological effect functions, in which $\mathbf{v}(i,W_{s})$ denotes the velocity vector field of the $\mathds{R}^{d}$-boundary $\pd\mathcal{I}_{s}$ for $s\in(t,T]$ on the classical Wiener space $\mathcal{C}_{0}([0,T];\mathds{R}^{m})$ and $\lrcorner$ denotes the interior product.\footnote{The interior product defines the contraction of the corresponding differential form with respect to $\mathbf{v}(i,W_{s})$ on $\mathcal{C}_{0}([0,T];\mathds{R}^{m})$. For its detailed definitions and properties see, e.g., \cite[Tu, 2011]{T}.}
\end{theorem}

As a remark, if the index set $\mathcal{I}$ happens to be reduced to the finite subset $J$ via (\ref{2.2.6}), nothing will change in Theorem \ref{thm:4} if the duality pairings over $\mathcal{I}$ are replaced by one in the Euclidean space $\mathds{R}^{\mathrm{card}J}$ (in which case $w$ is none but a fixed ($\mathrm{card}J$)-vector); correspondingly, the boundary $\pd\mathcal{I}$ is to be replaced by $\pd J\equiv J$. Such relations are well understood with the ordering cone $\mathcal{K}$, as Figure \ref{fig:2} has clarified.

A key observation from Theorem \ref{thm:4} is that the optimal investment policy $\Pi^{\ast}$ admits the following canonical decomposition:
\begin{equation}\label{5.1.2}
  \Pi^{\ast}=\cl_{\mathds{L}^{1}}\Big\{(\Pi^{(\theta)}+\Pi^{(H)}+\Pi^{(P)}+\Pi^{(\maltese)}|w): w\in\mathcal{K}^{\dag},\;\sup_{t\in[0,T]}\|w(t)\|_{1}=1\Big\}
\end{equation}
upon setting, for $t\in[0,T]$,
\begin{align}\label{5.1.3}
  \Pi^{(\theta)}_{t}&=\xi^{-1}_{t}\E\bigg[\int^{T}_{t}\xi_{s}\gamma_{\mathcal{I}_{s}}(\eta\xi_{s}P_{s}|w(s))\dd s+\xi_{T}\Gamma(\eta\xi_{T}|w)\bigg|\mathscr{F}_{t}\bigg](\sigma^{\intercal}_{t})^{-1}\theta_{t}, \nonumber\\
  \Pi^{(H)}_{t}&=\xi^{-1}_{t}(\sigma^{\intercal}_{t})^{-1}\E\bigg[\int^{T}_{t}\xi_{s} \big(\gamma_{\mathcal{I}_{s}}(\eta\xi_{s}P_{s}|w(s))-\Psi_{\mathcal{I}_{s}}(\eta\xi_{s}P_{s}|w(s))\big)H_{\xi,t,s}\dd s \nonumber\\
  &\qquad+\xi_{T}\bigg(\Gamma(\eta\xi_{T}|w)-(U')^{-1}\bigg(\frac{\eta\xi_{T}T}{\int^{T}_{0}\langle w(s),\1\rangle_{\mathcal{I}_{s}}\dd s}\bigg)\bigg)H_{\xi,t,T}\bigg|\mathscr{F}_{t}\bigg], \nonumber\\
  \Pi^{(P)}_{t}&=\xi^{-1}_{t}(\sigma^{\intercal}_{t})^{-1}\E\bigg[\int^{T}_{t}\xi_{s}\big(H_{P,t,s}+\sigma^{\intercal}_{P,t}\big) \mathrm{diag}(P_{s})(\psi_{\mathcal{I}_{s}}(\eta\xi_{s}P_{s}|w(s)) \nonumber\\
  &\qquad-\gamma_{\mathcal{I}_{s}}(\eta\xi_{s}P_{s}|w(s)))\dd s\bigg|\mathscr{F}_{t}\bigg], \nonumber\\
  \Pi^{(\maltese)}_{t}&=-\xi^{-1}_{t}(\sigma^{\intercal}_{t})^{-1}\E\bigg[\int^{T}_{t}\xi_{s}\upsilon(t,s|w(s))\dd s+\xi_{T}\Upsilon(t,T|w)\bigg|\mathscr{F}_{t}\bigg],
\end{align}
which stand for, in sequence, a mean--variance portfolio, a market risk-hedging portfolio, a commodity price risk-hedging portfolio, and an indecisiveness risk-hedging portfolio. Without the last two components, each selector of the (set-valued) optimal investment policy coincides with what the classical optimal portfolio decomposition gives under time-invariant complete preferences (see, e.g., \cite[Detemple et al., 2003]{DGR}).

The effect of preference incompleteness on the optimal investment policy is threefold. First, it alters the components of the optimal consumption policies, which depend not only on the utility elements but on their fluctuating patterns as well. This again highlights the dynamic intransitivity of preferences (\cite[Nishimura and Ok, 2016]{NO}) and such alteration, which may be associated with many external factors (\cite[Elias, 1982]{E3} and \cite[Patalano and Wengrowitz, 2007]{PW1}) as discussed before, is even prone to generating new functionals of the Brownian motion in addition to the state price density $\xi$ (as in Example 2 and Example 3). Second, preference incompleteness may lead to new portfolio (sub)components (those of $\Pi^{(\maltese)}$) which capture the investor's motive for hedging risks arising from his time-varying indecisiveness, and such risks diminish if preference fluctuations do not reflect future uncertainty -- or mathematically, if the index set $\mathcal{I}$ is deterministic. In the meantime, the other portfolio component, $\Pi^{(P)}$, signals a need to hedge potential risks associated with the relative valuation of commodities. Third, preference incompleteness generates multiplicity in optimal investment, as the optimal portfolio (\ref{5.1.2}) becomes a parameterized set-valued process in general.

As we have seen from Subsection \ref{ss:3.2}, in cases where preference independency is clear, it is sufficient to work with multi-utility that is additive in consumption, i.e., multi-utility whose elements are of the form
\begin{equation}\label{5.1.4}
  u_{i}(t,c)=\langle\alpha_{i}(t),\breve{u}(c)\rangle_{n}\equiv\sum^{n}_{j=1}\alpha_{ij}(t)\breve{u}_{j}(c_{j}),\quad i\in\mathcal{I}
\end{equation}
for a sequence of suitable functions $\{\breve{u}_{j}:j\in\mathds{N}\cap[1,n]\}\subseteq\mathcal{C}^{\infty}(\mathds{R}^{n}_{+};\mathds{R})$ and time-dependent coefficients $\alpha_{ij}$'s. Furthermore, if one is only concerned with a single utility parameter ($d=1$), with $\mathcal{R}\subseteq\mathds{R}$, then $\mathcal{I}_{t}\in\mathcal{P}(\mathds{R})$ becomes a random interval for $t\in[0,T]$. We then have the following relatively simpler result.

\begin{corollary}\label{cor:1}
If the multi-utility $u$ is consumption-additive taking the form of (\ref{5.1.4}) and $d=1$, then the optimal investment policy can be written\footnote{In the third line of (\ref{5.1.5}), $H_{P,j,t,s}$ and $\sigma^{\intercal}_{P,j,t}$ refer to the $j$th columns of $H_{P,t,}$ and $\sigma^{\intercal}_{P,t}$, respectively. Note that the letter $j$ is reserved for indexing the $n$ consumption goods from the very beginning.}
\begin{align}\label{5.1.5}
  \Pi^{\ast}_{t}&=\cl_{\mathds{L}^{1}}\Bigg\{\xi^{-1}_{t}\E\Bigg[\int^{T}_{t}\xi_{s}\sum^{n}_{j=1}P_{j,s}\gamma_{j}(\eta\xi_{s}P_{j,s}|w(s))\dd s+\xi_{T}\Gamma(\eta\xi_{T}|w)\Bigg|\mathscr{F}_{t}\Bigg](\sigma^{\intercal}_{t})^{-1}\theta_{t} \nonumber\\
  &\qquad-\xi^{-1}_{t}(\sigma^{\intercal}_{t})^{-1}\E\Bigg[\int^{T}_{t}\xi_{s}\Bigg(\sum^{n}_{j=1} P_{j,s}\bigg((\breve{u}'_{j})^{-1}\bigg(\frac{\eta\xi_{s}P_{j,s}}{\langle w(s),\alpha_{j}(s)\rangle_{\mathcal{I}_{s}}}\bigg)-\gamma_{j}(\eta\xi_{s}P_{j,s}|w(s))\bigg)H_{\xi,t,s} \nonumber\\
  &\qquad-\sum^{n}_{j=1}P_{j,s}\bigg((\breve{u}'_{j})^{-1}\bigg(\frac{\eta\xi_{s}P_{j,s}}{\langle w(s),\alpha_{j}(s)\rangle_{\mathcal{I}_{s}}}\bigg)-\gamma_{j}(\eta\xi_{s}P_{j,s}|w(s))\bigg)\big(H_{P,j,t,s}+\sigma^{\intercal}_{P,j,t}\big) \nonumber\\
  &\qquad+\eta\xi_{s}\sum^{n}_{j=1}P^{2}_{j,s}((\breve{u}'_{j})^{-1})'\bigg(\frac{\eta\xi_{s}P_{j,s}}{\langle w(s),\alpha_{j}(s)\rangle_{\mathcal{I}_{s}}}\bigg) \nonumber\\
  &\qquad\times\frac{w_{\breve{Y}_{+,s}}(s)\alpha_{\breve{Y}_{+,s}j}(s)(\mathcal{D}_{t}\breve{Y}_{+,s})^{\intercal} \mathds{1}_{\mathcal{R}}(\breve{Y}_{+,s})-w_{\breve{Y}_{-,s}}(s)\alpha_{\breve{Y}_{-,s}j}(s)(\mathcal{D}_{t}\breve{Y}_{-,s})^{\intercal} \mathds{1}_{\mathcal{R}}(\breve{Y}_{-,s})}{\langle w(s),\alpha_{j}(s)\rangle^{2}_{\mathcal{I}_{s}}}\Bigg)\dd s \nonumber\\
  &\qquad+\xi_{T}\bigg(\bigg((U')^{-1}\bigg(\frac{\eta\xi_{T}T}{\int^{T}_{0}\langle w(s),\1\rangle_{\mathcal{I}_{s}}\dd s}\bigg)-\Gamma(\eta\xi_{T}|w)\bigg)H_{\xi,t,T}+\eta\xi_{T}T \nonumber\\
  &\qquad\times((U')^{-1})'\bigg(\frac{\eta\xi_{T}T}{\int^{T}_{0}\langle w(s),\1\rangle_{\mathcal{I}_{s}}\dd s}\bigg)\frac{1}{\big(\int^{T}_{0}\langle w(s),\1\rangle_{\mathcal{I}_{s}}\dd s\big)^{2}} \nonumber\\
  &\qquad\times\int^{T}_{t}\big(w_{\breve{Y}_{+,s}}(s)(\mathcal{D}_{t}\breve{Y}_{+,s})^{\intercal}\mathds{1}_{\mathcal{R}}(\breve{Y}_{+,s}) -w_{\breve{Y}_{-,s}}(s)(\mathcal{D}_{t}\breve{Y}_{-,s})^{\intercal}\mathds{1}_{\mathcal{R}}(\breve{Y}_{-,s})\big)\dd s\bigg)\Bigg|\mathscr{F}_{t}\Bigg]: \nonumber\\
  &\quad w(s)\in\mathcal{K}^{\dag}_{s},\;\forall s\in[t,T];\;\sup_{s\in[0,T]}\|w(s)\|_{1}>0,\;\PP\text{-a.s.}\Bigg\},\quad t\in[0,T],
\end{align}
where
\begin{equation*}
  \gamma_{j}(\eta\xi P_{j}|w)=-\frac{\eta\xi P_{j}}{\langle w,\alpha_{j}\rangle_{\mathcal{I}}}((\breve{u}'_{j})^{-1})'\bigg(\frac{\eta\xi P_{j}}{\langle w,\alpha_{j}\rangle_{\mathcal{I}}}\bigg),\quad j\in\mathds{N}\cap[1,n]
\end{equation*}
are consumption good-specific risk tolerance functions and $\{\breve{Y}_{\pm}\}=\pd\breve{\mathcal{I}}$, $\breve{\mathcal{I}}$ being the unrefined index set such that $\mathcal{I}=\mathcal{R}\cap\breve{\mathcal{I}}$ in (\ref{2.2.1}).
\end{corollary}

Minor modifications of the formula (\ref{5.1.5}) can be made in the same way as in the case of (\ref{5.1.1}) of Theorem \ref{thm:4} if (\ref{2.2.6}) is applied to transform $\mathcal{I}$ into $J$.

\subsection{Procedures and examples revisited}\label{ss:5.2}

For how the optimal investment policy $\Pi^{\ast}$ should be computed in practice, we outline some procedures again, under the assumption of smooth utility (Assumption \ref{as:3}). At the outset, if the multi-utility $u$ is consumption-additive in the sense of (\ref{5.1.4}) and there is only one parameter ($d=1$) of interest, one can simply apply Corollary \ref{cor:1}, which only requires finding the two endpoints $\breve{Y}_{\pm}$ making up the boundary $\pd\breve{\mathcal{I}}$ of the unrefined index set, while the global parameter range $\mathcal{R}\subseteq\mathds{R}$ is predetermined. In light of the general dynamics (\ref{2.2.1}), such endpoints are easy to identify from the dynamic support of $\breve{\mathcal{I}}$.

\begin{figure}[H]
  \centering
  \includegraphics[scale=0.21]{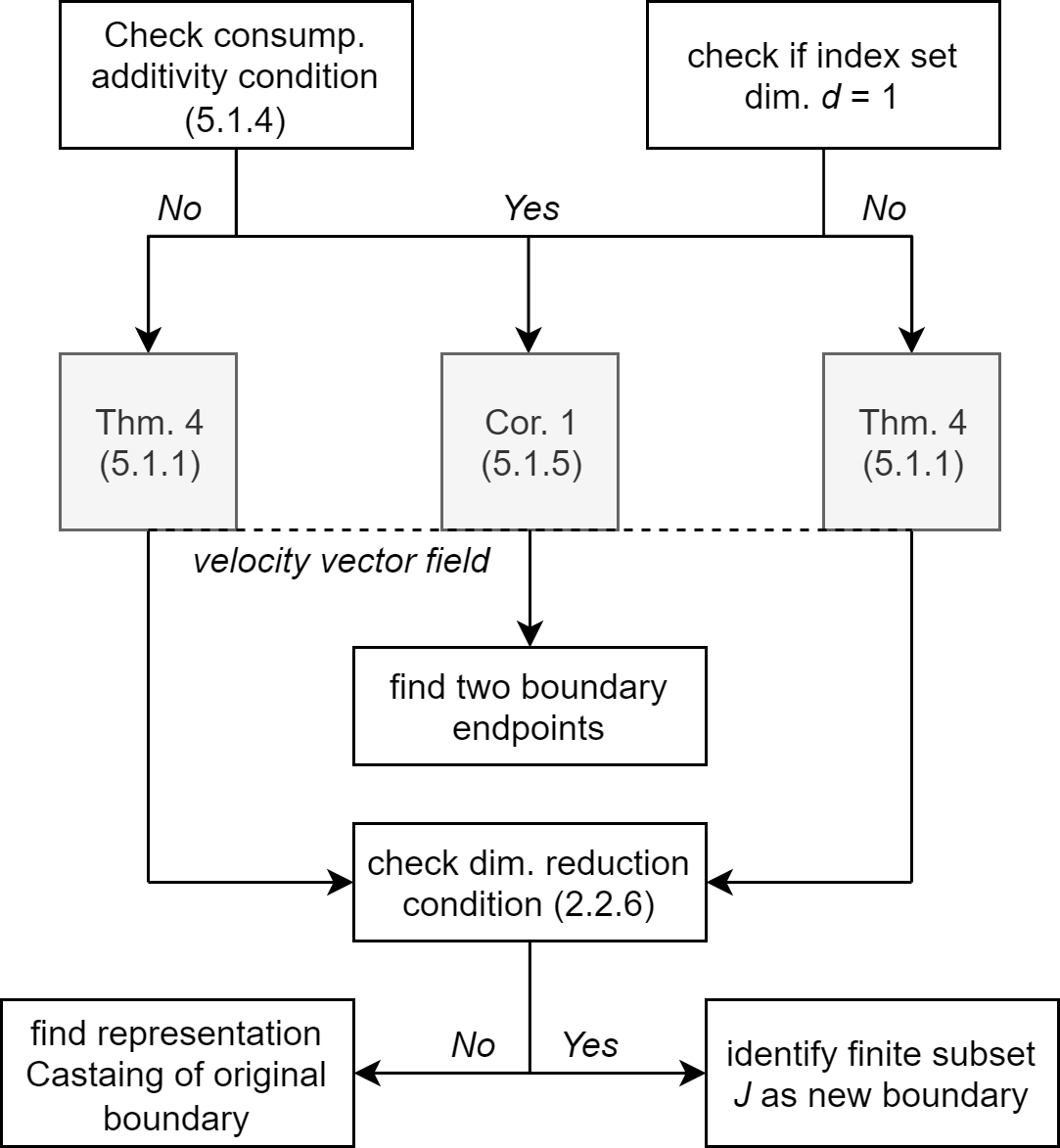}\\
  \caption{Procedures for computing optimal investment policies}
  \label{fig:6}
\end{figure}

If either condition fails, one has to apply Theorem \ref{thm:4}. If $d\geq2$, it requires constructing a sequence $\big\{Y_{1,h_{1}(k)}+\bar{Y}^{\mathcal{t}}_{2,k,\nu_{h_{2}(k)}}+\bar{Y}_{3,k,\nu_{h_{3}(k)}}:k\in\mathds{N}_{++}\big\}\subseteq\mathbb{D}^{1,2}$ before the Malliavin derivatives can be computed. At bottom, one is asked to identify a representation Castaing of $\pd\mathcal{I}_{t}$ for a given $t\in[0,T]$ which is integrably bounded and then provide a corresponding approximation. In fact, the process of finding such sequences will not be so arduous as one may think at first glance, provided that $\pd\mathcal{I}$ exhibits certain levels of geometric regularity, e.g., is a (regular) $d$-polytope. On the other hand, highly irregular boundaries of an index set are generally avoidable because preference changes in different channels should not be arbitrarily convoluted concerning the hardship in associating them with empirical evidence. We will leave detailed explanations of these computational issues to Section \ref{s:6} along with proposed simulation methods for set-valued stochastic processes. As before, we give a flowchart in Figure \ref{fig:6} to highlight these necessary steps.

We are now prepared to complete the three examples by computing the optimal investment policies. Recall that $m=1$ and the assumption that $D\equiv0$ and $P\equiv\1$ made to analyze the net effects of incomplete preferences. An immediate implication is that $H_{P,t,}\equiv\0$ for any fixed $t\in[0,T]$ and the commodity price risk-hedging portfolio component $\Pi^{(P)}$ in (\ref{5.1.3}) diminishes.

\vspace{0.1in}

\textbf{Example 1 (continued).}\quad We have $d=1$ and $n=2$. It is clear that with constant $(r,\theta)$, $H_{\xi,t,}\equiv0$ for any fixed $t\in[0,T]$. Also, we note that the multi-utility is consumption-additive in Case (I) and Case (II) but not in Case (III), and so Corollary \ref{cor:1} only applies to the first two cases. In Case (I), in particular, since the optimality conditions along with the budget set have implied that $(w^{1/p}_{1}+w^{1/p}_{2})/\eta^{1/p}=X_{0}\rho_{p}(r,\theta)/(e^{\rho_{p}(r,\theta)T}-1)$, the risk tolerance function is free of $w$ and is easily verified to be
\begin{equation*}
  \gamma_{\{1,2\}}(\eta\xi)=\frac{\rho_{p}(r,\theta)X_{0}}{p\xi^{1/p}(e^{\rho_{p}(r,\theta)T}-1)}.
\end{equation*}
Hence, we obtain the single-valued optimal investment policy
\begin{equation}\label{5.2.1}
  \Pi^{\ast}_{t}=\bigg\{\frac{X_{0}\theta(e^{\rho_{p}(r,\theta)(T-t)}-1)}{p\sigma\xi^{1/p}_{t}(e^{\rho_{p}(r,\theta)T}-1)}\bigg\},\quad t\in[0,T],
\end{equation}
which gives a single mean--variance portfolio, as there is no randomness in the market coefficients. In a similar fashion, one can show that in Case (II), the optimal investment policy is the singleton given by (\ref{5.2.1}). This outcome basically tells the investor to concentrate on a precise scale of his portfolio Sharpe ratio.

In Case (III), however, the optimal total consumption expenditure is not single-valued, and the risk tolerance function is necessarily (strictly) parameterized as well,
\begin{equation*}
  \gamma_{\{\varkappa_{1},\varkappa_{2}\}}(\eta\xi|w)=\frac{2\eta\xi(1-p)\psi(\eta\xi|w)^{2p+1}}{p(1-p)\psi(\eta\xi|w)^{p} +(1-2p)(w_{1}\varkappa_{1}+w_{2}\varkappa_{2}) \psi(\eta\xi|w)},
\end{equation*}
where $\psi(\eta\xi|w)$ is recalled to solve (\ref{3.2.5}). Therefore, the optimal investment policy is a set-valued process given by
\begin{align}\label{5.2.2}
  \Pi^{\ast}_{t}&=\cl_{\mathds{L}^{1}}\bigg\{\frac{\theta\eta}{\sigma\xi_{t}}\int^{T}_{t} \E\bigg[\frac{2(1-p)\xi^{2}_{s}\psi(\eta\xi_{s}|w)^{2p+1}}{p(1-p)\psi(\eta\xi_{s}|w)^{p}+(1-2p)(w_{1}\varkappa_{1}+w_{2}\varkappa_{2}) \psi(\eta\xi_{s}|w)}\bigg|\mathscr{F}_{t}\bigg]\dd s: \nonumber\\
  &\qquad\eta\text{ given};\;w\in\mathds{R}^{2}_{+},\;\|w\|_{1}=1\bigg\},\quad t\in[0,T],
\end{align}
which is a collection of mean--variance portfolios (refer to (\ref{5.1.3})); i.e., the investor now has room, however small, to vary the weight to put on the portfolio Sharpe ratio at any point in time. \hfill{\scriptsize $\blacklozenge$}

\vspace{0.1in}

\textbf{Example 2 (continued).}\quad Since the specified multi-utility is consumption-additive and $d=1$, we use Corollary \ref{cor:1} and obtain immediately $\breve{u}_{1}(c_{1})=(c^{1-p}_{1}-1)/(1-p)$, $\breve{u}_{2}(c_{2})=(c^{1-p}_{2}-1)/(1-p)$, and $\alpha_{ij}(t)=e^{-\beta t}(\chi_{i}\mathds{1}_{\{1\}}(j)+\mathds{1}_{\{2\}}(j))$; recalling applying (\ref{2.2.6}) in (\ref{3.2.6}), we obtain the risk tolerance functions
\begin{equation*}
  \gamma_{1}(\eta\xi|w)=\frac{1}{p}\bigg(\frac{w_{1}\chi_{0}+w_{2}\chi_{\lambda W^{\uparrow}+1}}{\eta\xi e^{\beta\imath}}\bigg)^{1/p},\quad \gamma_{2}(\eta\xi|w)=\frac{1}{p}(\eta\xi e^{\beta\imath})^{-1/p}.
\end{equation*}
Again, with $(r,\theta)$ being constant, $H_{\xi,t,}\equiv0$, and as noted before, $\mathcal{R}=\mathds{R}_{+}$, $\breve{Y}_{+}=\lambda W^{\uparrow}+1$ and $\breve{Y}_{-}=\lambda W^{\downarrow}$. The Malliavin derivatives of these processes are well established (see, e.g., \cite[Renaud and Remillard, 2006]{RR2}) thanks to the Clark--Ocone formula, and we have
\begin{equation*}
  \mathcal{D}_{t}\breve{Y}_{+,s}\mathds{1}_{\mathds{R}_{+}}(\breve{Y}_{+,s})=\lambda\mathrm{erfc}\frac{W^{\uparrow}_{t}-W_{t}}{\sqrt{2(s-t)}} \quad\text{and}\quad \mathcal{D}_{t}\breve{Y}_{-,s}\mathds{1}_{\mathds{R}_{+}}(\breve{Y}_{-,s})=0,\quad s\in[t,T].
\end{equation*}
After specifying (\ref{5.1.5}) accordingly the optimal investment policy for the investor reads
\begin{align}\label{5.2.3}
  \Pi^{\ast}_{t}&=\cl_{\mathds{L}^{1}}\Bigg\{\frac{\theta}{p\sigma\eta^{1/p}\xi_{t}}\int^{T}_{t}\E\bigg[\frac{\xi^{1-1/p}_{s}}{e^{\beta s/p}}\bigg((w_{1}\chi_{0}+w_{2}\chi_{\lambda W^{\uparrow}_{s}+1})^{1/p}+1\bigg)\bigg|\mathscr{F}_{s}\bigg]\dd s \nonumber\\
  &\qquad+\frac{\lambda w_{2}}{p\sigma\eta^{1/p}\xi_{t}}\int^{T}_{t}\E\bigg[\frac{\xi^{1-1/p}_{s}}{e^{\beta s/p}}\bigg(\frac{\chi'_{\lambda W^{\uparrow}_{s}+1}}{(w_{1}\chi_{0}+w_{2}\chi_{\lambda W^{\uparrow}_{s}+1})^{1-1/p}}+1\bigg)\mathrm{erfc}\frac{W^{\uparrow}_{t}-W_{t}}{\sqrt{2(s-t)}}\bigg|\mathscr{F}_{t}\bigg]\dd s: \nonumber\\
  &\quad\eta\text{ given};\;w\in\mathds{R}^{2}_{+},\;\|w\|_{1}=1\Bigg\},\quad t\in[0,T],
\end{align}
which again can be evaluated via simulation, though an alternative integral representation is also available by using the conditional joint density of $(W^{\uparrow}_{s},W_{s})|\mathscr{F}_{t}$, $s\in[t,T]$, based on the Markov property. In this setup, by (\ref{5.1.3}) the optimal investment policy is necessarily a set-valued process combining a mean--variance portfolio with a hedging demand for indecisiveness risk. In other words, the investor needs to take account of socialization-driven preference fluctuations which can affect his opportunity set, in addition to the risk--return tradeoff, and the weights that he can put on the two components are variable. \hfill{\scriptsize $\blacklozenge$}

\vspace{0.1in}

\textbf{Example 3 (continued).}\quad Since $d=2$, we apply Theorem \ref{thm:4}. Using that $\mathcal{I}=[0,\lambda W^{\uparrow}+1]\times[\sigma+1,\sigma+2]$ and the risk-linked function (\ref{3.2.8}), the consumption and bequest risk tolerance functions are parameterized and respectively found to be
\begin{align*}
  \gamma_{\mathcal{I}}(\eta\xi|w)&=\eta\xi e^{\beta\imath}\Bigg(\bigg\langle w,\frac{p_{(\cdot)_{2}}\chi_{(\cdot)_{1}}}{\vartheta^{-1}_{p,w_{2};\sigma}\big(\eta\xi e^{\beta\imath}/\langle w_{1},\chi\rangle_{[0,\lambda W^{\uparrow}+1]}\big)^{p_{(\cdot)_{2}}+1}}\bigg\rangle^{-1}_{\mathcal{I}}\\
  &\qquad+\bigg\langle w,\frac{p_{(\cdot)_{2}}}{\vartheta^{-1}_{p,w_{2};\sigma}\big(\eta\xi e^{\beta\imath}/\langle w_{1},\1\rangle_{[0,\lambda W^{\uparrow}+1]}\big)^{p_{(\cdot)_{2}}+1}}\bigg\rangle^{-1}_{\mathcal{I}}\Bigg)
\end{align*}
and
\begin{equation*}
  \Gamma(\eta\xi_{T}|w):=\frac{1}{p_{\circ}}\bigg(\frac{\int^{T}_{t}\|w(s)\|_{1}\dd s}{\eta\xi_{T}Te^{\beta T}}\bigg)^{1/p_{\circ}}.
\end{equation*}
Recalling from (\ref{3.2.7}) that $\sigma$ is now an exponential Ornstein--Uhlenbeck process parameterized by $\{\sigma_{0}>0,\kappa>0,\varsigma<0\}$, we have directly
\begin{equation*}
  H_{\xi,t,s}=-\varsigma\int^{s}_{t}\theta_{v}e^{-\kappa(v-t)}(\dd W_{v}+\theta_{v}\dd v)\quad\text{and}\quad\mathcal{D}_{t}\sigma_{s}=\varsigma\sigma_{s}e^{-\kappa(s-t)},\quad s\in[t,T].
\end{equation*}
Also, as $\pd\mathcal{I}$ is rectangle-valued, finding the required velocity vector field is routine, leading to
\begin{align*}
  \mathbf{v}(i,W_{s})&\equiv\mathbf{v}(i_{1},i_{2},W_{s})\\
  &=\lambda\mathrm{erfc}\frac{W^{\uparrow}_{t}-W_{t}}{\sqrt{2(s-t)}}\mathds{1}_{\{\lambda W^{\uparrow}_{s}+1\}\times[\sigma_{s}+1,\sigma_{s}+2]}(i)\frac{\pd}{\pd i_{1}}\\
  &\qquad+\varsigma\sigma_{s}e^{-\kappa(s-t)}\big(\mathds{1}_{[0,\lambda W^{\uparrow}_{s}+1]\times\{\sigma_{s}+2\}}(i)-\mathds{1}_{[0,\lambda W^{\uparrow}_{s}+1]\times\{\sigma_{s}+1\}}(i)\big)\frac{\pd}{\pd i_{2}},\quad s\in[t,T],
\end{align*}
and the corresponding interior products can be easily computed. The optimal investment policy is understandably convoluted in light of (\ref{5.1.1}); after some tedious calculations the following precise result can be given:
\begin{align}\label{5.2.4}
  \Pi^{\ast}_{t}&=\cl_{\mathds{L}^{1}}\Bigg\{\frac{\theta_{t}}{\sigma_{t}\xi_{t}}\E\Bigg[\int^{T}_{t}\eta\xi^{2}_{s}e^{\beta s}\Bigg(\Bigg(\int_{[0,\lambda W^{\uparrow}_{s}+1]\times[\sigma_{s}+1,\sigma_{s}+2]}\bar{w}_{i}p_{i_{2}}\chi_{i_{1}} \nonumber\\
  &\qquad\times\vartheta^{-1}_{p,\bar{w}_{2};\sigma_{s}}\bigg(\frac{\eta\xi_{s}e^{\beta s}}{\int^{\lambda W^{\uparrow}_{s}+1}_{0}\bar{w}_{1,i_{1}}\chi_{i_{1}}\dd i_{1}}\bigg)^{-p_{i_{2}}-1}\dd i\Bigg)^{-1} \nonumber\\
  &\qquad+\Bigg(\int_{[0,\lambda W^{\uparrow}_{s}+1]\times[\sigma_{s}+1,\sigma_{s}+2]}\bar{w}_{i}p_{i_{2}}\vartheta^{-1}_{p,\bar{w}_{2};\sigma_{s}}\bigg(\frac{\eta\xi_{s}e^{\beta s}}{\int^{\lambda W^{\uparrow}_{s}+1}_{0}\bar{w}_{1,i_{1}}\dd i_{1}}\bigg)^{-p_{i_{2}}-1}\dd i\Bigg)^{-1}\Bigg)\dd s \nonumber\\
  &\qquad+\frac{\xi^{1-1/p_{\circ}}_{T}}{p_{\circ}(\eta Te^{\beta T})^{1/p_{\circ}}}\bigg(\int^{T}_{0}\int_{[0,\lambda W^{\uparrow}_{s}+1]\times[\sigma_{s}+1,\sigma_{s}+2]}\bar{w}_{i}\dd i\dd s\bigg)^{1/p_{\circ}}\Bigg|\mathscr{F}_{t}\Bigg] \nonumber\\
  &\qquad+\frac{\varsigma}{\sigma_{t}\xi_{t}}\E\Bigg[\int^{T}_{t}\xi_{s}\Bigg(\vartheta^{-1}_{p,\bar{w}_{2};\sigma_{s}}\bigg(\frac{\eta\xi_{s}e^{\beta s}}{\int^{\lambda W^{\uparrow}_{s}+1}_{0}\bar{w}_{1,i_{1}}\chi_{i_{1}}\dd i_{1}}\bigg)+\vartheta^{-1}_{p,\bar{w}_{2};\sigma_{s}}\bigg(\frac{\eta\xi_{s}e^{\beta s}}{\int^{\lambda W^{\uparrow}_{s}+1}_{0}\bar{w}_{1,i_{1}}\dd i_{1}}\bigg) \nonumber\\
  &\qquad\!\!-\eta\xi_{s}e^{\beta s}\Bigg(\Bigg(\int_{[0,\lambda W^{\uparrow}_{s}+1]\times[\sigma_{s}+1,\sigma_{s}+2]}\bar{w}_{i}p_{i_{2}}\chi_{i_{1}} \vartheta^{-1}_{p,\bar{w}_{2};\sigma_{s}}\bigg(\frac{\eta\xi_{s}e^{\beta s}}{\int^{\lambda W^{\uparrow}_{s}+1}_{0}\bar{w}_{1,i_{1}}\chi_{i_{1}}\dd i_{1}}\bigg)^{-p_{i_{2}}-1}\dd i\Bigg)^{-1} \nonumber\\
  &\qquad+\Bigg(\int_{[0,\lambda W^{\uparrow}_{s}+1]\times[\sigma_{s}+1,\sigma_{s}+2]}\bar{w}_{i}p_{i_{2}} \vartheta^{-1}_{p,\bar{w}_{2};\sigma_{s}}\bigg(\frac{\eta\xi_{s}e^{\beta s}}{\int^{\lambda W^{\uparrow}_{s}+1}_{0}\bar{w}_{1,i_{1}}\dd i_{1}}\bigg)^{-p_{i_{2}}-1}\dd i\Bigg)^{-1}\Bigg)\Bigg) \nonumber\\
  &\qquad\times\int^{s}_{t}\theta_{v}e^{-\kappa(v-t)}(\dd W_{v}+\theta_{v}\dd v)\dd s+\frac{\xi^{1-1/p_{\circ}}_{T}}{(\eta Te^{\beta T})^{1/p_{\circ}}}\bigg(1-\frac{1}{p_{\circ}}\bigg) \nonumber\\
  &\qquad\times\bigg(\int^{T}_{0}\int_{[0,\lambda W^{\uparrow}_{s}+1]\times[\sigma_{s}+1,\sigma_{s}+2]}\bar{w}_{i}\dd i\dd s\bigg)^{1/p_{\circ}} \int^{T}_{t}\theta_{s}e^{-\kappa(s-t)}(\dd W_{s}+\theta_{s}\dd s)\Bigg|\mathscr{F}_{t}\Bigg] \nonumber\\
  &\qquad+\frac{1}{\sigma_{t}\xi_{t}}\E\Bigg[\int^{T}_{t}\xi_{s}\Bigg(\Bigg(\int_{[0,\lambda W^{\uparrow}_{s}+1]\times[\sigma_{s}+1,\sigma_{s}+2]}\bar{w}_{i}p_{i_{2}}\chi_{i_{1}} \nonumber\\
  &\qquad\times\vartheta^{-1}_{p,\bar{w}_{2};\sigma_{s}}\bigg(\frac{\eta\xi_{s}e^{\beta s}}{\int^{\lambda W^{\uparrow}_{s}+1}_{0}\bar{w}_{1,i_{1}}\chi_{i_{1}}\dd i_{1}}\bigg)^{-p_{i_{2}}-1}\dd i\Bigg)^{-1}\Bigg(\lambda\mathrm{erfc}\frac{W^{\uparrow}_{t}-W_{t}}{\sqrt{2(s-t)}}\bar{w}_{1,\lambda W^{\uparrow}_{s}+1}\chi_{\lambda W^{\uparrow}_{s}+1} \nonumber\\
  &\qquad\times\int^{\sigma_{s}+2}_{\sigma_{s}+1}\bar{w}_{2,i_{2}}\vartheta^{-1}_{p,\bar{w}_{2};\sigma_{s}}\bigg(\frac{\eta\xi_{s}e^{\beta s}}{\int^{\lambda W^{\uparrow}_{s}+1}_{0}\bar{w}_{1,i_{1}}\chi_{i_{1}}\dd i_{1}}\bigg)^{-p_{i_{2}}}\dd i_{2} \nonumber\\
  &\qquad+\varsigma\sigma_{s}e^{-\kappa(s-t)}\int^{\lambda W^{\uparrow}_{s}+1}_{0}\bar{w}_{1,i_{1}}\chi_{i_{1}}\dd i_{1} \Bigg(\bar{w}_{2,\sigma_{s}+2}\vartheta^{-1}_{p,\bar{w}_{2};\sigma_{s}}\bigg(\frac{\eta\xi_{s}e^{\beta s}}{\int^{\lambda W^{\uparrow}_{s}+1}_{0}\bar{w}_{1,i_{1}}\chi_{i_{1}}\dd i_{1}}\bigg)^{-p_{\sigma_{s}+2}} \nonumber\\
  &\qquad-\bar{w}_{2,\sigma_{s}+1}\vartheta^{-1}_{p,\bar{w}_{2};\sigma_{s}}\bigg(\frac{\eta\xi_{s}e^{\beta s}}{\int^{\lambda W^{\uparrow}_{s}+1}_{0}\bar{w}_{1,i_{1}}\chi_{i_{1}}\dd i_{1}}\bigg)^{-p_{\sigma_{s}+1}}\Bigg)\Bigg) \nonumber\\
  &\qquad+\Bigg(\int_{[0,\lambda W^{\uparrow}_{s}+1]\times[\sigma_{s}+1,\sigma_{s}+2]}\bar{w}_{i}p_{i_{2}} \vartheta^{-1}_{p,\bar{w}_{2};\sigma_{s}}\bigg(\frac{\eta\xi_{s}e^{\beta s}}{\int^{\lambda W^{\uparrow}_{s}+1}_{0}\bar{w}_{1,i_{1}}\dd i_{1}}\bigg)^{-p_{i_{2}}-1}\dd i\Bigg)^{-1} \nonumber\\
  &\qquad\times\Bigg(\lambda\mathrm{erfc}\frac{W^{\uparrow}_{t}-W_{t}}{\sqrt{2(s-t)}}\bar{w}_{1,\lambda W^{\uparrow}_{s}+1} \int^{\sigma_{s}+2}_{\sigma_{s}+1}\bar{w}_{2,i_{2}}\vartheta^{-1}_{p,\bar{w}_{2};\sigma_{s}}\bigg(\frac{\eta\xi_{s}e^{\beta s}}{\int^{\lambda W^{\uparrow}_{s}+1}_{0}\bar{w}_{1,i_{1}}\dd i_{1}}\bigg)^{-p_{i_{2}}}\dd i_{2} \nonumber\\
  &\qquad+\varsigma\sigma_{s}e^{-\kappa(s-t)}\int^{\lambda W^{\uparrow}_{s}+1}_{0}\bar{w}_{1,i_{1}}\dd i_{1} \Bigg(\bar{w}_{2,\sigma_{s}+2}\vartheta^{-1}_{p,\bar{w}_{2};\sigma_{s}}\bigg(\frac{\eta\xi_{s}e^{\beta s}}{\int^{\lambda W^{\uparrow}_{s}+1}_{0}\bar{w}_{1,i_{1}}\dd i_{1}}\bigg)^{-p_{\sigma_{s}+2}} \nonumber\\
  &\qquad-\bar{w}_{2,\sigma_{s}+1}\vartheta^{-1}_{p,\bar{w}_{2};\sigma_{s}}\bigg(\frac{\eta\xi_{s}e^{\beta s}}{\int^{\lambda W^{\uparrow}_{s}+1}_{0}\bar{w}_{1,i_{1}}\dd i_{1}}\bigg)^{-p_{\sigma_{s}+1}}\Bigg)\Bigg)\Bigg)\dd s \nonumber\\
  &\qquad+\frac{\xi^{1-1/p_{\circ}}_{T}}{(\eta\xi_{T}Te^{\beta T})^{1/p_{\circ}}}\bigg(\int^{T}_{0}\int_{[0,\lambda W^{\uparrow}_{s}+1]\times[\sigma_{s}+1,\sigma_{s}+2]}\bar{w}_{i}\dd i\dd s\bigg)^{1/p_{\circ}} \nonumber\\
  &\qquad\times\int^{T}_{t}\bigg(\lambda\mathrm{erfc}\frac{W^{\uparrow}_{t}-W_{t}}{\sqrt{2(s-t)}}\bar{w}_{1,\lambda W^{\uparrow}_{s}+1}\int^{\sigma_{s}+2}_{\sigma_{s}+1}\bar{w}_{2,i_{2}}\dd i_{2} \nonumber\\
  &\qquad+\varsigma\sigma_{s}e^{-\kappa(s-t)}\bigg(\bar{w}_{2,\sigma_{s}+2}\int^{\lambda W^{\uparrow}_{s}+1}_{0}\bar{w}_{1,i_{1}}\dd i_{1}-\bar{w}_{2,\sigma_{s}+1}\int^{\lambda W^{\uparrow}_{s}+1}_{0}\bar{w}_{1,i_{1}}\dd i_{1}\bigg)\bigg)\dd s \nonumber\\
  &\qquad\times\bigg(\int^{T}_{0}\int_{[0,\lambda W^{\uparrow}_{s}+1]\times[\sigma_{s}+1,\sigma_{s}+2]}\bar{w}_{i}\dd i\dd s\bigg)^{-2}\Bigg|\mathscr{F}_{t}\Bigg]: \nonumber\\
  &\quad\eta\text{ given};\;\bar{w}\in\bar{\mathcal{K}}^{\dag},\;\|\bar{w}\|_{1}=1\Bigg\},\quad t\in[0,T],
\end{align}
which, despite lengthiness, is amenable to simulation techniques. Observably, the optimal investment policy is now a set-valued process consisting of a mean--variance portfolio, a market volatility risk-hedging portfolio, as well as an indecisiveness risk-hedging portfolio. In this case, market volatility generates a hedging demand in the latter two portfolio components, and indecisiveness risk also reflects socialization effects through perceived importance of the less essential good. Different from before, the investor should take into account fluctuations in his investment opportunities due to stochastic volatility itself as well as its influence on his risk aversion. \hfill{\scriptsize $\blacklozenge$}

\vspace{0.2in}

\section{Simulation and numerical experiments}\label{s:6}

In this section we discuss simulation techniques for set-valued stochastic processes, which are needed to realize the optimal consumption--investment policies. In accordance with the illustrative examples, particular attention will be drawn to infinitely smooth utility elements (Assumption \ref{as:3}), but our analysis will accommodate multi-utility spaces of arbitrary dimensions.

To employ Euler discretization we start with the finite time partition $P_{K}=\{t_{l|K}:l\in\mathds{N}\cap[0,K]\}$, $K\in\mathds{N}_{++}$, of the generic time interval $[0,t]$, $t\in(0,T]$. Then, we define for every $l\leq K$ and $q\in\{1,2,3\}$
\begin{equation}\label{6.1}
  \hat{I}^{(K)}_{q,t_{l|K}}:=I_{q,0}+\sum^{l-1}_{\iota=0}(t_{\iota+1|K}-t_{\iota|K}) f_{q,t_{\iota|K}}+\overline{\co}_{\mathds{L}^{2}}\Bigg\{\sum^{l-1}_{\iota=0}g_{q,k,t_{\iota|K}}(W_{t_{\iota+1|K}}-W_{t_{\iota|K}}): k\in\mathds{N}_{++}\Bigg\}
\end{equation}
and subsequently establish the approximate multi-utility index set as
\begin{equation}\label{6.2}
  \hat{\mathcal{I}}^{(K)}_{t}:=\mathcal{R}\cap\Bigg(\hat{I}^{(K)}_{1,t_{K|K}}+\bigcap^{(K-1)\wedge\hat{\mathcal{l}}}_{l=0}\hat{I}^{(K)}_{2,t_{l+1|K}} +\co_{\mathds{R}^{d}}\bigcup^{K-1}_{l=0}\hat{I}^{(K)}_{3,t_{l+1|K}}\Bigg),
\end{equation}
where
\begin{equation}\label{6.3}
  \hat{\mathcal{l}}:=\inf\Bigg\{l\in\mathds{N}\cap[1,K]:\mathrm{card}\bigcap^{K-1}_{l=0}\hat{I}^{(K)}_{2,t_{l+1}|K}=1\Bigg\}-1;
\end{equation}
note that $t_{K|K}\equiv t$. Then we can prove the following approximation result.

\begin{proposition}\label{pro:6}
For any $t\in[0,T]$ with corresponding partition $P_{K}\subsetneq[0,t]$, the index set process (\ref{2.2.1}), and the approximate (\ref{6.2}) with (\ref{6.1}) and (\ref{6.3}) it holds that
\begin{equation*}
  \lim_{K\rightarrow\infty}\E\big[\mathcal{d}_{\rm H}\big(\mathcal{I}_{t},\hat{\mathcal{I}}^{(K)}_{t}\big)\big]=0.
\end{equation*}
\end{proposition}

With Proposition \ref{pro:6} the temporal evolution of the investor's multi-utility index set may be visualized in dimensions $d\leq3$, for which numerical integration over $\mathcal{I}$ is to be carried out using generalized Gauss quadrature rules (see \cite[Ma et al., 1996]{MRW}); meanwhile, note that the global parameter space $\mathcal{R}$ is predetermined. Since all elements in the return process (\ref{2.1.1}), the risk-free rate $r$, and therefore the state price density $\xi$, are single-valued diffusion processes, classical Euler discretization methods apply to their simulations and we refer to \cite[Schoutens, 2003, \text{Sect.} 8]{S2}.

In the following, we conduct numerical experiments based on the optimal consumption--investment policies obtained in Example 1 through Example 3,\footnote{All implementation programs are written in Python 3.8 and run on a personal computer with an Intel(R) Core(TM) i5-7200U CPU @2.50GHz processor.} respectively. In particular, it is unrealistic to illustrate every single optimal policy given that there are infinitely many equally optimal; instead, we vary the weight functional $w$ in a ``sufficiently'' fine discrete subspace of the dual space from which it takes values, which will eventually generate a decent approximation of the entire set of optimal policies, with occasional singletons. In addition, to ease comparison, we fix the market parameter values $r=0.001$ and $\mu=0.02$, an investment horizon $T=1$ (year), and an initial wealth amount $X_{0}=100.00$. Other case-specific parameters are summarized in Table \ref{tab:1}.

\begin{table}[H]\footnotesize
  \centering
  \caption{Case-specific parameter values in Examples}
  \label{tab:1}
  \begin{tabular}{c|c|c}
    \hline
    Category & parameter & dual cone (approximation)\\ \hline
    Example 1 Case (I) & $\sigma=0.36$, $p=6$ & $\mathds{R}^{2}_{+}$ (100-partition) \\
    Example 1 Case (II) & $\sigma=0.36$, $p=6$, $\chi=3$ & $\mathds{R}^{2}_{+}$ (100-partition) \\
    Example 1 Case (III) & $\sigma=0.36$, $p=6$, $[\varkappa_{1},\varkappa_{2}]=[1,2]$ & $\mathds{R}^{2}_{+}$ (100-partition) \\ \hline
    Example 2 & $\sigma=0.36$, $p=6$, $\beta=0.05$, $\lambda=0.2$, $\chi=\mathrm{id}\wedge5$ & $\mathds{R}^{2}_{+}$ (100-partition) \\ \hline
    \multirow{2}*{Example 3} & $\sigma_{0}=0.36$, $\kappa=0.1$, $\varsigma=-0.8$, $\beta=0.05$, & $(\mathcal{C}_{\rm b}(\mathds{R}_{+}\times[1,\infty);\mathds{R}_{+}))^{\dag}$ \\ & $\lambda=0.2$, $p_{\circ}=5$, $\chi=p/3=\mathrm{id}\wedge5$ & (100-$w_{1}$-partition, $w_{2}=\delta_{\{2.2\}}$) \\
    \hline
  \end{tabular}
\end{table}

\begin{figure}[H]
  \centering
  \includegraphics[width=2.3in]{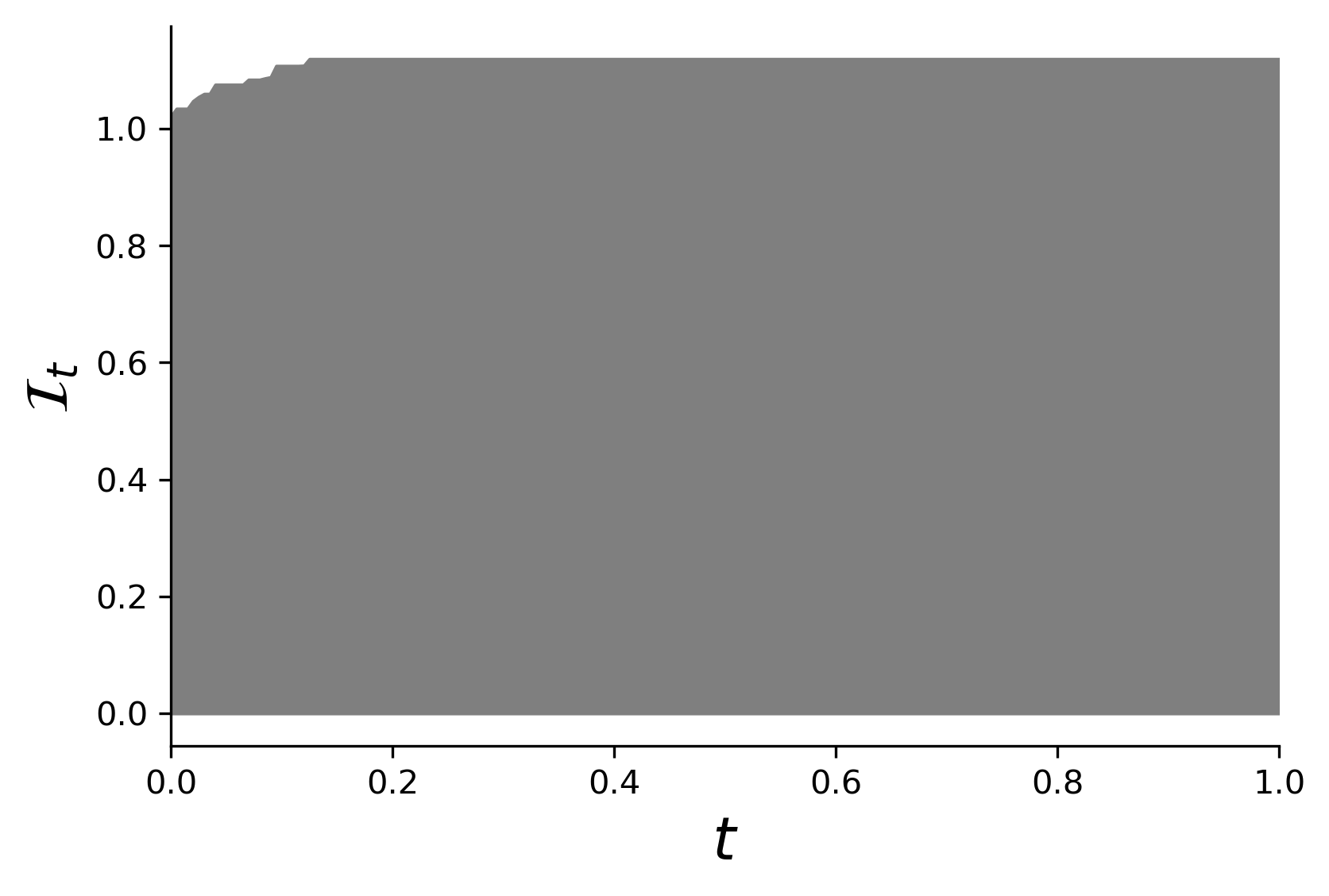}
  \caption{Realized multi-utility representation index set process in Example 2}
  \label{fig:8}
\end{figure}

\begin{figure}[H]
  \centering
  \includegraphics[width=2.3in]{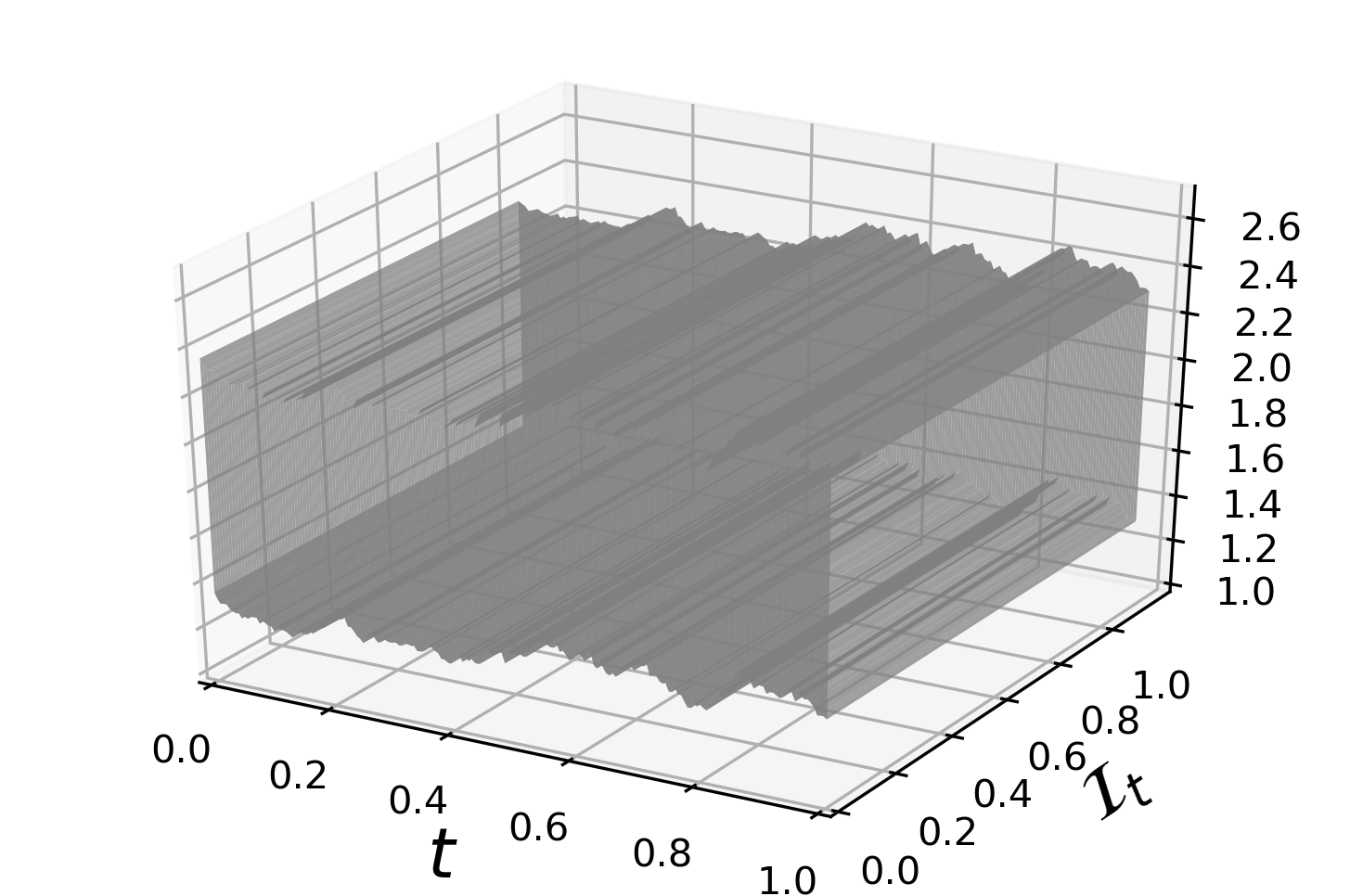}
  \caption{Realized multi-utility representation index set process in Example 3}
  \label{fig:11}
\end{figure}

Recall that in Example 1, the index set process $\mathcal{I}$ is constant, while in Example 2 and Example 3 it is stochastic with values in $\mathrm{Cl}(\mathds{R})$ and $\mathrm{Cl}(\mathds{R}^{2})$, respectively. Based on a fixed realization $(\omega_{t})_{t\in[0,T]}$ of the universal coordinate process, we plot the set-valued sample paths of $\mathcal{I}$ in Figure \ref{fig:8} and Figure \ref{fig:11}, respectively.

\begin{figure}[H]
  \centering
  \includegraphics[width=2in]{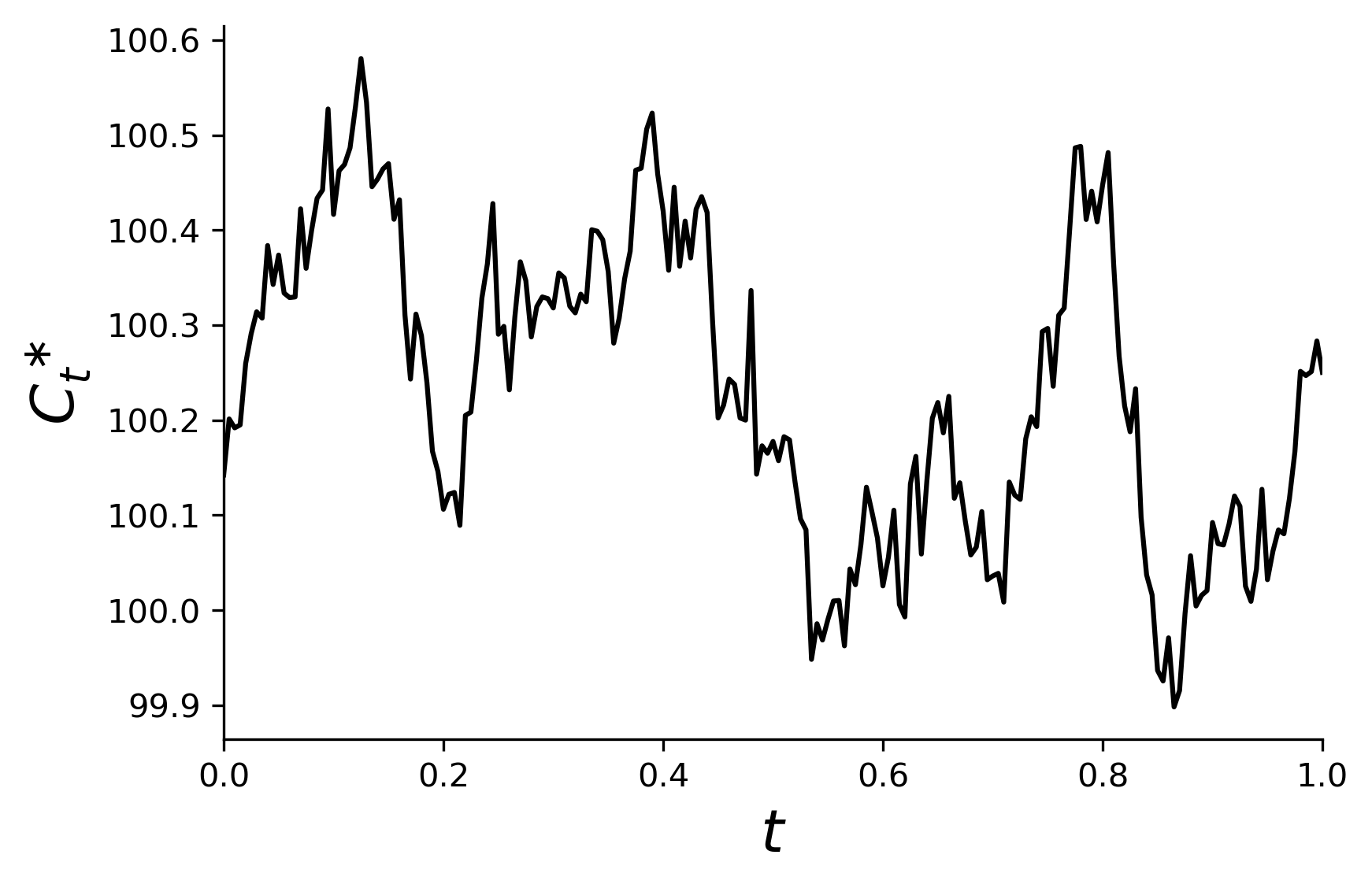}
  \includegraphics[width=2in]{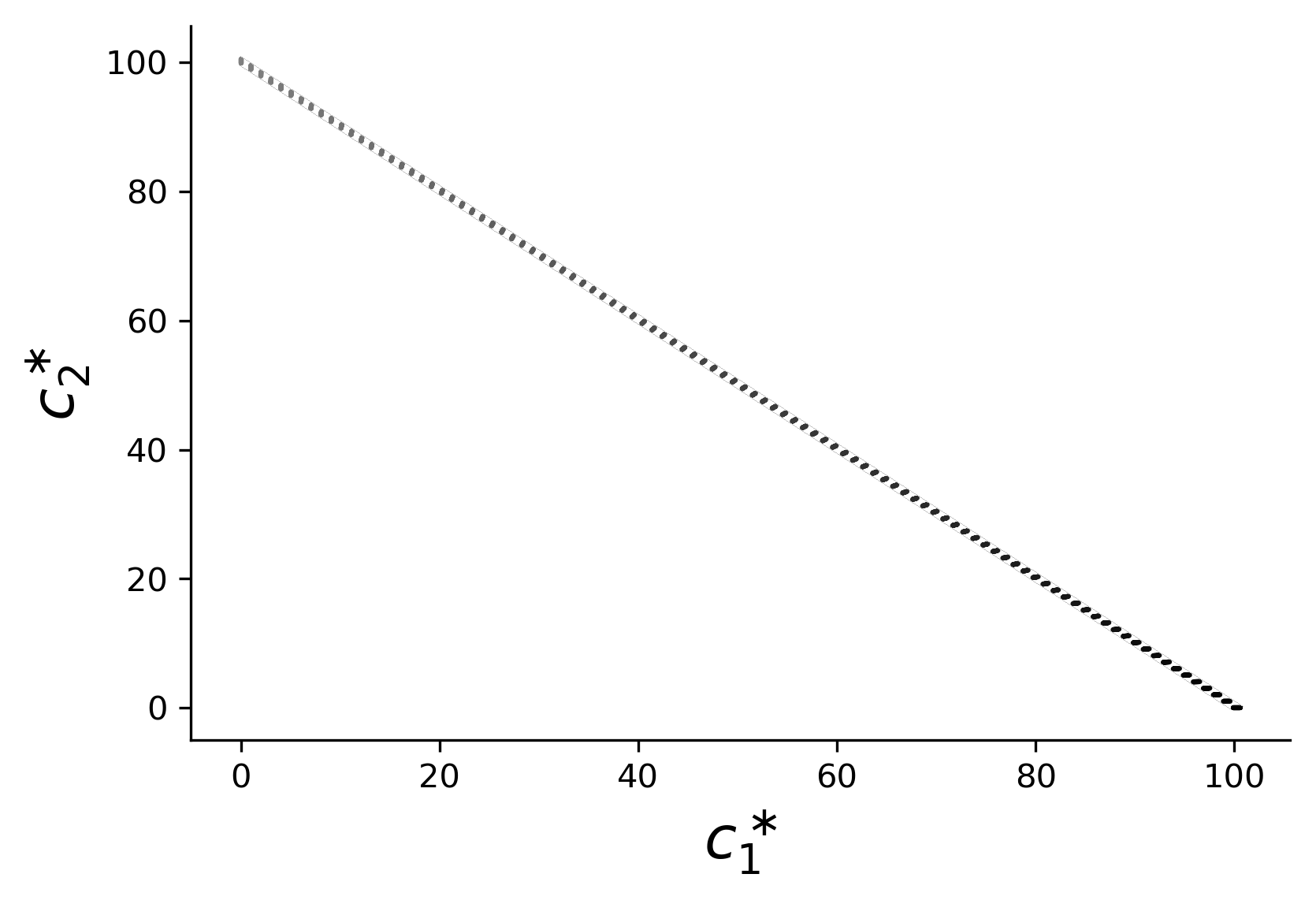}
  \includegraphics[width=2in]{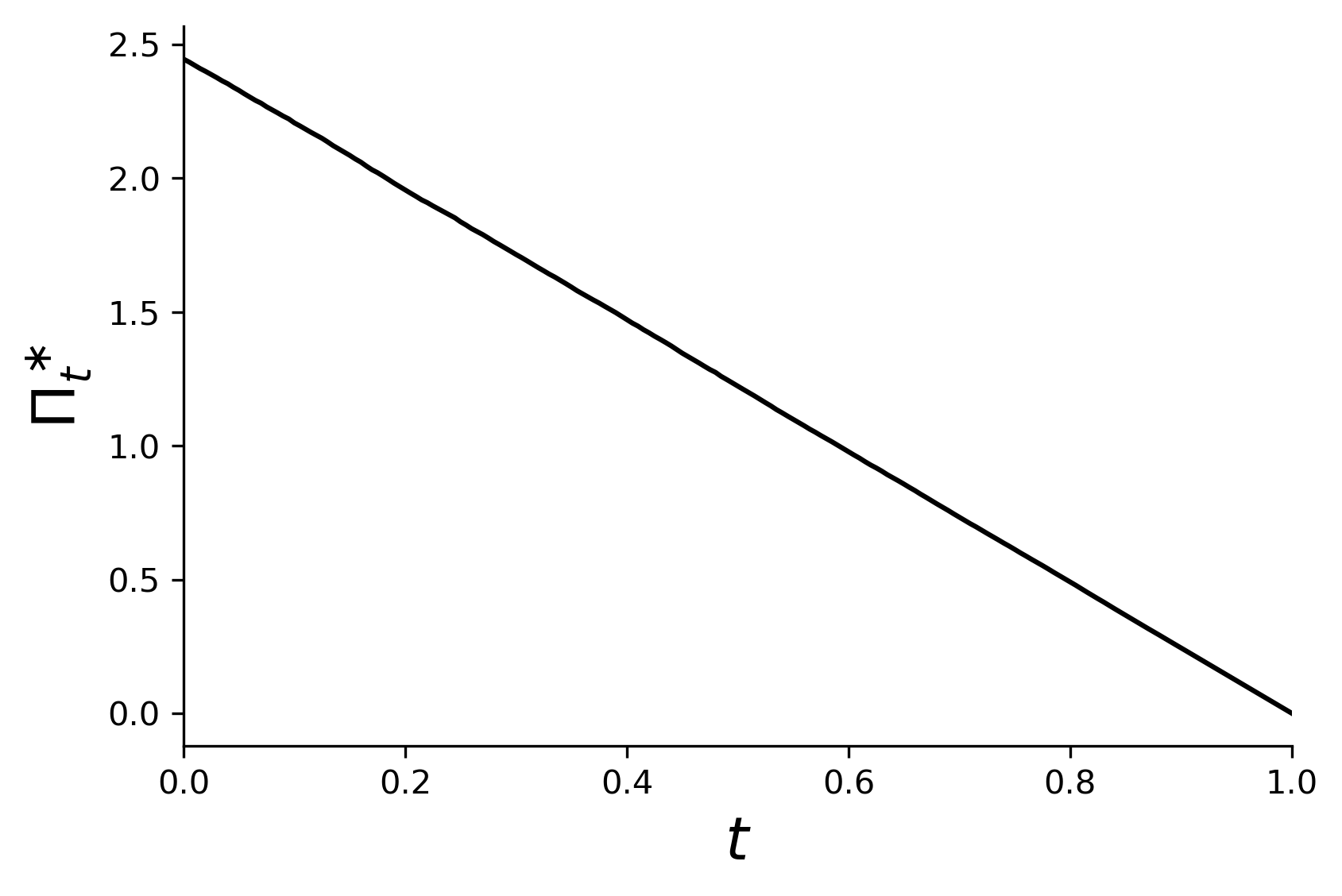}
  \caption*{Case (I)}
  \includegraphics[width=2in]{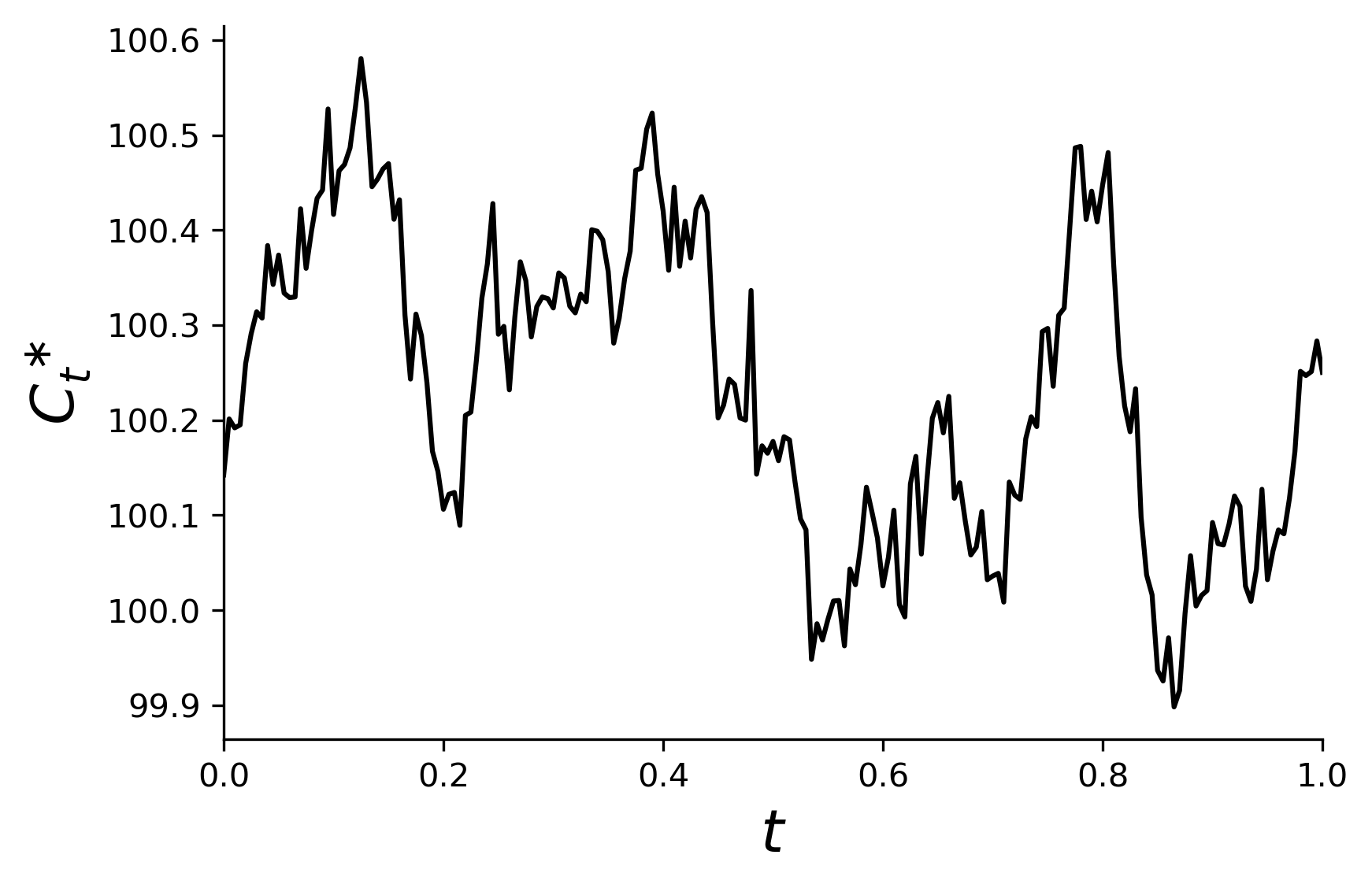}
  \includegraphics[width=2in]{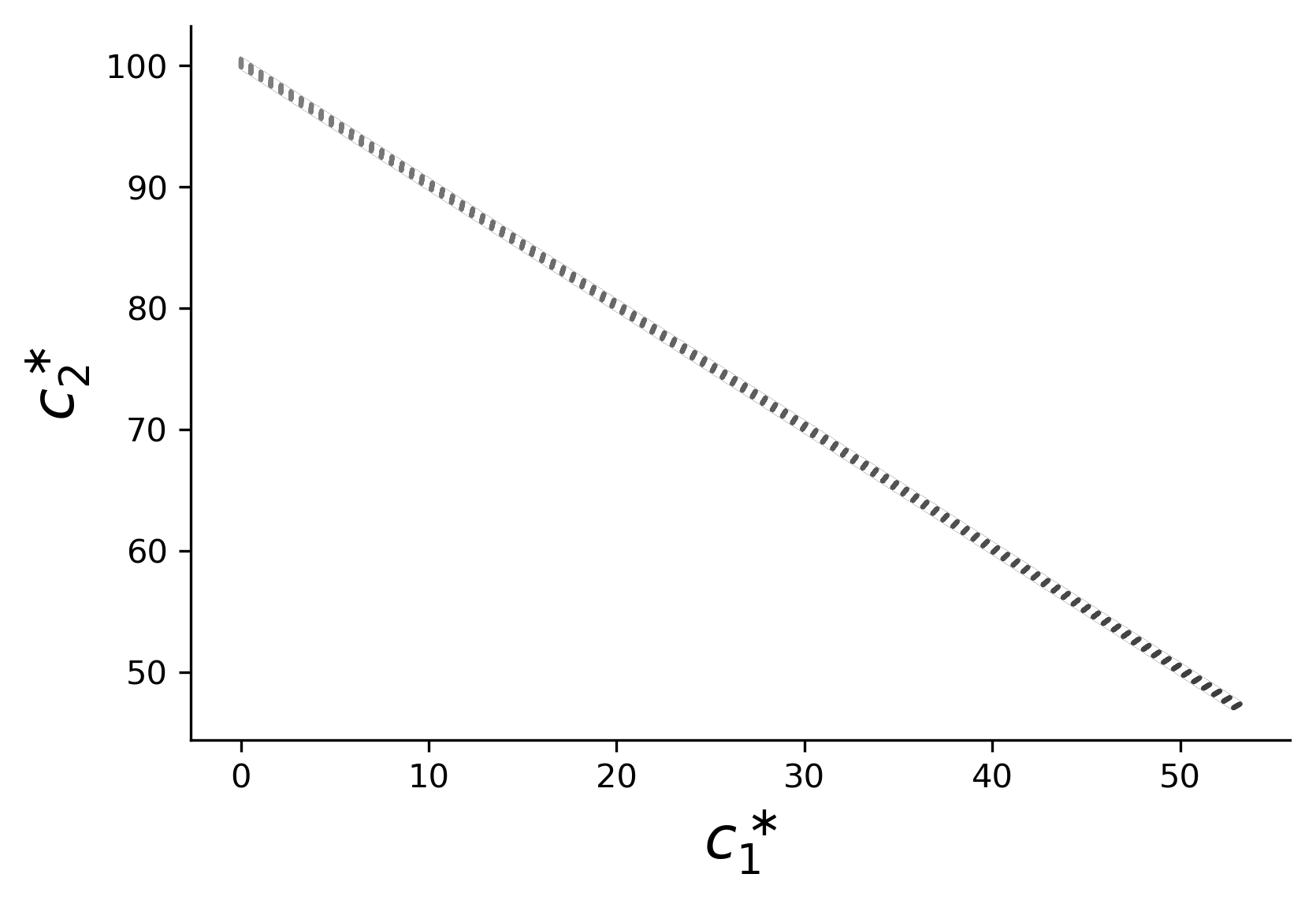}
  \includegraphics[width=2in]{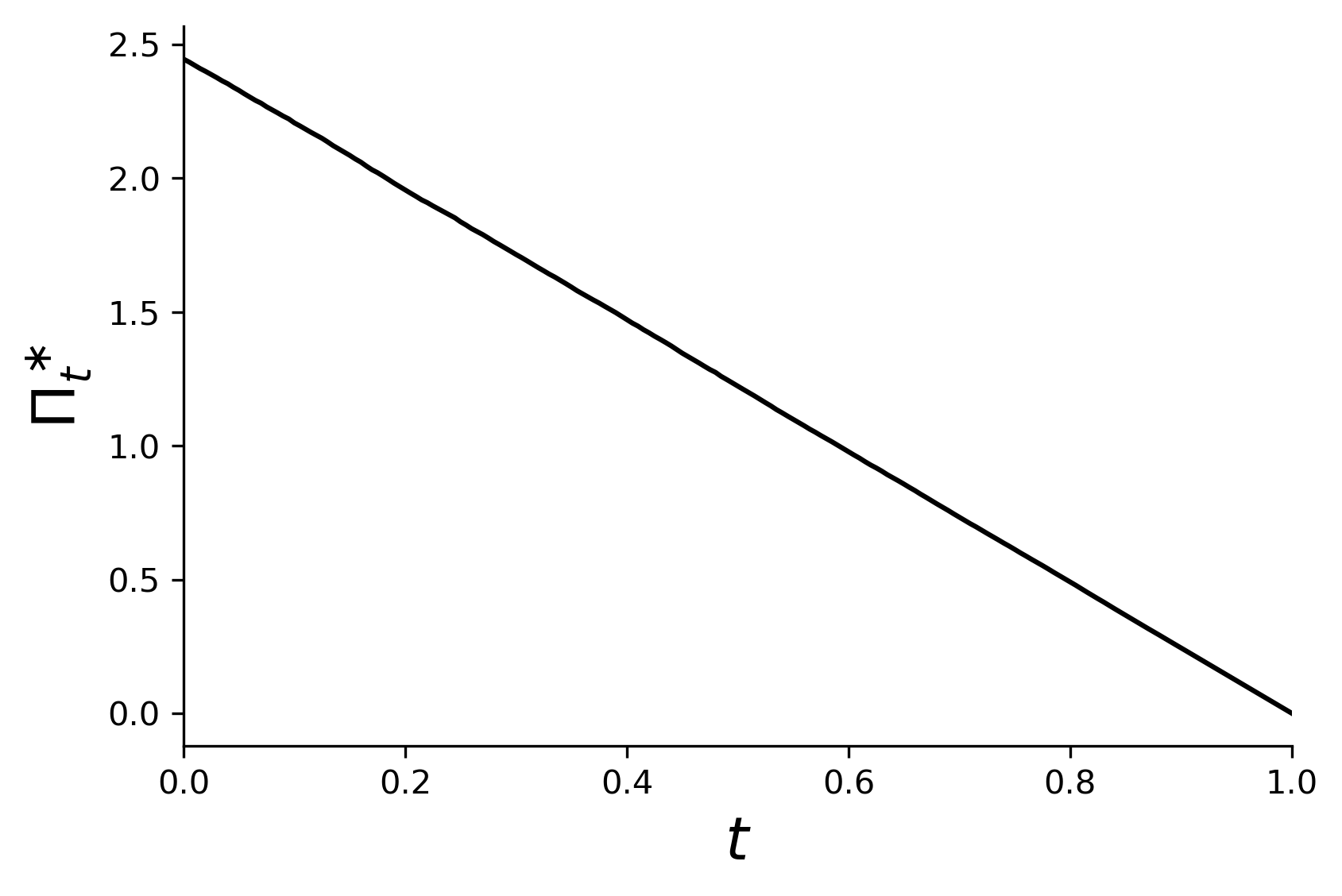}
  \caption*{Case (II)}
  \includegraphics[width=2in]{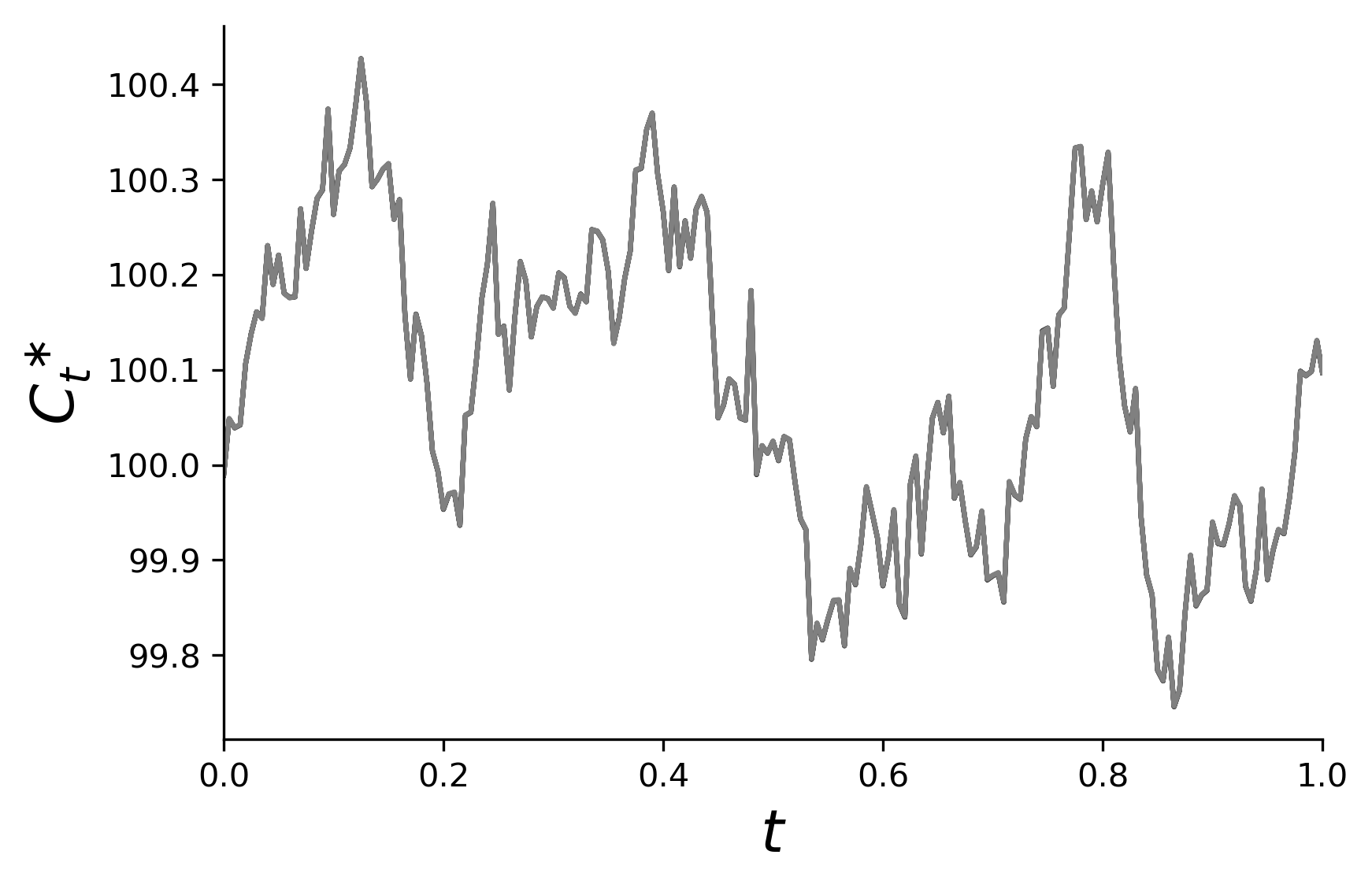}
  \includegraphics[width=2in]{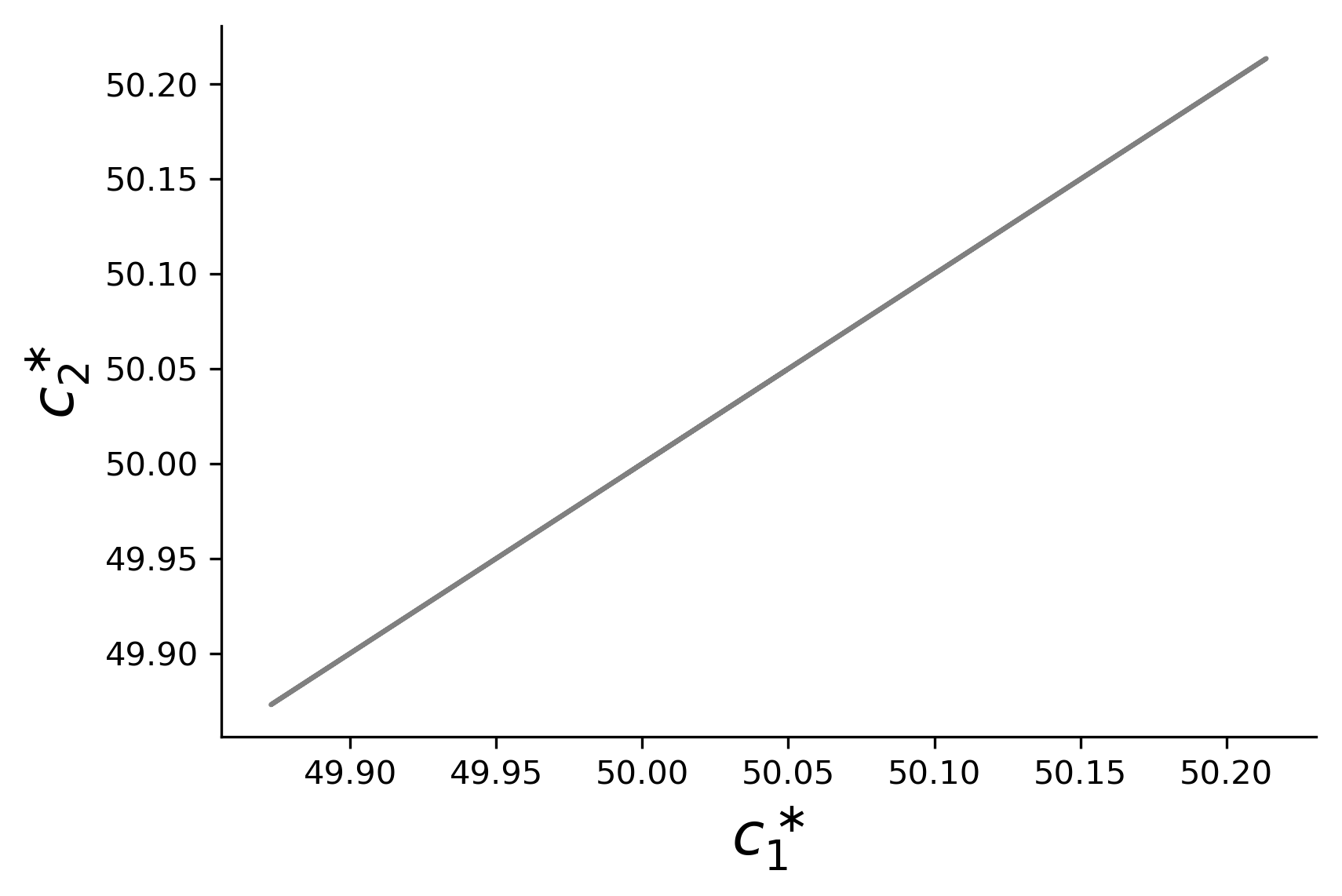}
  \includegraphics[width=2in]{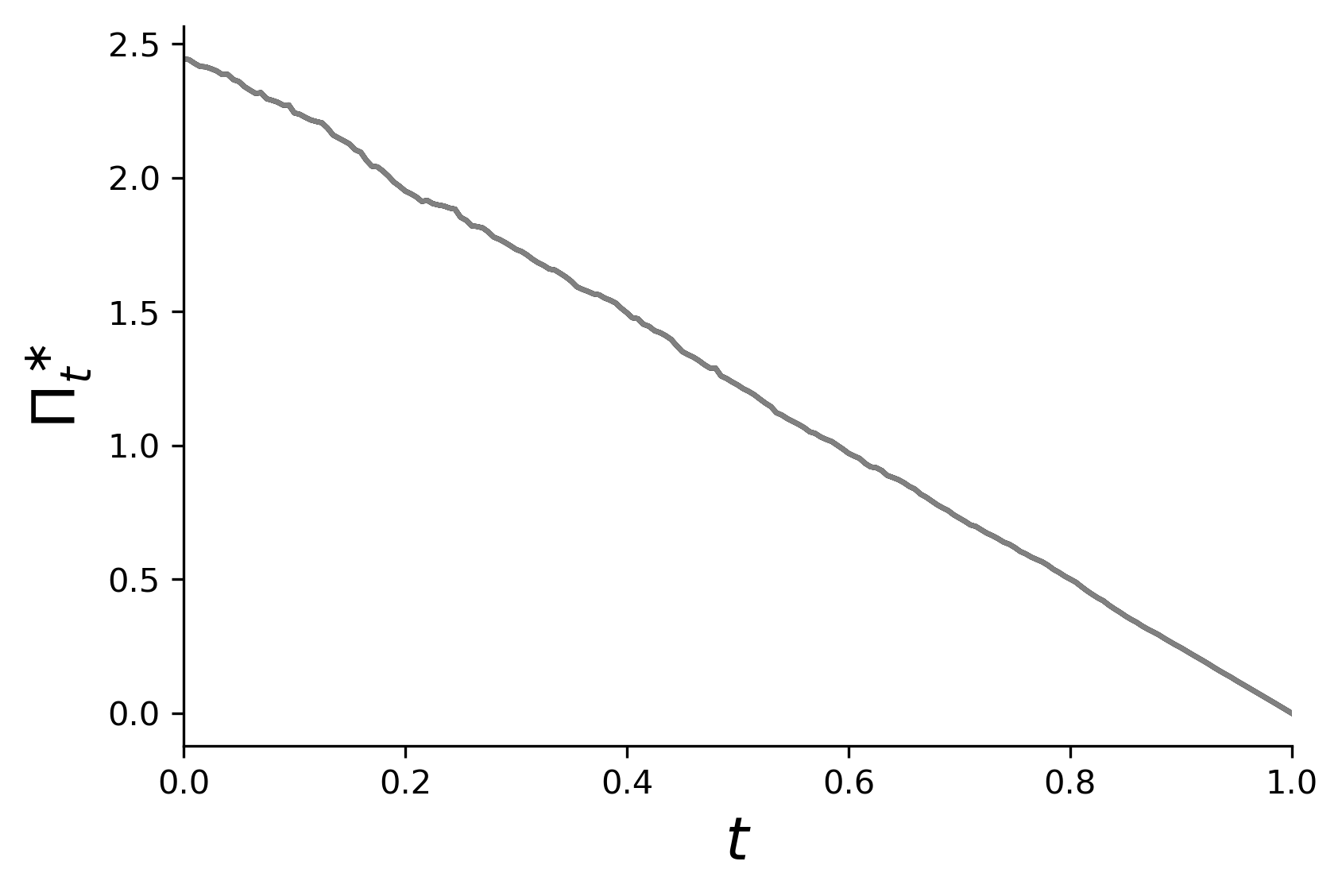}
  \caption*{Case (III)}
  \caption{Realized optimal consumption and investment policies in Example 1}
  \label{fig:7}
\end{figure}

All numerical expectations that appear in the solved policies are evaluated based on uniform time quadratures of size 1/200 (in the unit of years) and 100 samples of independent simulations, as well as the same realization $\omega$. For Example 1, the corresponding realized optimal consumption and investment policies in all three Cases are demonstrated in three separate panels in Figure \ref{fig:7} -- one for the sample paths of optimal total consumption expenditure (left), one for the coordinates formed by optimal consumption elements (center), and the other for the parameterized sample paths of optimal investment (right). For Example 2 and Example 3, we also add two panels to illustrate the stochastic consumption coordinates and the new optimal portfolio decomposition to show the effects of time-varying preferences.

In Case (I) and Case (II), the optimal total consumption expenditure and the optimal investment are both single-valued, while those in Case (III) are (properly) set-valued, despite unnoticeable differences in its elements. On the contrary, the coordinates of optimal consumption elements are floating in Case (I) and Case (II) in preset ranges reflecting the degree of indecisiveness, while those in Case (III) are fixed at one-one because of their uniformly equal weights at optimum. The optimal investment policies are all of mean--variance type and the investor liquidates all of his hedging positions at the end of his investment horizon.

\begin{figure}[H]
  \centering
  \includegraphics[width=2in]{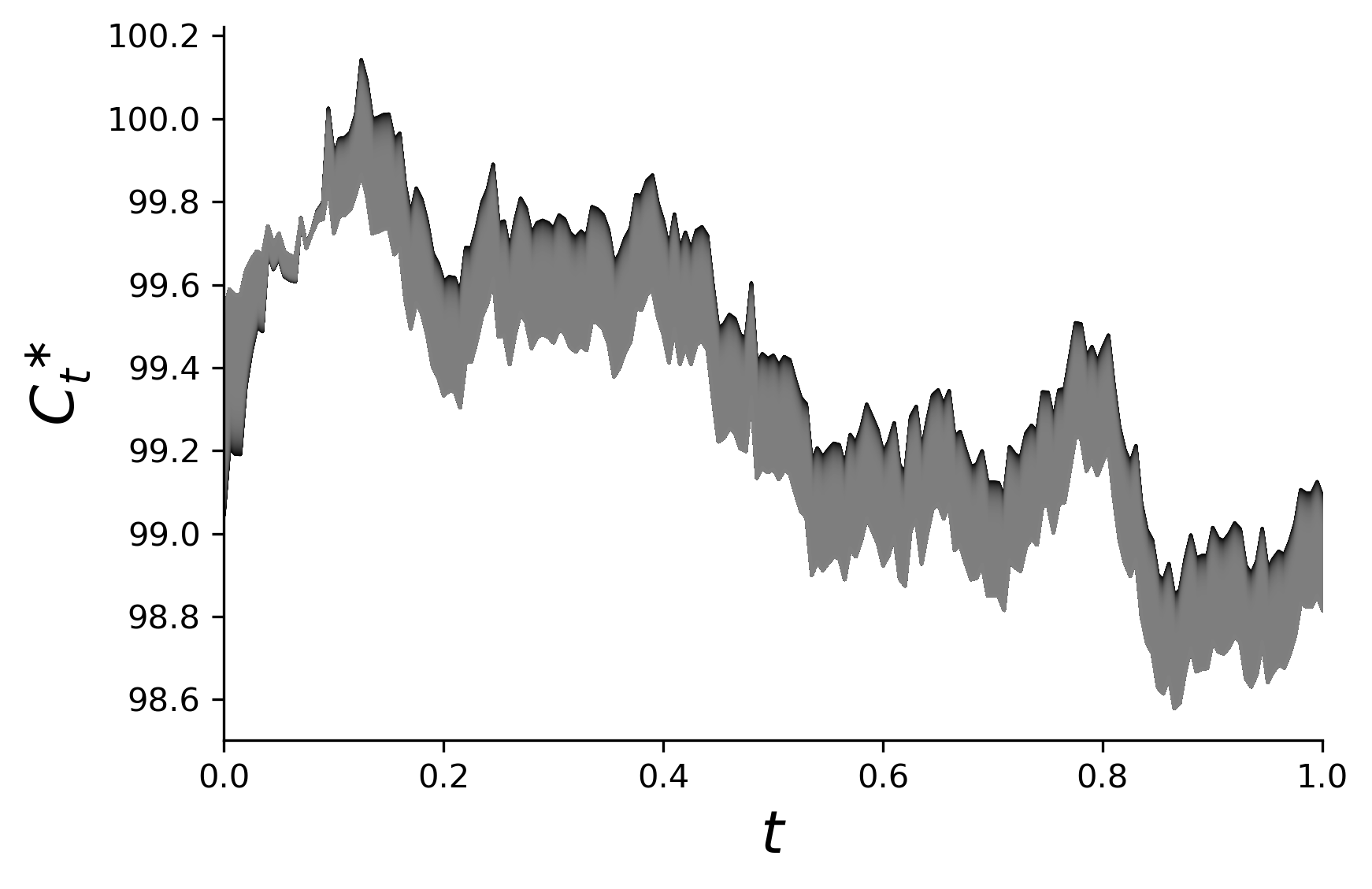}
  \includegraphics[width=2in]{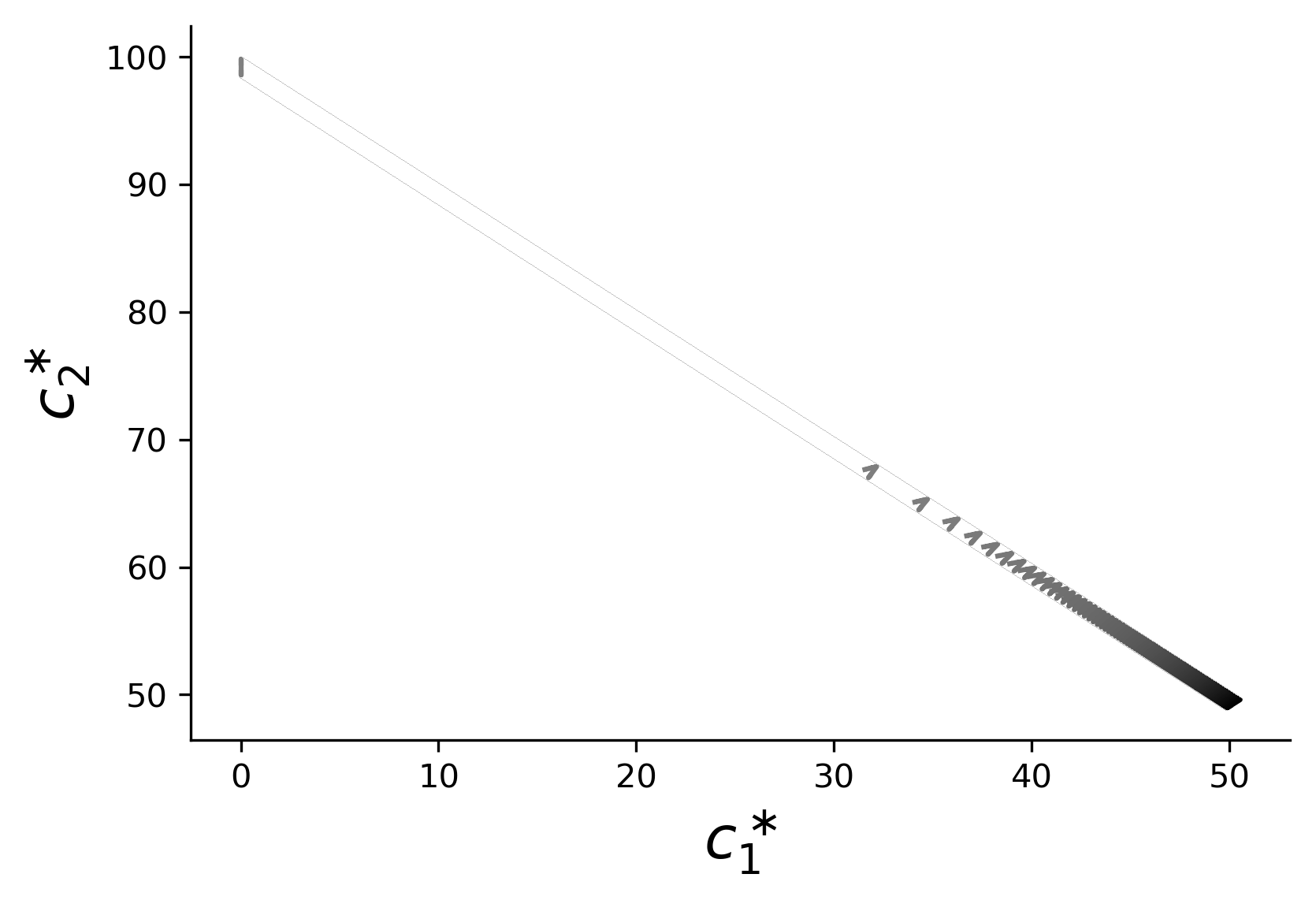}
  \includegraphics[width=2in]{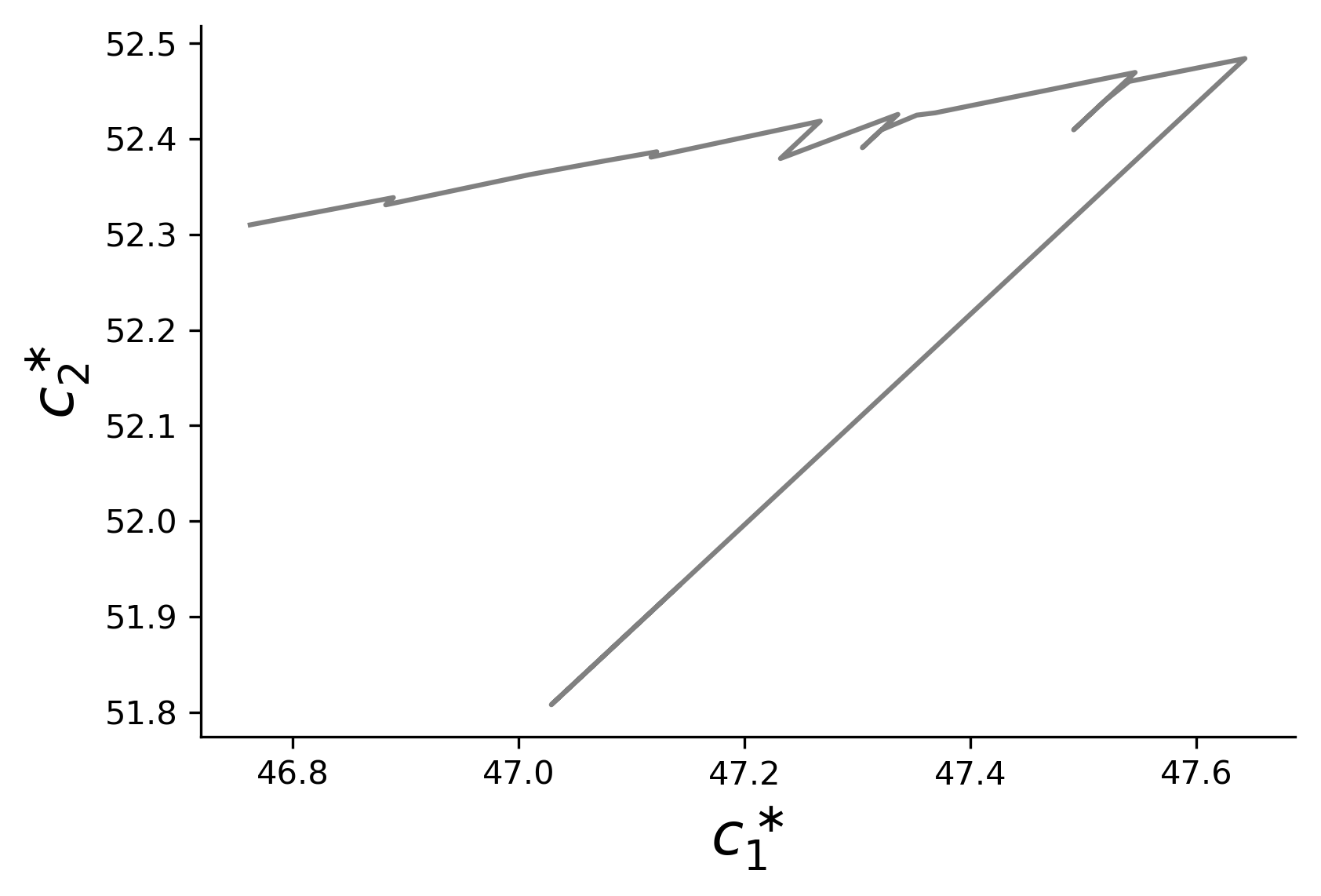}
  \caption{Realized optimal consumption policies (with zoom-in view) in Example 2}
  \label{fig:9}
\end{figure}

\begin{figure}[H]
  \centering
  \includegraphics[width=2in]{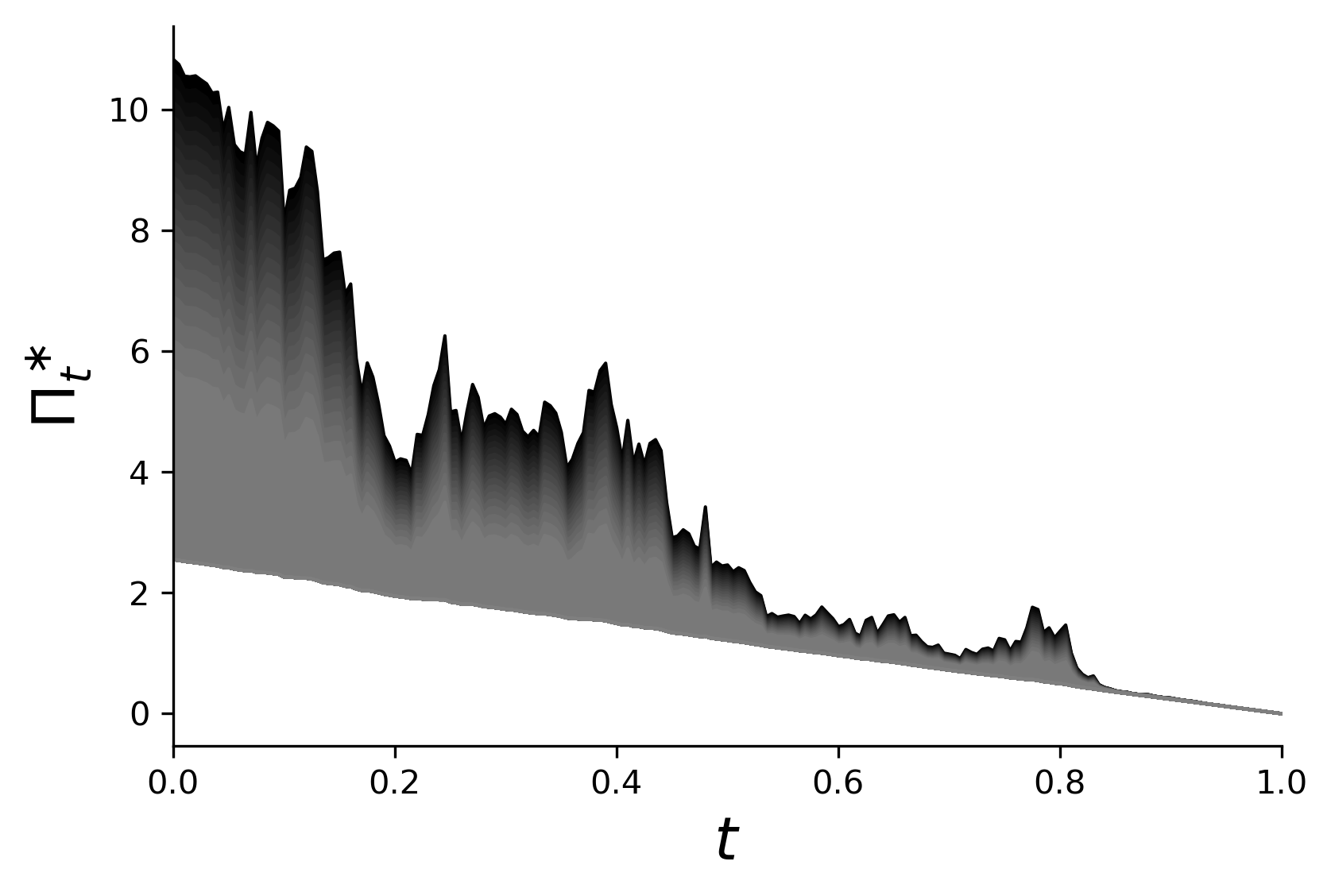}
  \includegraphics[width=2in]{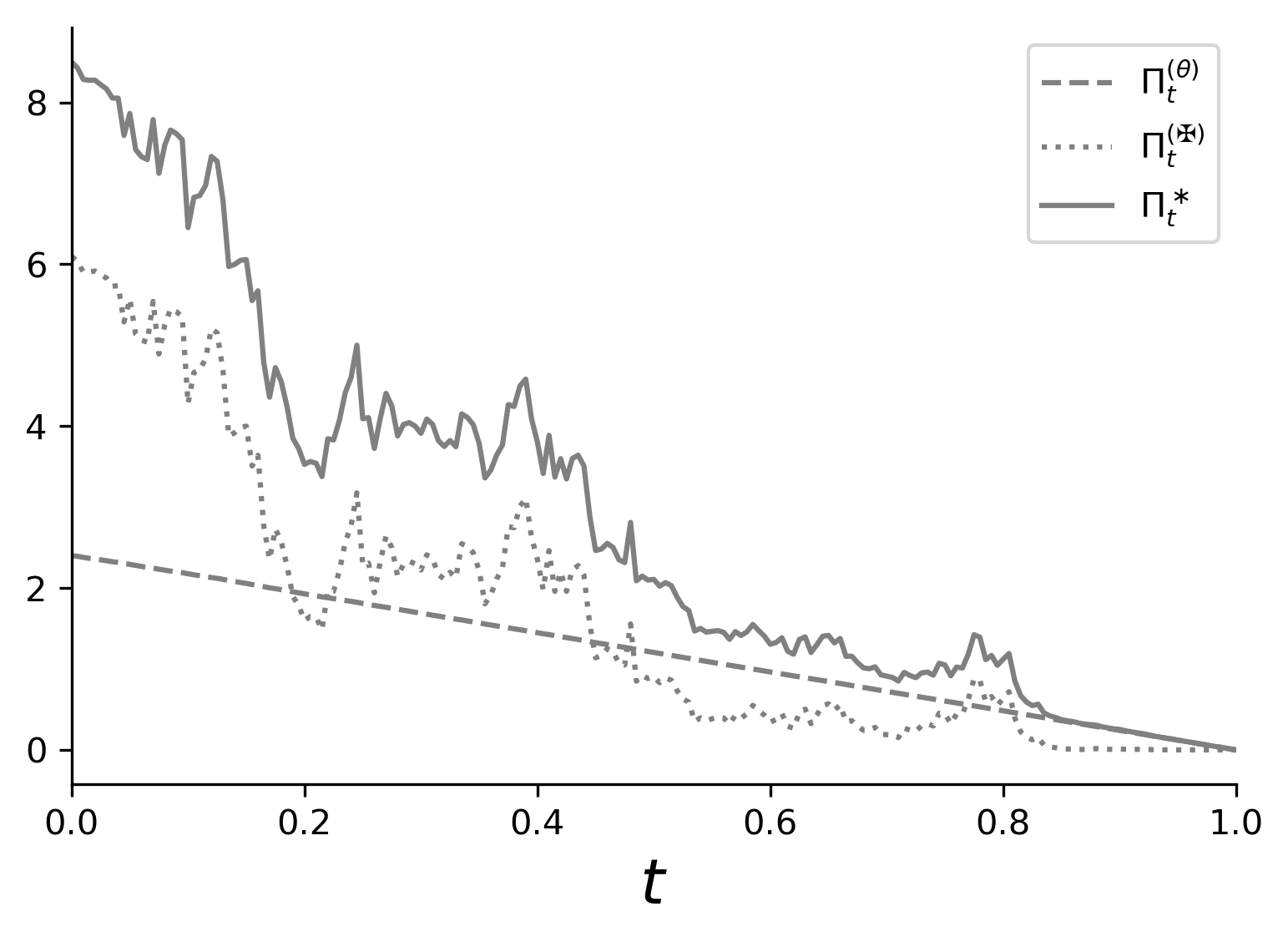}
  \caption{Realized optimal investment policies (with portfolio decomposition) in Example 2}
  \label{fig:10}
\end{figure}

We can clearly see that in Example 2, the optimal total consumption expenditure and the optimal investment are both set-valued, caused by the perception of an increasing degree of attention, or effectively, an enlarged rate of substitution. An attendant consequence is stochastic coordinates between the optimal consumption elements for any fixed weight functional $w$. In this case, the optimal investment policy is decomposed into a mean--variance portfolio and an indecisiveness risk-hedging portfolio. The effect is of course to temporally increase the size of the investor's hedging position due to preference fluctuations.

\begin{figure}[H]
  \centering
  \includegraphics[width=2in]{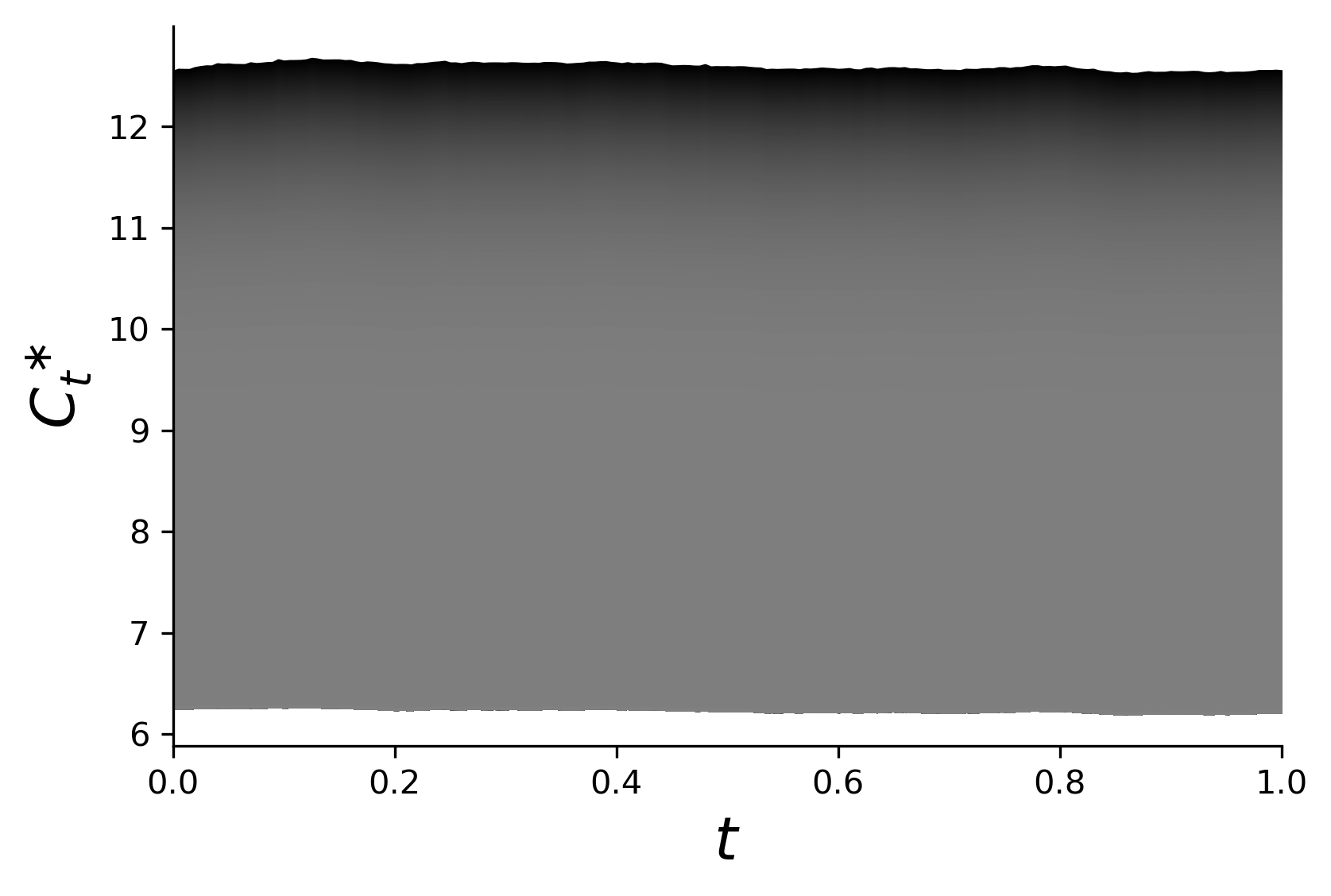}
  \includegraphics[width=2in]{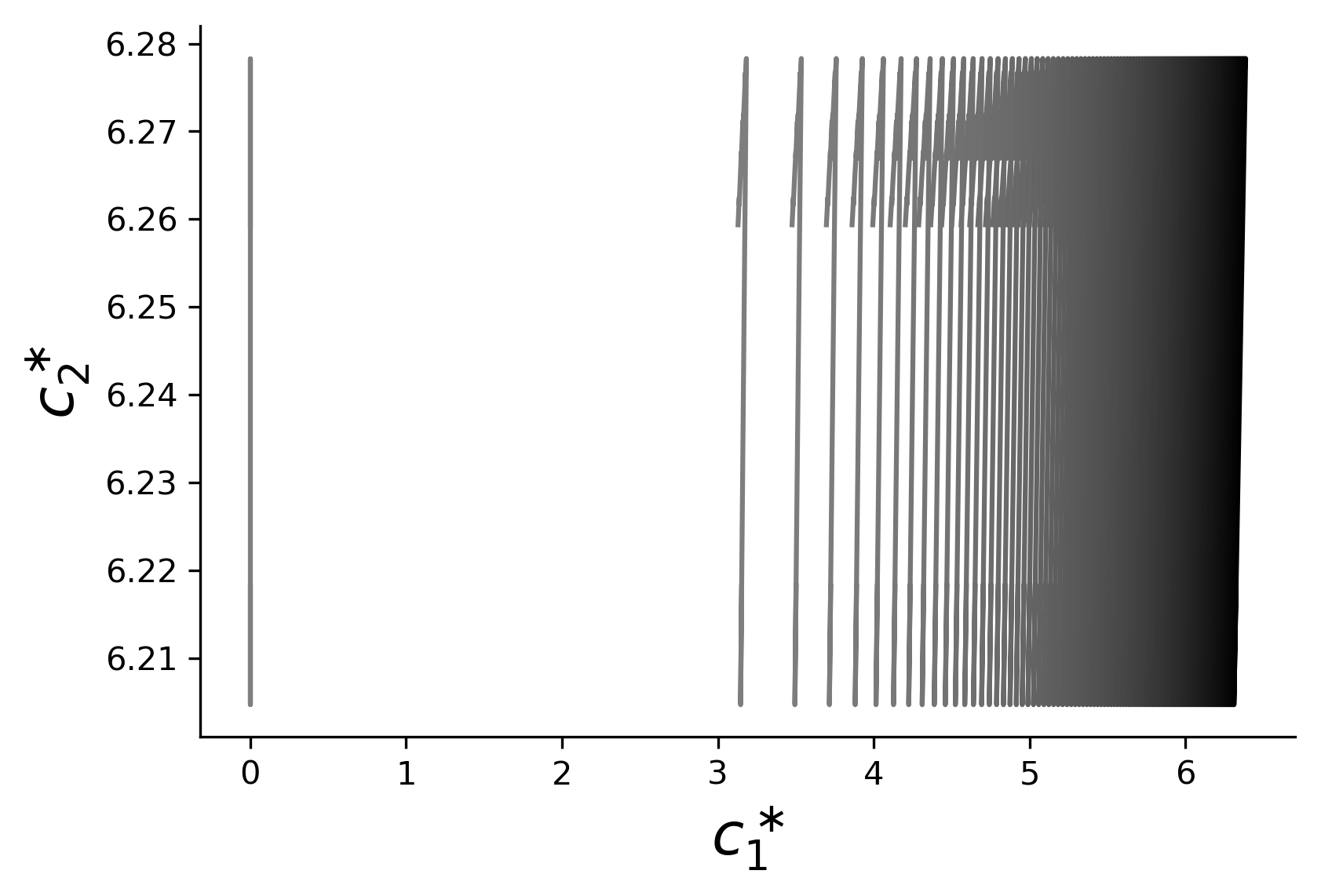}
  \includegraphics[width=2in]{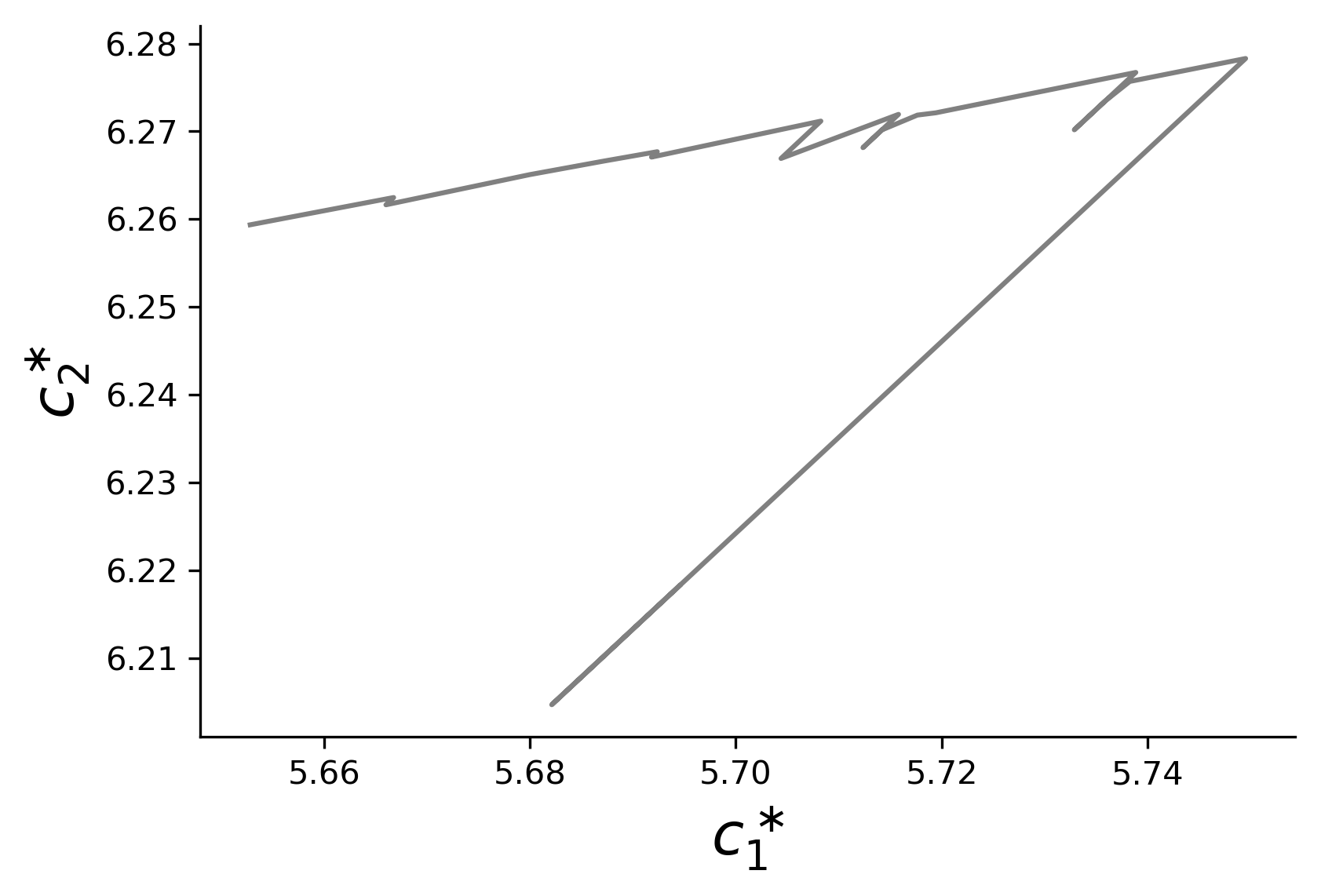}
  \caption{Realized optimal consumption policies (with zoom-in view) in Example 3}
  \label{fig:12}
\end{figure}

\begin{figure}[H]
  \centering
  \includegraphics[width=2in]{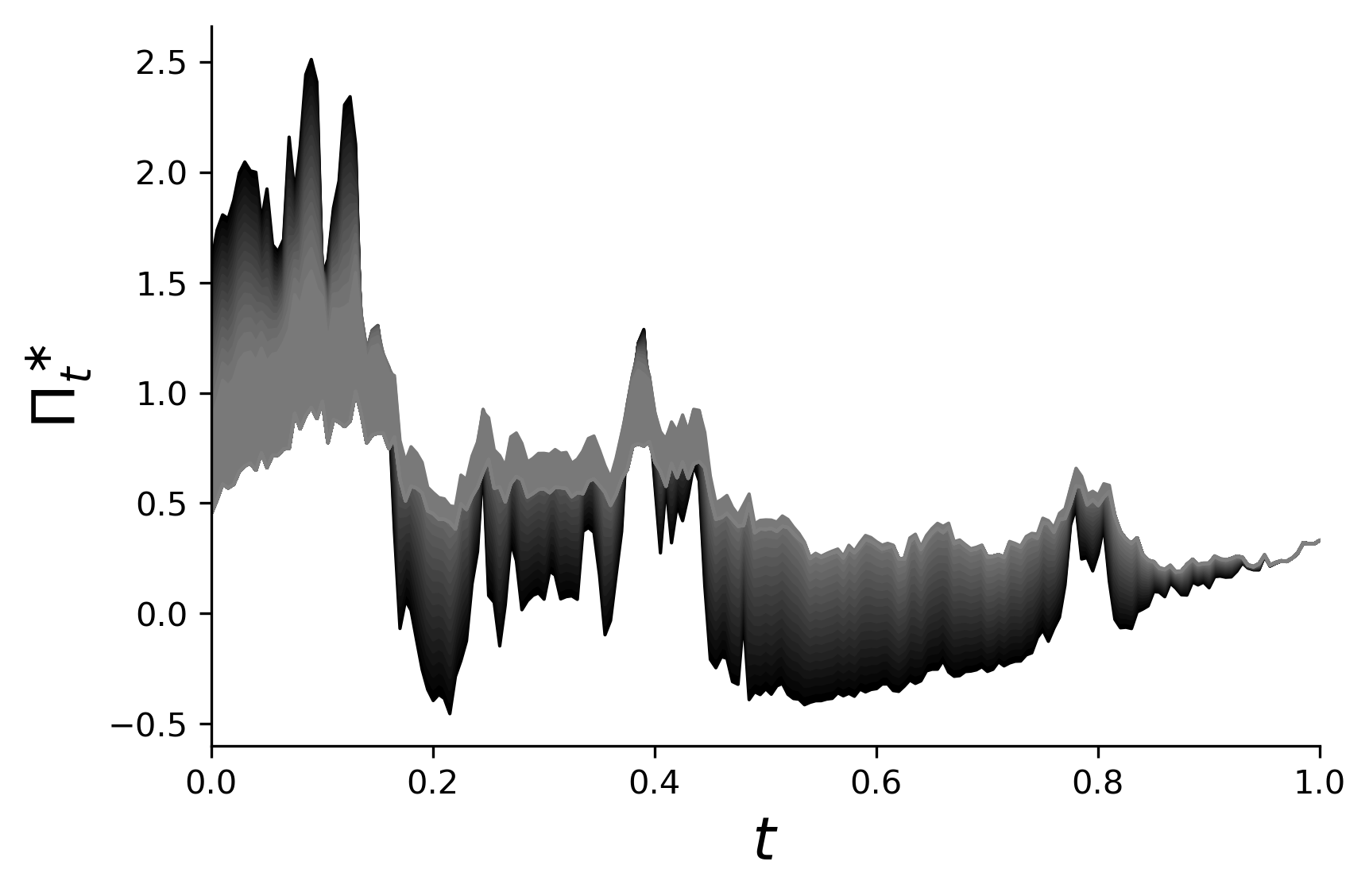}
  \includegraphics[width=2in]{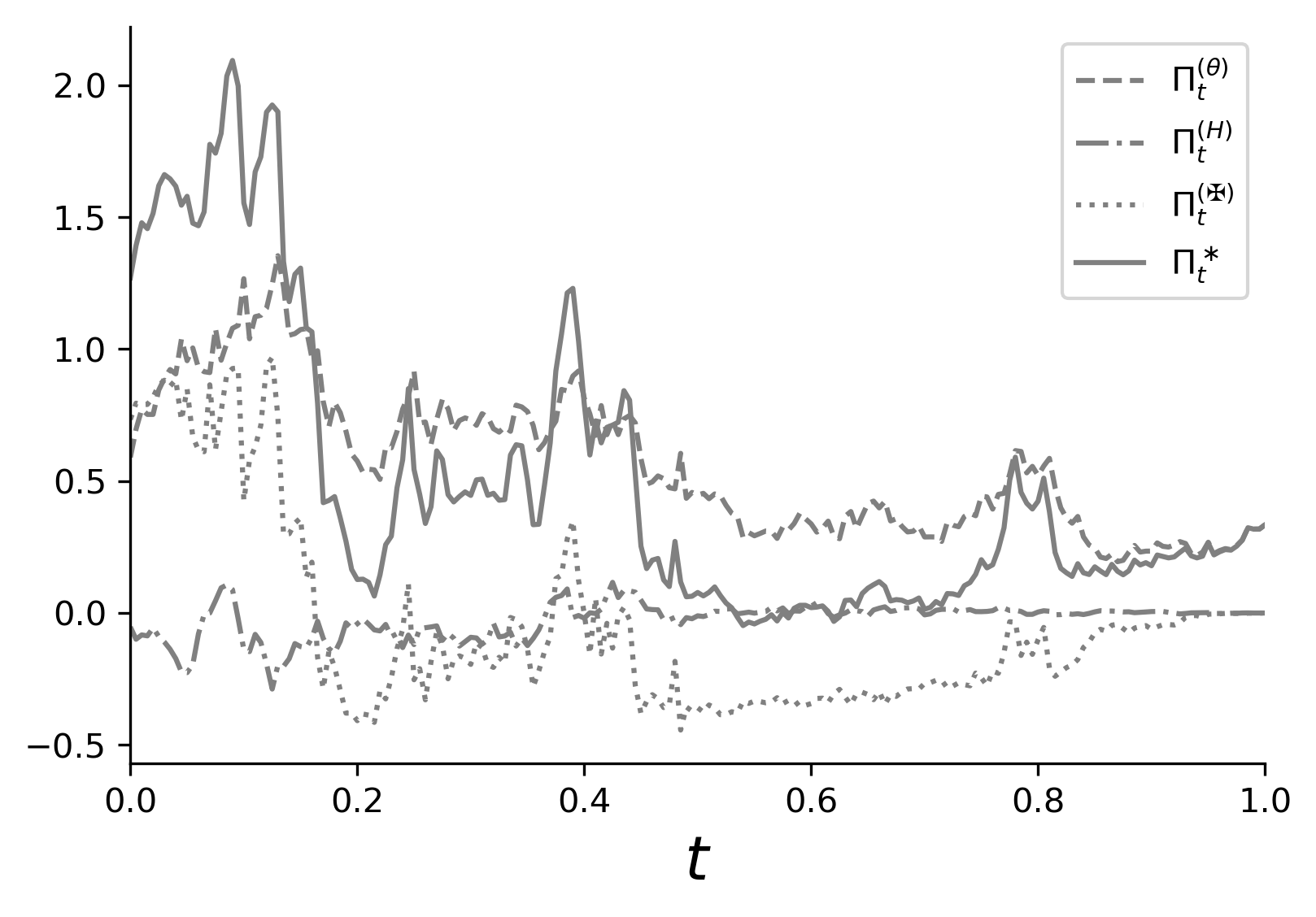}
  \caption{Realized optimal investment policies (with portfolio decomposition) in Example 3}
  \label{fig:13}
\end{figure}

Unfortunately, it is not possible to visualize, using only finitely many pages, all shapes of optimal consumption--investment policies in Example 3 because of infinite dimensions. What Figure \ref{fig:12} and Figure \ref{fig:13} have illustrated for instance are the policies given degenerate second dimension of the weight functional $w_{2}$ (recall (\ref{3.2.10}) and Table \ref{tab:1}), which only constitutes a proper subset of the true solution set. In this subcase, the investor concentrates on the subjective lower bound of market volatility ($\varepsilon=0.2$) and volatility-driven risk aversion has effectively precise values. Nevertheless, the optimal policies, all being set-valued, are vastly different from those in Example 2 (compare Figure \ref{fig:10} and Figure \ref{fig:11}) because of the inclusion of bequest utility and stochastic volatility. As expected from previous analysis, each optimal investment policy is broken down to a mean--variance portfolio, a market risk-hedging portfolio, and an indecisiveness risk-hedging portfolio, the latter two being temporally negative, which however have led to a further increase in the investor's (magnitude) temporal hedging position. Since the investor now values bequest, his hedging positions are not liquidated towards the end of the investment horizon. By experimenting with other non-degenerate forms of $w_{2}$ many interesting consumption--investment policies can be envisaged, all of which are equally optimal and for which indecisiveness can also happen in risk aversion.

\vspace{0.2in}

\section{Conclusions}\label{s:7}

In this paper we have profoundly studied martingale solutions to optimal consumption--investment choices when preferences are incomplete and fluctuating in the wake of indecisiveness in tastes. The market has a canonical setup with one risk-free asset, $m$ risky assets, and $n$ consumption goods, whose dynamics are driven by general diffusion, as well as $d$ fluctuating preference parameters with (possibly) inexact values. In adherence to the outline in Section \ref{s:1}, we have analyzed several important external psychological factors that have been empirically evidenced in proposing to model time-varying incomplete preferences by a stochastic index set following a $d$-dimensional set-valued It\^{o} process with two transformed monotone components. The structure is remarkably flexible enough to incorporate randomness in patience, degrees of risk aversion, and degrees of attention, all of which may be consumption good-specific and change according to market characteristics and socialization effects.

After significant modifications of the Gass--Saaty scalarization method, a unified approach towards characterizing the full set of optimal consumption--bequest policies is proposed by defining weight functionals in a stochastic dual cone, while the associated optimal investment policy is achievable via exploiting a multi-valued version of Malliavin calculus, in conjunction with techniques from stochastic geometry. Depending on various factors, such as the design of multi-utility and the time variation of the underlying index set, optimal total consumption expenditure and investment may or may not be set-valued. Originally, this is in direct relation to the degree of indecisiveness, which does not have to completely vanish to result in single-valued policies. In contrast, good-specific consumption policies are most likely set-valued under optimality in such a setting, highlighting a preference for deliberate randomization under incomparability or conflicts. Another key conclusion is a new portfolio decomposition (mentioning (\ref{5.1.2}) and (\ref{5.1.3})) where commodity price and indecisiveness risks also bring about two hedging components, which tend to add to the amplitude of the aggregate hedging demand, beside a standard mean--variance portfolio and a market risk-hedging portfolio.

From the demonstrations in Section \ref{s:6}, implementing the corresponding optimal consumption and investment policies in practice is arguably plain sailing. For optimal consumption, one only needs to pick a weight functional $w$ from the corresponding dual cone, with unit (total variation or Taxicab) norm, and then compute the policy $(c^{\ast},X^{\ast}_{T}|w)$ in the closed solution set $\mathcal{S}^{\ast}$ (in (\ref{3.1.3})) with the (modified) Gass--Saaty method. The key steps consist in solving the optimality conditions (\ref{3.2.1}) and detailed procedures have been presented in Section \ref{ss:3.2}, or Figure \ref{fig:2}. We remark that the weight functional is economically meaningful in terms of averaging the conflicting utility elements governing taste imprecision, rather than an entirely artificial parameter, and it provides the flexibility needed to recover all equally optimal policies, with only one-time computation, and can be customized from time to time. Once the optimal consumption policies are written in feedback form, the optimal investment policy can be subsequently computed using Theorem \ref{thm:4} or Corollary \ref{cor:1} as an element in the formula (\ref{5.1.1}) or (\ref{5.1.5}). Importantly, the choice of the weight functional $w$ must be the same as that used for optimal consumption, in order to ensure achievement of the desired optimality. The main difference, however, is that while any limit point of parameterized optimal consumption policies always coincides with the policy subject to the limit of the parameterizing weight functionals, this is in not guaranteed for optimal investment policies, which fact is taken care of by $\mathds{L}^{1}$-closure, and in consequence one may end up with limiting (or utopian) investment policies that do not go with any weight functionals within the dual cone, but such limiting situations would be hard to manage compared with using one single weight functional.

Our study has also shed light upon several interesting directions for future research. First, an attempt could be made to coalesce imprecise beliefs (about the market model) into imprecise tastes and then developing methods to solve a multi-utility maximization problem that is only partially robust, as mentioned since Section \ref{s:1}. As a simple example, one can allow the market dynamics (\ref{2.1.1}) to be ambiguous, apart from conflicts in comparing consumption goods, subject to a preselect pool of probability measures. With the mean--variance frontier illustration in Section \ref{s:1}, a problem of this nature would trigger maximization of the following objective function:
\begin{equation}\label{7.1}
  \inf_{\PP\in\P}\bigg(\bigg\{\Pi^{\intercal}\mu^{\PP}-\mathds{R}^{2}_{+}, -\frac{1}{2}\big\|(\sigma^{\PP})^{\intercal}\Pi\big\|^{2}_{2}-\mathds{R}^{2}_{+}\bigg\}\bigg) =\bigcap_{\PP\in\P}\bigg(\bigg\{\Pi^{\intercal}\mu^{\PP},-\frac{1}{2}\big\|(\sigma^{\PP})^{\intercal}\Pi\big\|^{2}_{2}\bigg\} -\mathds{R}^{2}_{+}\bigg),
\end{equation}
where the superscript $\PP$ indicates the probability measure under which expectations are taken for the risky returns. In (\ref{7.1}), the infimum is taken over a convex space $\P$ of equivalent probability measures each admitting a bijective correspondence with the belief-specific coefficients $\big(\mu^{\PP},\sigma^{\PP}\big)$. In contrast to only imprecise beliefs (e.g., \cite[Liang and Ma, 2020, \text{Sect.} 2]{LM1}), the infimum is taken for a multi-valued quantity with respect to the natural ordering cone $\mathds{R}^{2}_{+}$ and should be understood in a set-valued sense (similar to (\ref{4.2})). Studying problems resembling (\ref{7.1}) will have appealing consequences in consideration of ``probability-utility pairs'' proposed in \cite[Nau, 2006, \text{Sect.} 1.4]{N1}. For the second direction, one could introduce other sources of randomness such as jump risks and memory effects into the market dynamics (\ref{2.1.1}) to better explain stylized facts. Although the possibility of obtaining martingale solutions to Merton's consumption--investment problem in the presence of price jumps has been considered in the literature (see, e.g., \cite[Michelbrink and Le, 2012]{ML}), one would require a set-valued analog of (fractionally integrated) L\'{e}vy--It\^{o} processes to generalize the multi-utility index set dynamics (\ref{2.2.1}), which has recently been developed in \cite[Xia, 2025]{X}; with these tools, preferences can have much more variability with jumps and memory. Thirdly, as mentioned since Section \ref{ss:2.1}, an ongoing study is devoted to habit-driven indecisiveness to deal with the mere-exposure effect (\cite[Bornstein, 1989]{B2}), which will naturally complement the externality-based results in this paper. Last but not least, general equilibrium models beyond doubt deserve a profound study subject to individual rationality under time-varying incomplete preferences, while the multi-utility representation index set is naturally endogenized; as said in Section \ref{s:1}, the present paper takes the first step towards the ultimate goal to analyze and test incompleteness and time variation in economic agents' preferences on the dynamic ranges of asset prices as well as their coexistence.

\clearpage

\appendix
\gdef\thesection{Appendix \Alph{section}}

\renewcommand{\theequation}{A.\arabic{equation}}

\section{Mathematical proofs}\label{A}

\textbf{Proposition \ref{pro:1}}
\begin{proof}
First note that each set-valued process $I_{q}$ for $q\in\{1,2,3\}$ as defined in (\ref{2.2.3}) is clearly $\mathbb{F}$-non-anticipating. For every $t\in[0,T]$ each $I_{q,t}$ is compact convex-valued in $\mathds{R}^{d}$, $\PP$-a.s., because the integral functionals $\int^{t}_{0}f_{q,s}\dd s$ and $\int^{t}_{0}\overline{\co}_{\mathds{L}^{2}}G_{q,s}\dd W_{s}$ are both closed convex subsets of $\mathds{L}^{1}_{\mathbb{F}}([0,T]\times\Omega;\mathds{R}^{d})$.

According to the selection theorems for Aumann stochastic integrals and set-valued It\^{o} integrals (e.g., see \cite[Li and Li, 2009, \text{Thm.} 3]{LL2} and \cite[Kisielewicz, 2020, \text{Thm.} 5.6.2]{K3}), since $I_{q,t}$, $q\in\{1,2,3\}$, is for every $t\in[0,T]$ $\mathbb{F}$-non-anticipating and ($\PP$-a.s.) convex-valued, there exists a sequence $\{Y_{q,k}:k\in\mathds{N}_{++}\}$ of $d$-dimensional non-anticipating It\^{o} processes\footnote{Such processes, when considered as elements of $\mathds{L}^{1}_{\mathbb{F}}([0,T]\times\Omega,\mathds{R}^{d})$ or $\mathds{L}^{2}_{\mathbb{F}}([0,T]\times\Omega,\mathds{R}^{d})$, are called the representation Castaing of the process $I_{q}$, $q\in\{1,2,3\}$, which concept is closely related to measurability of stochastic multifunctions.} such that $I_{q,t}=\cl_{\mathds{R}^{d}}\{Y_{q,k,t}:k\in\mathds{N}_{++}\}$, $\PP$-a.s., for every $t\in[0,T]$. Since $Y_{q,k}$'s have $\PP$-\text{a.s.} continuous sample paths, we are permitted to write for every $t\in[0,T]$
\begin{align*}
  \mathcal{I}_{t}&=\cl_{\mathds{R}^{d}}\Bigg(\cl_{\mathds{R}^{d}}\{Y_{1,k,t}:k\in\mathds{N}_{++}\} +\bigcap_{s\in[0,t\wedge\mathcal{t}]}\cl_{\mathds{R}^{d}}\{Y_{2,k,s}:k\in\mathds{N}_{++}\}\\
  &\qquad+\overline{\co}_{\mathds{R}^{d}}\bigcup_{s\in[0,t]\cap\mathds{Q}}\cl_{\mathds{R}^{d}}\{Y_{3,k,s}:k\in\mathds{N}_{++}\}\Bigg), \quad\PP\text{-a.s.}
\end{align*}
For the second term, note that $\{\mathcal{t}\leq t\}=\big\{\mathrm{card}\bigcap_{s\in[0,t]}I_{2,s}\leq1\big\}$ from (\ref{2.2.2}), and then applying Carath\'{e}odory's theorem to the third term gives that $\mathcal{I}_{t}$ is $\mathcal{B}([0,T])\otimes\mathcal{F}$-measurable. As $Y_{q,k}$'s are also $\mathbb{F}$-adapted, it holds that each $Y_{q,k,t}$ is $\mathscr{F}_{t}$-measurable for every $t\in[0,T]$, and then it is easy to establish that $\mathcal{I}_{q,t}$ is also $\mathscr{F}_{t}$-measurable from the last equation, and thus the adapted-ness of $\mathcal{I}$ with respect to $\mathbb{F}$ as well. In this case, $\mathcal{t}$ is also verified to be an $\mathbb{F}$-stopping time. Therefore, $\mathcal{I}$ is $\mathbb{F}$-non-anticipating.

Next, we observe for every $t\in[0,T]$ that
\begin{equation*}
  \mathcal{d}_{\rm H}\bigg(\overline{\co}_{\mathds{R}^{d}}\bigcup_{s\in[0,t]}I_{3,s},\{\0\}\bigg)\leq\sup_{s\in[0,t]}\mathcal{d}_{\rm H}(I_{3,s},\{\0\}),\quad\PP\text{-a.s.},
\end{equation*}
but by the construction (\ref{2.2.1}) and the properties of the Hausdorff distance it follows that, $\PP$-a.s.,
\begin{align*}
  \mathcal{d}_{\rm H}(\mathcal{I}_{t},\{\0\})&\leq\mathcal{d}_{\rm H}(I_{1,t},\{\0\})+\mathcal{d}_{\rm H}(I_{2,0},\{\0\})+\sup_{s\in[0,t]}\mathcal{d}_{\rm H}(I_{3,s},\{\0\})\\
  &\leq\sum^{3}_{q=1}\mathcal{d}_{\rm H}(I_{q,0},\{\0\})+\mathcal{d}_{\rm H}\bigg(\int^{t}_{0}f_{1,s}\dd s,\{\0\}\bigg)+\sup_{s\in[0,t]}\mathcal{d}_{\rm H}\bigg(\int^{t}_{0}f_{3,s}\dd s,\{\0\}\bigg)\\
  &\qquad+(\sqrt{d}+1)\bigg(\mathcal{d}_{\rm H}\bigg(\int^{t}_{0}G_{1,s}\dd W_{s},\{\0\}\bigg)+\sup_{s\in[0,t]}\mathcal{d}_{\rm H}\bigg(\int^{s}_{0}G_{3,v}\dd W_{v},\{\0\}\bigg)\bigg),
\end{align*}
where in the last inequality we have also used \cite[Kisielewicz, 2020, \text{Thm.} 5.2.1]{K3} to take the closed convex hulls outside the integrands along with Starr's corollary to the Shapley-Folkman theorem (originated from \cite[Starr, 1969, \text{Appx.} 2]{S3}). Since $\mathcal{I}$ is $\mathbb{F}$-non-anticipating and $I_{3}$ is ($\PP$-a.s.) continuous, so that the positive-valued process $\mathcal{d}_{\rm H}\big(\int^{\imath}_{0}G_{3,s}\dd W_{s},\{\0\}\big)$ is a submartingale (\cite[Kisielewicz, 2020, \text{Corol.} 5.5.1]{K3}), we can take expectations to obtain
\begin{align*}
  &\quad\E[\mathcal{d}_{\rm H}(\mathcal{I}_{t},\{\0\})]\\
  &\leq\sum^{3}_{q=1}\mathcal{d}_{\rm H}(I_{q,0},\{\0\})+\E\bigg[\mathcal{d}_{\rm H}\bigg(\int^{t}_{0}f_{1,s}\dd s,\{\0\}\bigg)\bigg]+\E\bigg[\sup_{s\in[0,t]}\mathcal{d}_{\rm H}\bigg(\int^{s}_{0}f_{3,v}\dd v,\{\0\}\bigg)\bigg]\\
  &\qquad+(\sqrt{d}+1)\Bigg(\E\bigg[\mathcal{d}^{2}_{\rm H}\bigg(\int^{t}_{0}G_{1,s}\dd W_{s},\{\0\}\bigg)\bigg]^{1/2}+2\E\bigg[\mathcal{d}^{2}_{\rm H}\bigg(\int^{t}_{0}G_{3,s}\dd W_{s},\{\0\}\bigg)\bigg]^{1/2}\Bigg)\\
  &\leq\sum^{3}_{q=1}\mathcal{d}_{\rm H}(I_{q,0},\{\0\})+\E\bigg[\mathcal{d}_{\rm H}\bigg(\int^{t}_{0}f_{1,s}\dd s,\{\0\}\bigg)\bigg]+t\E\Big[\sup_{s\in[0,t]}\mathcal{d}_{\rm H}(f_{3,s},\{\0\})\Big]\\
  &\qquad+(\sqrt{d}+1)\Bigg(\Bigg(\int^{t}_{0}\sum^{\infty}_{k=1}\E\big[\|g_{1,k,s}\|^{2}_{\mathrm{F}}\big]\dd s\Bigg)^{1/2}+2\Bigg(\int^{t}_{0}\sum^{\infty}_{k=1}\E\big[\|g_{3,k,s}\|^{2}_{\mathrm{F}}\big]\dd s\Bigg)^{1/2}\Bigg)\\
  &<\infty,
\end{align*}
where the properties of the Hausdorff distance, the $\PP$-\text{a.s.} continuity of the Aumann integrals, Doob's maximal inequality, H\"{o}lder's inequality, \cite[Kisielewicz, 2020, \text{Corol.} 5.4.3]{K3}, and the conditions listed in (\ref{2.2.4}) have been used, in proper order. It is hence concluded that $\mathcal{I}$ is integrably bounded (i.e., it has integrable Hausdorff distances; see the mention in \cite[Malinowski, 2013, \text{Sect.} 2.1]{M1}).

In addition, fix any $t\in[0,T]$ and pick an arbitrary $\epsilon>0$. We observe that in the limit as $\epsilon\searrow0$, $\PP$-a.s.,
\begin{align*}
  \sup_{z\in\mathcal{I}_{t+\epsilon}}\mathrm{dist}(z,\mathcal{I}_{t})&\leq\mathcal{d}_{\rm H}(I_{1,t+\epsilon},I_{1,t})+\sup_{v\in[0,\epsilon]}\sup_{z\in I_{2,t+v}}\mathrm{dist}(z,I_{2,t})+\sup_{v\in[0,\epsilon]}\sup_{z\in I_{3,t+v}}\mathrm{dist}(z,I_{3,t})\\
  &\leq\mathcal{d}_{\rm H}(I_{1,t+\epsilon},I_{1,t})+2\sup_{v\in[0,\epsilon]}\max\{\mathcal{d}_{\rm H}(I_{2,t+v},I_{2,t}),\mathcal{d}_{\rm H}(I_{3,t+v},I_{3,t})\}\rightarrow0,
\end{align*}
where the limit follows from the ($\PP$-a.s.) continuity of $I_{q}$ for all $q\in\{1,2,3\}$, and hence $\mathcal{I}$ is Hausdorff upper semi-continuous. According to \cite[Kisielewicz, 1991, \text{Thm.} 3.8]{K1}, since $\mathcal{I}_{t}\subseteq\mathds{R}^{d}$ is ($\PP$-a.s.) compact and convex for every $t\in[0,T]$, it follows that $\mathcal{I}$ is upper semi-continuous. Repeating the steps for $\sup_{z\in\mathcal{I}_{t}}\mathrm{dist}(z,\mathcal{I}_{t+\epsilon})$ we obtain the Hausdorff lower semi-continuity, which implies lower semi-continuity, and hence the $\PP$-\text{a.s.} continuity of $\mathcal{I}$.
\end{proof}

\medskip

\noindent\textbf{Proposition \ref{pro:2}}
\begin{proof}
Note that the multi-utility defined in (\ref{2.2.5}) can be naturally extended to the (entire) space $\mathds{R}^{d}$ as follows,
\begin{equation}\label{A.1}
  \bar{u}(t,c|\mathcal{I}_{t})=\{u_{i}(t,c)\mathds{1}_{\mathcal{I}_{t}}(i):i\in\mathds{R}^{d}\},\quad t\in[0,T],
\end{equation}
Then in light of Proposition \ref{pro:1} $\bar{u}(\imath,c|\mathcal{I})$ as a process is automatically understood to be $\mathbb{F}$-non-anticipating,\footnote{For notational convenience, whenever the time variable is trivially understood for a given stochastic process it is either suppressed or represented by the identity (temporal) map $\imath$.} and because each $u_{i}$ is c\`{a}dl\`{a}g in time, $\mathcal{I}$ is ($\PP$-a.s.) continuous, and $U$ is deterministic, we have actually
\begin{equation*}
  \bigcup_{s\in[0,t]}\prod_{i\in\mathcal{I}_{s}}\im(u_{i}(s,\cdot)+U) =\bigcup_{s\in[0,t]\cap\mathds{Q}}\prod_{i\in\mathds{R}^{d}}\im(u_{i}(s,\cdot)\mathds{1}_{\mathcal{I}_{s}}(i)+U) \ni\0,
\end{equation*}
which is also measurable with respect to $\mathcal{B}([0,T])\otimes\mathcal{F}$ and $\mathscr{F}_{t}$ when considered for a fixed $t\in[0,T]$ by Carath\'{e}odory's theorem. Therefore, by taking $\mathcal{K}_{t}$ to be a closed convex cone within the right-hand side at any time $t$ the required measurability is established.
\end{proof}

\medskip

\noindent\textbf{Proposition \ref{pro:3}}
\begin{proof}
By definition, let $c,c',c''\in\mathds{R}^{n}_{+}$. Then $c\succeq_{t}c$ immediately because $u(t,c)-u(t,c)=\0\in\mathcal{K}_{t}$, signifying reflexivity. Also, $c\succeq_{t}c'$ if and only if $u(t,c)-u(t,c')\in\mathcal{K}_{t}$ and $c'\succeq_{t}c''$ if and only if $u(t,c')-u(t,c'')\in\mathcal{K}_{t}$, which implies that
\begin{equation*}
  u(t,c)-u(t,c'')=u(t,c)-u(t,c')+u(t,c')-u(t,c'')\in\mathcal{K}_{t},
\end{equation*}
so that $\succeq_{t}$ is transitive. To see that $\succeq_{t}$ is incomplete in general it suffices to take the natural ordering cone $\mathcal{K}_{t}=\mathcal{C}_{\rm b}(\mathcal{I}_{t};\mathds{R}_{+})$ , and then if $c\nsucceq_{t}c'$, $u(t,c)-u(t,c')\in\big(\mathcal{C}_{\rm b}(\mathcal{I}_{t};\mathds{R}_{+})\big)^{\complement}\nsubseteq\mathcal{C}_{\rm b}(\mathcal{I}_{t};\mathds{R}_{-})=-\mathcal{K}_{t}$, not implying $c'\succeq_{t}c$.
\end{proof}

\medskip

\noindent\textbf{Theorem \ref{thm:1}}
\begin{proof}
For assertion (i), a straightforward application of It\^{o}'s formula to the discounted wealth process $\xi X$ gives that
\begin{equation}\label{A.2}
  \int^{t}_{0}\xi_{s}C_{s}\dd s+\xi_{t}X_{t}=\int^{t}_{0}\langle\xi_{s}\Pi_{s},\sigma_{s}\dd W_{s}\rangle_{m}-\int^{t}_{0}\langle\xi_{s}X_{s}\theta_{s},\dd W_{s}\rangle_{m},\quad t\in[0,T].
\end{equation}
The left-hand side is obviously nonnegative if $(c,\Pi)\in\mathfrak{A}(X_{0})$ while the right-hand side constitutes a continuous $\mathbb{F}$-local martingale. Therefore, it is a familiar result (see, e.g., \cite[Lyasoff, 2017, \text{Chap.} IX \text{Ex.} 9.68]{L2}) that the process $\int^{\imath}_{0}\xi_{s}C_{s}\dd s+\xi X$ is a supermartingale, which implies
\begin{equation*}
  \E\bigg[\int^{T}_{0}\xi_{s}C_{s}\dd s+\xi_{T}X_{T}\bigg]\leq X_{0},
\end{equation*}
and hence $(c,X_{T})\in\mathfrak{B}(X_{0})$.

Conversely, for (ii), we note that the process $\E\big[\int^{T}_{0}\xi_{s}C_{s}\dd s+\xi_{T}X_{T}\big|\mathscr{F}_{\imath}\big]=:M$ is an $\mathbb{F}$-martingale, so that by martingale representation
\begin{equation}\label{A.3}
  M_{t}=M_{0}+\int^{t}_{0}\langle\phi_{s},\dd W_{s}\rangle_{m},\quad t\in[0,T].
\end{equation}
Since $\sigma$ is invertible, we choose the investment policy as follows based on (\ref{2.3.1}),
\begin{equation}\label{A.4}
  \Pi=X(\sigma^{\intercal})^{-1}\theta+\xi^{-1}(\sigma^{\intercal})^{-1}\phi,
\end{equation}
which plugged into (\ref{A.2}) yields that
\begin{equation*}
  \xi_{t}X_{t}\geq\E\bigg[\int^{T}_{t}\xi_{s}C_{s}\dd s\bigg|\mathscr{F}_{t}\bigg]\geq0.
\end{equation*}
As $\xi$ is strictly positive, this establishes as desired $X\geq0$ so that $(c,\Pi)\in\mathfrak{A}(X_{0})$.
\end{proof}

\medskip

\noindent\textbf{Theorem \ref{thm:2}}
\begin{proof}
First, by the definition of $w$ we can take $\bar{w}$ to be an extended Radon measure over $\bar{\mathcal{I}}$ such that for every $t\in[0,T]$ the stochastic restriction\footnote{In reduced (finite) dimensions, such measure restriction is not needed since $\mathrm{card}J$ is fixed over time.}
\begin{equation}\label{A.5}
  w(t)=\bar{w}\upharpoonright_{\mathcal{I}_{t}}
\end{equation}
is in force $\PP$-a.s., hence enabling the following correspondence:
\begin{align}\label{A.6}
  V(c,X_{T}|w)&=\int^{T}_{0}\E\bigg[\bigg\langle w(t),u(t,c_{t})+\frac{U(X_{T})}{T}\bigg\rangle_{\mathcal{I}_{t}}\bigg]\dd t \nonumber\\
  &=\int^{T}_{0}\bigg\langle\bar{w},\E\bigg[u(t,c_{t})+\frac{U(X_{T})}{T}\bigg]\bigg\rangle_{\bar{\mathcal{I}}}\dd t \nonumber\\
  &=\bigg\langle\bar{w},\E\bigg[\int^{T}_{0}\bigg(u(t,c_{t})+\frac{U(X_{T})}{T}\bigg)\dd t\bigg]\bigg\rangle_{\bar{\mathcal{I}}} \nonumber\\
  &=\langle\bar{w},V(c,X_{T})\rangle_{\bar{\mathcal{I}}},
\end{align}
where the first equality follows from linearity and the second and third use the integrable bounded-ness of $\mathcal{I}$ (see Proposition \ref{pro:1}).

Note that the budget set is by construction a compact convex subset of $\mathfrak{C}_{n}\times\mathds{L}^{1}_{\mathcal{F}}(\Omega;\mathds{R}_{+})$. Given the concavity properties in Assumption \ref{as:1} and the mentioned closed-ness of $u(t,\mathds{R}^{n}_{+})$, dominated convergence readily ensures that the criterion space $V(\mathfrak{B}(X_{0}))$ is a closed convex subset of $\mathcal{C}_{\rm b}(\bar{\mathcal{I}};\mathds{R})$. Also, if $(c^{\ast},X^{\ast}_{T})$ is a $\bar{\mathcal{K}}$-maximal solution of the multi-criteria problem (\ref{2.3.5}), then by definition we have $(V(c^{\ast},X^{\ast}_{T})+\Int\bar{\mathcal{K}})\cap V(\mathfrak{B}(X_{0}))=\emptyset$. This forms the ground for us to employ the Hahn--Banach separation theorem to establish the existence of a (continuous) linear functional $\bar{w}\in\bar{\mathcal{K}}^{\dag}$ (the dual cone of $\bar{\mathcal{K}}$) with (total variation) norm $\|\bar{w}\|_{1}>0$ such that
\begin{equation*}
  \langle\bar{w},V(c^{\ast},X^{\ast}_{T})+k\rangle_{\bar{\mathcal{I}}}>\langle\bar{w},V(c,X_{T})\rangle_{\bar{\mathcal{I}}},
\end{equation*}
for all $(c,X_{T})\in\mathfrak{B}(X_{0})$ and $k\in\Int\bar{\mathcal{K}}$. Together with the correspondences (\ref{A.5}) and (\ref{A.6}), this signifies the maximality of $(c^{\ast},X^{\ast}_{T}|w)$ for (\ref{3.1.1}), under the condition that there exists some $\tau\in[0,T]$ such that $\|w(\tau)\|_{1}>0$ $\PP$-a.s., or equivalently, $\sup_{t\in[0,T]}\|w(t)\|_{1}>0$, hence taking care of assertion (i).

For assertion (ii), by virtue of contradiction suppose that $(c^{\ast},X^{\ast}_{T})$ is not weakly $\bar{\mathcal{K}}$-maximal for (\ref{2.3.5}). Then there exist $(\hat{c},\hat{X}_{T})\in\mathfrak{B}(X_{0})$ and $\hat{k}\in\Int\bar{\mathcal{K}}$ such that $V(\hat{c},\hat{X}_{T})=V(c^{\ast},X^{\ast}_{T})+\hat{k}$. Therefore, for $\bar{w}$ defined as above, since $\bar{w}\in\bar{\mathcal{K}}^{\dag}$ and $\bar{w}\neq\0$, we have $\langle\bar{w},V(\hat{c},\hat{X}_{T})\rangle_{\bar{\mathcal{I}}}>\langle\bar{w},V(c^{\ast},X^{\ast}_{T})\rangle_{\bar{\mathcal{I}}}$, which by (\ref{A.6}) is in contradiction to the maximality of $(c^{\ast},X^{\ast}_{T})$ for (\ref{3.1.1}).
\end{proof}

\medskip

\noindent\textbf{Proposition \ref{pro:4}}
\begin{proof}
According (\ref{3.1.2}), since $w$ is a linear functional, the single-criterion problem (\ref{3.1.1}) stays unchanged up to scaling $w$ by a positive constant. The required result hence follows straight from Theorem \ref{thm:2}.
\end{proof}

\medskip

\noindent\textbf{Theorem \ref{thm:3}}
\begin{proof}
Fixing a $t\in[0,T]$, by the fact that $u(t,\cdot)\in\mathfrak{U}_{\mathcal{I}_{t}}$ we have that $u(t,\cdot)$ is $\mathcal{C}_{\rm b}(\mathcal{I}_{t};\mathds{R})$-concave, and so is $\tilde{u}(t,\cdot)$ by definition. Also, $\mathcal{K}_{t}\neq\mathcal{C}_{\rm b}(\mathcal{I}_{t};\mathds{R})$ by pointedness, so that $\tilde{u}(t,\cdot)$ is a proper multifunction. Hence, we can apply \cite[Hamel et al., 2015, \text{Thm.} 5.8]{HHLRS} to conclude that $\tilde{u}^{\ddag}(t,\cdot)=\cl_{\mathcal{C}_{\rm b}}\tilde{u}(t,\cdot)$, which leads to the first assertion because $u(t,\mathds{R}^{n}_{+})$ is closed in $\mathcal{C}_{\rm b}(\mathcal{I}_{t};\mathds{R})$ given the continuity of every utility element $u_{i}(t,\cdot)$, $i\in\mathcal{I}_{t}$, assumed for (\ref{2.2.5}). The second assertion follows immediately from Definition \ref{def:5}.
\end{proof}

\medskip

\noindent\textbf{Proposition \ref{pro:5}}
\begin{proof}
By the pointedness of $\bar{\mathcal{K}}$, we have $\bar{\mathcal{K}}\neq\mathcal{C}_{\rm b}(\bar{\mathcal{I}};\mathds{R})$, so that $\tilde{V}$ must be a proper multifunction. Since $\tilde{V}$ takes values in the complete lattice $\mathcal{L}(\mathcal{C}_{\rm b}(\bar{\mathcal{I}};\mathds{R}),\bar{\mathcal{K}})$, by \cite[Hamel and L\"{o}hne, 2014, \text{Thm.} 6.1]{HL} we have that strong duality holds, meaning that solving the dual problem (\ref{4.6}) is equivalent to doing the primal problem (\ref{4.1}).

The budget set $\mathfrak{B}(X_{0})$ gives one single-valued constraint, from which it follows that
\begin{align*}
  \bigcap_{\substack{y\in(\mathfrak{C}_{n}\times\mathds{L}^{1}_{\mathcal{F}}(\Omega;\mathds{R}_{+}))^{\dag},\\ \bar{\lambda}\in\bar{\mathcal{K}}^{\dag}}}\mathcal{H}(y,\bar{\lambda}) &=\bigcap_{\eta\geq0}\sup_{(c,X_{T})\in\mathfrak{C}_{n}\times\mathds{L}^{1}_{\mathcal{F}}(\Omega;\mathds{R}_{+})}L((c,X_{T}),\eta)\\
  &=\inf_{\eta\geq0}\sup_{(c,X_{T})\in\mathfrak{C}_{n}\times\mathds{L}^{1}_{\mathcal{F}}(\Omega;\mathds{R}_{+})}L((c,X_{T}),\eta),
\end{align*}
where
\begin{equation*}
  L((c,X_{T}),\eta):=\tilde{V}(c,X_{T})+\eta\bigg(X_{0}-\E\bigg[\int^{T}_{0}\xi_{t}C_{t}\dd t+\xi_{T}X_{T}\bigg]\bigg),\quad\eta\geq0
\end{equation*}
is a reduced Lagrangian function of (\ref{4.5}), also valued in $\mathcal{L}(\mathcal{C}_{\rm b}(\bar{\mathcal{I}};\mathds{R}),\bar{\mathcal{K}})$. It remains to show that transforming the static ordering cone $\bar{\mathcal{K}}$ into the dynamic counterpart $\mathcal{K}$ makes no difference. Indeed, for every $t\in[0,T]$, if $z\in\tilde{u}(t,c)\equiv u(t,c)-\mathcal{K}_{t}$, then by construction $z\in u(t,c)-\bar{\mathcal{K}}$ as well, together with the spatial extension (\ref{A.1}). Conversely, suppose that $z\in u(t,c)-\bar{\mathcal{K}}$, and then after adopting the spatial extension that
\begin{equation*}
  \tilde{u}(t,c)=u(t,c)-\mathcal{K}_{t}=\{\{u_{i}(t,c)\mathds{1}_{\mathcal{I}_{t}}(i)-k\mathds{1}_{\mathcal{K}_{t}}(k):i\in\bar{\mathcal{I}}\}: k\in\bar{\mathcal{K}}\},\quad t\in[0,T]
\end{equation*}
we see that $u(t,c)-\bar{\mathcal{K}}\setminus\mathcal{K}_{t}=\tilde{u}(t,c)$, so that $z\in\tilde{u}(t,c)$ still. Therefore, we can apply Theorem \ref{thm:3} to write
\begin{align*}
  L((c,X_{T}),\eta)&=\eta X_{0}+\E\bigg[\int^{T}_{0}(\tilde{u}(t,c_{t})-\eta\xi_{t}C_{t})\dd t+(U(X_{T})-\eta\xi_{T}X_{T})\bigg]\\
  &\subseteq\eta X_{0}+\E\bigg[\int^{T}_{0}(-\tilde{u})^{\dag}(t,-\eta\xi_{t}\1,\0)\dd t+(-U)^{\dag}(-\eta\xi_{T})\bigg],
\end{align*}
from which the required result follows.
\end{proof}

\medskip

\noindent\textbf{Theorem \ref{thm:4}}
\begin{proof}
We begin by proving the existence of required Malliavin derivatives, whose operator is denoted as $\mathcal{D}$. In particular, we write by $\mathbb{D}^{1,2}\equiv\mathbb{D}^{1,2}(W,[0,T])$ the collection of all Malliavin-differentiable $\mathcal{F}$-measurable functionals in $\mathds{L}^{2}_{\mathcal{F}}(\Omega)$.\footnote{For a concrete introduction to Malliavin calculus we refer readers to \cite[Nualart, 1995]{N2}, or to \cite[Lyasoff, 2017, \text{Sect.} 14.1]{L2} for a r\'{e}sum\'{e}, which we suppress here.}

By \cite[Nualart, 1995, \text{Lemma} 2.2.2]{N2} we have immediately $\xi_{t}\in\mathbb{D}^{1,2}$ for every $t\in[0,T]$. However, we need to prove the Malliavin differentiability of the index set $\mathcal{I}$. Since $\mathbb{D}^{1,2}$ is dense in $\mathds{L}^{2}_{\mathcal{F}}(\Omega;\mathds{R}^{d})$, for each $q\in\{1,2,3\}$ there exists a sequence $\{Y_{q,k}:k\in\mathds{N}_{++}\}$ of ($d$-dimensional) Malliavin-differentiable It\^{o} processes such that $I_{q,t}=\cl_{\mathds{R}^{d}}\{Y_{q,k}:k\in\mathds{N}_{++}\}$, $\PP$-a.s., for every $t\in[0,T]$.

Then, for any generic $t\in[0,T]$, consider a time partition $P_{K}:=\{t_{l}:l\in\mathds{N}\cap[0,K]\}$ of $[0,t]$, with $\{t_{K}:K\in\mathds{N}_{++}\}\equiv[0,t]\cap\mathds{Q}$. For an arbitrary unit normal $\nu\in\mathds{R}^{d}$, for $q=3$ we define the extremal variables $\bar{Y}_{3,k,\nu}$ and $\bar{Y}_{3,k,K,\nu}$ as the solutions to the scalarized maximization problems\footnote{Maximization can also be understood as being vector-valued in the sense of Definition \ref{def:4} where the ordering cone is exactly defined by $\nu$.} $\sup_{s\in[0,t]}\langle Y_{3,k,s},\nu\rangle_{d}$ and $\sup_{l\in[1,K]}\langle Y_{3,k,t_{l}},\nu\rangle_{d}$, respectively, for any $K\in\mathds{N}_{++}$. It is easy to see that the map from $\{Y_{3,k,t_{l}}:l\in\mathds{N}\cap[0,K]\}\in\mathds{R}^{d(K+1)}$ to the solution of $\sup_{s\in[0,t]}\langle Y_{3,k,s},\nu\rangle_{d}$ and $\sup_{l\in[0,K]}\langle Y_{3,k,t_{l}},\nu\rangle_{d}$ is Lipschitz-continuous in $\mathds{R}^{d}$ for any $\nu$ with maximization taken over a finite set. Thus, according to \cite[Nualart, 1995, \text{Prop.} 1.2.3]{N2} $\bar{Y}_{3,k,K,\nu}$'s are all elements of $\mathbb{D}^{1,2}$. Also, we have $\bar{Y}_{3,k,K,\nu}\rightarrow\bar{Y}_{3,k,\nu}$ in $\mathds{L}^{2}_{\mathbb{F}}(\Omega,\mathds{R}^{d})$ for every $k$. Since each $Y_{3,k}$ is an It\^{o} process, by (\ref{2.2.4}) we also have
\begin{equation*}
  \E\big[\|\mathcal{D}_{t_{K}}\bar{Y}_{3,k,K,\nu}\|^{2}_{\mathrm{F}}\big]\leq\E\Big[\sup_{s\in[0,t]}\|\mathcal{D}_{s}Y_{3,k,s}\|^{2}_{\mathrm{F}}\Big] \leq\sup_{k\in\mathds{N}_{++}}\E\Big[\sup_{s\in[0,t]}\|g_{3,k,s}\|^{2}_{\mathrm{F}}\Big]<\infty.
\end{equation*}
Recall that the coefficient processes $g_{3,k}$'s are by assumption continuous. Then, using \cite[Nualart, 1995, \text{Lemma} 1.2.3]{N2} it follows that $\bar{Y}_{3,k,\nu}\in\mathbb{D}^{1,2}$. This holds for every $k\in\mathds{N}_{++}$. Similarly, for $q=2$ we can work with the extremal variables $\bar{Y}_{2,k,\nu}$ and $\bar{Y}_{2,k,K,\nu}$ as the solutions to $\sup_{s\in[0,t]}\langle Y_{2,k,s},-\nu\rangle_{d}$ and $\sup_{l\in[0,K]}\langle Y_{2,k,t_{l}},-\nu\rangle_{d}$, respectively, for any $K\in\mathds{N}_{++}$ to build up to $\bar{Y}_{2,k,\nu}\in\mathbb{D}^{1,2}$ for every $k\in\mathds{N}_{++}$ as well.

Since $\mathcal{I}$ is compact-valued, it suffices to consider its boundary (taken in $\mathds{R}^{d}$). Indeed, by (\ref{2.2.3}) and Proposition \ref{pro:1}, we know that $I_{q,t}$'s and $\mathcal{I}_{t}$ all admit $\PP$-\text{a.s.} boundaries for every $t\in[0,T]$, as well as that the processes $\bigcap_{s\in[0,\imath]}I_{2,s}$ and $\bigcup_{s\in[0,\imath]}I_{3,s}$ are respectively set-decreasing and set-increasing and $\mathbb{F}$-non-anticipating. Hence, there exist collections $\{\nu_{h_{q}(k)}:k\in\mathds{N}_{++}\}$ of $\mathds{R}^{d}$-unit normals for some $h_{q}:\mathds{N}_{++}\mapsto\mathds{N}_{++}$, $q\in\{1,2,3\}$, such that for every $t\in[0,T]$
\begin{equation*}
  \pd\Bigg(I_{1,t}+\bigcap_{s\in[0,t\wedge\mathcal{t}]}I_{2,s}+\bigcup_{s\in[0,t]}I_{3,s}\Bigg)=\cl_{\mathds{R}^{d}}\big\{Y_{1,h_{1}(k),t} +\bar{Y}_{2,k,\nu_{h_{2}(k)},t\wedge\mathcal{t}}+\bar{Y}_{3,k,\nu_{h_{3}(k)},t}:k\in\mathds{N}_{++}\big\},
\end{equation*}
$\PP$-a.s., but with the unrefined index set (recall Figure \ref{fig:1}) $\breve{\mathcal{I}}_{t}=\overline{\co}_{\mathds{R}^{d}}\big(I_{1,t}+\bigcap_{s\in[0,t\wedge\mathcal{t}]}I_{2,s}+\bigcup_{s\in[0,t]}I_{3,s}\big)$ we have the constructional decomposition $\pd\breve{\mathcal{I}}_{t}=B_{1,t}\cup B_{2,t}$
where $B_{1,t}\subseteq\pd\big(I_{1,t}+\bigcap_{s\in[0,t\wedge\mathcal{t}]}I_{2,s}+\bigcup_{s\in[0,t]}I_{3,s}\big)$ and $B_{2,t}\cap\big(I_{1,t}+\bigcap_{s\in[0,t\wedge\mathcal{t}]}I_{2,s}+\bigcup_{s\in[0,t]}I_{3,s}\big)=\emptyset$, $\PP$-a.s., for every $t\in[0,T]$. Then there is a subsequence $\{k_{l}\}\subseteq\mathds{N}_{++}$ such that
\begin{equation*}
  B_{1,t}=\cl_{\mathds{R}^{d}}\big\{Y_{1,h_{1}(k_{l}),t}+\bar{Y}_{2,k_{l},\nu_{h_{2}(k_{l})},t\wedge\mathcal{t}}+\bar{Y}_{3,k_{l}, \nu_{h_{3}(k_{l})},t}\big\},\quad\PP\text{-a.s.},
\end{equation*}
and we have $B_{1,t}\in\mathbb{D}^{1,2}$, from which $B_{2,t}\in\mathbb{D}^{1,2}$ follows by Carath\'{e}odory's theorem. Putting these together, we have that $\pd\breve{\mathcal{I}}_{t}\subseteq\mathbb{D}^{1,2}$, $\forall t\in[0,T]$, i.e, it is Malliavin-differentiable. For every $t\in[0,T]$, since $\mathcal{R}$ is by definition closed, convex, and deterministic, the (refined) index set $\mathcal{I}_{t}=\mathcal{R}\cap\breve{\mathcal{I}}_{t}$ is clearly compact and convex (upon non-emptiness). Therefore, its boundary $\pd\mathcal{I}_{t}$ exists and is a subset of $\mathbb{D}^{1,2}$ as well, and there exists a velocity vector field $\mathbf{v}(i,W_{t})$ of $\pd\mathcal{I}_{t}\ni i$ on $\mathcal{C}_{0}([0,T];\mathds{R}^{m})$.

Now we turn to computing the Malliavin derivative of the variable $\phi_{t}$ in the martingale representation (\ref{A.3}), or in more detail,\footnote{As a reminder, in this proof the optimal (aggregated) consumption--bequest policy $(C^{\ast},X^{\ast}_{T})$ is understood to be the single-valued ones, i.e., those parameterized by a given $w$.}
\begin{equation*}
  \phi^{\ast}_{t}=\E\bigg[\mathcal{D}_{t}\bigg(\int^{T}_{0}\xi_{s}C^{\ast}_{s}\dd s+\xi_{T}X^{\ast}_{T}\bigg)\bigg|\mathscr{F}_{t}\bigg],\quad t\in[0,T].
\end{equation*}
First, from the identity (\ref{A.4}) the last equation gives
\begin{align}\label{A.7}
  \Pi^{\ast}_{t}&=\xi^{-1}_{t}\E\bigg[\int^{T}_{t}\xi_{s}C^{\ast}_{s}\dd s+\xi_{T}X^{\ast}_{T}\bigg|\mathscr{F}_{t}\bigg](\sigma^{\intercal}_{t})^{-1}\theta_{t} \nonumber\\
  &\qquad+\xi^{-1}_{t}(\sigma^{\intercal}_{t})^{-1}\E\bigg[\mathcal{D}_{t}\int^{T}_{0}\xi_{s}C^{\ast}_{s}\dd s+\mathcal{D}_{t}\bigg(\xi_{T}(U')^{-1}\bigg(\frac{\eta\xi_{T}T}{\int^{T}_{0}\langle w(s),\1\rangle_{\mathcal{I}_{s}}\dd s}\bigg)\bigg)\bigg|\mathscr{F}_{t}\bigg]^{\intercal} \nonumber\\
  &=:\xi^{-1}_{t}\E\bigg[\int^{T}_{t}\xi_{s}C^{\ast}_{s}\dd s+\xi_{T}X^{\ast}_{T}\bigg|\mathscr{F}_{t}\bigg](\sigma^{\intercal}_{t})^{-1}\theta_{t} \nonumber\\
  &\qquad+\xi^{-1}_{t}(\sigma^{\intercal}_{t})^{-1}\E\bigg[\int^{T}_{t}A(t,s)\dd s+B(t,T)\bigg|\mathscr{F}_{t}\bigg]^{\intercal},\quad t\in[0,T].
\end{align}
With $s\geq t$ fixed, the derivative of the state price density follows as
\begin{equation}\label{A.8}
  \mathcal{D}_{t}\xi_{s}=-\xi_{s}\bigg(\int^{s}_{t}\mathcal{D}_{t}r_{v}\dd v+\int^{s}_{t}(\dd W^{\intercal}_{v}+\theta^{\intercal}_{v}\dd v)\mathcal{D}_{t}\theta_{v}+\theta^{\intercal}_{t}\bigg)=:-\xi_{s}(H^{\intercal}_{\xi,t,s}+\theta^{\intercal}_{t}),
\end{equation}
which exists provided that $r_{v},\theta_{v}\in\mathbb{D}^{1,2}$ for $v\in[t,s]$, and then we have
\begin{equation}\label{A.9}
  A(t,s)=(\mathcal{D}_{t}\xi_{s})C^{\ast}_{s}+\xi_{s}(\mathcal{D}_{t}C^{\ast}_{s}) =-\xi_{s}(H^{\intercal}_{\xi,t,s}+\theta^{\intercal}_{t})C^{\ast}_{s}+\xi_{s}\mathcal{D}_{t}C^{\ast}_{s}
\end{equation}
and
\begin{align}\label{A.10}
  B(t,T)&=(\mathcal{D}_{t}\xi_{T})X^{\ast}_{T}+\xi_{T}(\mathcal{D}_{t}X^{\ast}_{T}) \nonumber\\
  &=-\xi_{T}(H^{\intercal}_{\xi,t,T}+\theta^{\intercal}_{t})X^{\ast}_{T}-\xi_{T}((U')^{-1})'\bigg(\frac{\eta\xi_{T}T}{\int^{T}_{0}\langle w(s),\1\rangle_{\mathcal{I}_{s}}\dd s}\bigg) \nonumber\\
  &\qquad\times\bigg(\frac{\eta\xi_{T}T(H^{\intercal}_{\xi,t,T}+\theta^{\intercal}_{t})}{\int^{T}_{0}\langle w(s),\1\rangle_{\mathcal{I}_{s}}\dd s}+\eta\xi_{T}T\varrho_{t,T}\bigg).
\end{align}
To gain some insight into the Malliavin derivative of $C^{\ast}_{s}$, we apply the operator $\mathcal{D}_{t}$ on both sides of (\ref{3.2.1}) evaluated at time $s\geq t$ to obtain
\begin{equation*}
  \eta\mathcal{D}_{t}(\xi_{s}P_{j,s})=\mathcal{D}_{t}\int_{\mathcal{I}_{s}}w_{i}(s)u^{(j)}(s,c^{\ast}_{s})\dd i,
\end{equation*}
which is only nontrivial for $\|w(s)\|_{1}>0$. If $\mathcal{I}_{s}$ is of full dimensionality, since $\pd\mathcal{I}_{s}\subseteq\mathbb{D}^{1,2}$ exists, we employ the Leibniz--Reynolds transport theorem (see, e.g., \cite[Flanders, 1973]{F1}) to write from the last equation
\begin{align}\label{A.11}
  \mathcal{D}_{t}\int_{\mathcal{I}_{s}}w_{i}(s)u^{(j)}(s,c^{\ast}_{s})\dd i&=\int_{\mathcal{I}_{s}}w_{i}(s)\mathcal{D}_{t}(u^{(j)}_{i}(s,c^{\ast}_{s}))\dd i \nonumber\\
  &\qquad+\int_{\pd\mathcal{I}_{s}}\mathbf{v}(i,W_{s})\lrcorner(w_{i}(s)u^{(j)}(s,c^{\ast}_{s})\dd i),
\end{align}
where the interior product $\lrcorner$ is understood to act on the differential $d$-form $w_{i}(s)u^{(j)}_{i}(s,c^{\ast}_{s})\dd i$. Then, we observe that
\begin{equation*}
  \mathcal{D}_{t}\big(u^{(j)}_{i}(s,c^{\ast}_{s})\big)=\sum^{n}_{j'=1}(u^{(j)})^{(j')}(s,c^{\ast}_{s})(\mathcal{D}_{t}c^{\ast}_{j',s}).
\end{equation*}
This leads with (\ref{A.11}) to the following system:
\begin{align*}
  &\sum^{n}_{j'=1}(\mathcal{D}_{t}c^{\ast}_{j',s})\big\langle w(s),(u^{(j)})^{(j')}(s,c^{\ast}_{s})\big\rangle_{\mathcal{I}_{s}}+\eta\xi_{s}P_{j,s}(H^{\intercal}_{\xi,t,s}+\theta^{\intercal}_{t}) -\eta\xi_{s}\mathcal{D}_{t}P_{j,s} \\
  &\qquad+\int_{\pd\mathcal{I}_{s}}\mathbf{v}(i,W_{s})\lrcorner(w_{i}(s)u^{(j)}(s,c^{\ast}_{s})\dd i)=0,\quad j\in\mathds{N}\cap[1,n],
\end{align*}
from which we define the Hessian matrix $\varTheta_{s}:=\big(\big\langle w(s),(u^{(j)})^{(j')}(s,c^{\ast}_{s})\big\rangle_{\mathcal{I}_{s}}\big)_{j,j'}\in\mathds{R}^{n\otimes n}$. For $\|w(s)\|_{1}>0$, by restricted redundancy (Assumption \ref{as:1}) and the Inada conditions (Assumption \ref{as:3}), $\varTheta_{s}$ is invertible, and we may hence write more concisely
\begin{equation}\label{A.12}
  \mathcal{D}_{t}c^{\ast}_{s}=-\varTheta^{-1}_{s}\eta\xi_{s}\big(P_{s}(H^{\intercal}_{\xi,t,s}+\theta^{\intercal}_{t})-\mathcal{D}_{t}P_{s}\big) -\varTheta^{-1}_{s}\bigg(\int_{\pd\mathcal{I}_{s}}\mathbf{v}(i,W_{s})\lrcorner\big(w_{i}(s)u^{(j)}_{i}(s,c^{\ast}_{s})\dd i\big)\bigg)_{j},
\end{equation}
which is an $(n\otimes m)$-matrix, but $c^{\ast}_{s}=\psi_{\mathcal{I}_{s}}(\eta\xi_{s}P_{s}|w_{i}(s)))$, and with the commodity price vector $P$ we have
\begin{equation*}
  \mathcal{D}_{t}C^{\ast}_{s}=\mathcal{D}_{t}\langle P_{s},c^{\ast}_{s}\rangle_{n}=P^{\intercal}_{s}\mathcal{D}_{t}c^{\ast}_{s}+c^{\ast\intercal}_{s}\mathcal{D}_{t}P_{s}.
\end{equation*}
From routine calculations, the SDE (\ref{2.1.2}) has the strong solution (with time variable $s$)
\begin{equation*}
  P_{s}=\mathrm{diag}(P_{0})\exp\bigg(\int^{s}_{0}\bigg(\mu_{P,v}-\frac{1}{2}\big(\|\sigma_{P,j,v}\|^{2}_{2}\big)_{j\in\mathds{N}\cap[1,n]}\bigg)\dd v+\int^{s}_{0}\sigma_{P,v}\dd W_{v}\bigg),
\end{equation*}
where the exponential is understood as a componentwise operation, and then it follows that
\begin{align}\label{A.13}
  \mathcal{D}_{t}P_{s}&=\mathrm{diag}(P_{s})\bigg(\int^{s}_{t}\mathcal{D}_{t}\mu_{P,v}\dd v+\int^{s}_{t}\big((\dd W^{\intercal}_{v}-\sigma^{\intercal}_{P,j,v}\dd v)\mathcal{D}_{t}\sigma_{P,j,v}\big)_{j\in\mathds{N}\cap[1,n]}\dd v+\sigma_{P,t}\bigg) \nonumber\\
  &=:\mathrm{diag}(P_{s})\big(H^{\intercal}_{P,t,s}+\sigma_{P,t}\big),
\end{align}
provided $\mu_{P,v},\sigma_{P,v}\in\mathbb{D}^{1,2}$ for $v\in[t,s]$ as well, which is also $(n\otimes m)$-matrix-valued. Thus, the $A(t,s)$ term is established after simplifications. In the same vein, for $B(t,T)$ we have
\begin{equation}\label{A.14}
  \varrho_{t,T}=-\mathcal{D}_{t}\bigg(\int^{T}_{0}\langle w(s),\1\rangle_{\mathcal{I}_{s}}\dd s\bigg)^{-1}=\frac{\mathcal{D}_{t}\int^{T}_{0}\langle w(s),\1\rangle_{\mathcal{I}_{s}}\dd s}{\big(\int^{T}_{0}\langle w(s),\1\rangle_{\mathcal{I}_{s}}\dd s\big)^{2}}.
\end{equation}
For the Malliavin derivative of the integral $\int^{T}_{0}\langle w(s),\1\rangle_{\mathcal{I}_{s}}\dd s$, since $w$ is over $\mathcal{I}$ being integrably bounded (Proposition \ref{pro:1}), Fubini's theorem is applicable for us to write
\begin{equation*}
  \mathcal{D}_{t}\int^{T}_{0}\langle w(s),\1\rangle_{\mathcal{I}_{s}}\dd s=\int^{T}_{t}\mathcal{D}_{t}\int_{\mathcal{I}_{s}}w_{i}(s)\dd i\dd s=\int^{T}_{t}\int_{\pd\mathcal{I}_{s}}\mathbf{v}(i,W_{s})\lrcorner(w_{i}(s)\dd i)\dd s,
\end{equation*}
where the second equality follows by the Leibniz--Reynolds transport theorem again. Combining (\ref{A.7}), (\ref{A.8}), (\ref{A.9}), (\ref{A.10}), (\ref{A.12}) and (\ref{A.14}) and rearranging terms, we obtain the optimal investment policy parameterized by a given $w$.

To complete the proof it remains to notice that the conditional expectation of a closed set-valued random variable is defined as the closure of the conditional expectations of every measurable selector of such a random variable (see, e.g., \cite[Kisielewicz, 2020, \text{Corol.} 3.4.1]{K3}). In our context, this closed set-valued random variable reads for any fixed $t\in[0,T]$
\begin{equation*}
  \xi^{-1}_{t}\bigg(\int^{T}_{t}\xi_{s}C^{\ast}_{s}\dd s+\xi_{T}X^{\ast}_{T}\bigg)(\sigma^{\intercal}_{t})^{-1}\theta_{t}+\xi^{-1}_{t}\bigg(\int^{T}_{t}A(t,s)\dd s+B(t,T)\bigg)
\end{equation*}
according to (\ref{A.7}), whose selectors are precisely those parameterized by $w$.
\end{proof}

\medskip

\noindent\textbf{Corollary \ref{cor:1}}
\begin{proof}
Given (\ref{5.1.4}), we have immediately from (\ref{3.2.1})
\begin{equation*}
  C^{\ast}=\bigg\langle P,(\breve{u}')^{-1}\bigg(\frac{\eta\xi P}{\langle w,\alpha\rangle_{\mathcal{I}}}\bigg)\bigg\rangle_{n} =\sum^{n}_{j=1}P_{j}(\breve{u}'_{j})^{-1}\bigg(\frac{\eta\xi P_{j}}{\int_{\mathcal{I}}w_{i}\alpha_{ij}\dd i}\bigg),
\end{equation*}
where in the first equality $(\breve{u}')^{-1}$ is understood as a componentwise operation. Thus, for any $t\leq s$ within $[0,T]$,
\begin{align*}
  \mathcal{D}_{t}C^{\ast}_{s}&=\sum^{n}_{j=1}\bigg((\breve{u}'_{j})^{-1}\bigg(\frac{\eta\xi_{s}P_{j,s}}{\int_{\mathcal{I}}w_{i}(s)\alpha_{ij}(s)\dd i}\bigg)\mathcal{D}_{t}P_{j,s}+P_{j,s}\big((\breve{u}'_{j})^{-1}\big)'\bigg(\frac{\eta\xi_{s}P_{j,s}}{\int_{\mathcal{I}_{s}}w_{i}(s)\alpha_{ij}(s) \dd i}\bigg)\\
  &\qquad\times\bigg(\frac{\eta\mathcal{D}_{t}(\xi_{s}P_{j,s})}{\int_{\mathcal{I}_{s}}w_{i}(s)\alpha_{ij}(s)\dd i}-\frac{\eta\xi_{s}P_{j,s}\mathcal{D}_{t}\int_{\mathcal{I}_{s}}w_{i}(s)\alpha_{ij}(s)\dd i}{\big(\int_{\mathcal{I}_{s}}w_{i}(s)\alpha_{ij}(s)\dd i\big)^{2}}\bigg)\bigg),
\end{align*}
where $\mathcal{D}_{t}P_{j,s}$ has been derived in (\ref{A.13}). Also, since $d=1$, we have with the notations in the proof of Theorem \ref{thm:4} that the unit normal $\nu$ reduces to the points $\pm1$; then, for every $t\in[0,T]$,
\begin{align*}
  \breve{\mathcal{I}}_{t}&=\Big[\inf_{k\in\mathds{N}_{++}}(Y_{1,k,t}+\bar{Y}_{2,k,-1,t}+\bar{Y}_{3,k,-1,t}), \sup_{k\in\mathds{N}_{++}}(Y_{1,k,t}+\bar{Y}_{2,k,1,t}+\bar{Y}_{3,k,1,t})\Big]\\
  &=\Big[\inf_{k\in\mathds{N}_{++}}\Big(Y_{1,k,t}+\sup_{s\in[0,t]}Y_{2,k,s}+\inf_{s\in[0,t]}Y_{3,k,s}\Big), \sup_{k\in\mathds{N}_{++}}\Big(Y_{1,k,t}+\inf_{s\in[0,t]}Y_{2,k,s}+\sup_{s\in[0,t]}Y_{3,k,s}\Big)\Big]\\
  &=:[\breve{Y}_{-},\breve{Y}_{+}],\quad\PP\text{-a.s.},
\end{align*}
where we recall that $\{Y_{q,k,t}:k\in\mathds{N}_{++}\}\subseteq\mathbb{D}^{1,2}\cap I_{q,t}$ for any $q\in\{1,2,3\}$ and $\breve{Y}_{\pm}\in\mathbb{D}^{1,2}$. Then, using the Leibniz--Reynolds transport theorem again, we are led to the same equation as in (\ref{A.11}), which reduces in one dimension to
\begin{align*}
  \mathcal{D}_{t}\int_{\mathcal{R}\cap[\breve{Y}_{-,s},\breve{Y}_{+,s}]}w_{i}(s)\alpha_{ij}(s)\dd i&=w_{\breve{Y}_{+,s}}(s)\alpha_{\breve{Y}_{+,s}\;j}(s)\mathcal{D}_{t}\breve{Y}_{+,s}\mathds{1}_{\mathcal{R}}(\breve{Y}_{+,s})\\
  &\qquad-w_{\breve{Y}_{-,s}}(s)\alpha_{\breve{Y}_{-,s}\;j}(s)\mathcal{D}_{t}\breve{Y}_{-,s}\mathds{1}_{\mathcal{R}}(\breve{Y}_{-,s})
\end{align*}
and is nothing but the classical Leibniz rule applied to an integral over a stochastic domain. After simplifying terms we arrive at the desired formula.
\end{proof}

\medskip

\noindent\textbf{Proposition \ref{pro:6}}
\begin{proof}
By the properties of the Hausdorff distance we first observe that
\begin{align*}
  D_{3}&:=\mathcal{d}_{\rm H}\Bigg(\overline{\co}_{\mathds{R}^{d}}\bigcup_{s\in[0,t]}I_{3,s},\co_{\mathds{R}^{d}}\bigcup^{K-1}_{l=0}\hat{I}^{(K)}_{3,t_{l+1|K}}\Bigg)\\
  &\leq\mathcal{d}_{\rm H}\Bigg(\cl_{\mathds{R}^{d}}\bigcup_{s\in[0,t]}I_{3,s},\bigcup_{s\in[0,t]}I_{3,s}\Bigg)\\
  &\leq\mathcal{d}_{\rm H}\Bigg(\bigcup^{K-1}_{l=0}I_{3,t_{l+1|K}}\cup\cl_{\mathds{R}^{d}}\bigcup_{s\in[0,t]\setminus\{t_{l+1|K}:l\in\mathds{N}\cap[0,K-1]\}}I_{3,s}, \bigcup^{K-1}_{l=0}\hat{I}^{(K)}_{3,t_{l+1|K}}\Bigg)\\
  &\leq\mathcal{d}_{\rm H}\Bigg(\bigcup^{K-1}_{l=0}I_{3,t_{l+1|K}},\bigcup^{K-1}_{l=0}\hat{I}^{(K)}_{3,t_{l+1|K}}\Bigg)\\
  &\qquad+\mathcal{d}_{\rm H}\Bigg(\bigcup^{K-1}_{l=0}I_{3,t_{l+1|K}},\cl_{\mathds{R}^{d}}\bigcup_{s\in[0,t]\setminus\{t_{l+1|K}:l\in\mathds{N}\cap[0,K-1]\}}I_{3,s}\Bigg)\\
  &\leq\sum^{K-1}_{l=0}\mathcal{d}_{\rm H}\big(I_{3,t_{l+1|K}},\hat{I}^{(K)}_{3,t_{l+1|K}}\big)+\mathcal{d}_{\rm H}\Bigg(\bigcup^{K-1}_{l=0}I_{3,t_{l+1|K}},\cl_{\mathds{R}^{d}}\bigcup_{s\in[0,t]\setminus\{t_{l+1|K}:l\in\mathds{N}\cap[0,K-1]\}}I_{3,s}\Bigg).
\end{align*}
Using the $\PP$-\text{a.s.} continuity of the set-valued process $\bigcup_{s\in[0,\imath]}I_{3,s}$ (Proposition \ref{pro:1}) the second Hausdorff distance in the last inequality vanishes in the limit as $K\rightarrow\infty$, $\PP$-a.s. With similar arguments we have
\begin{align*}
  D_{2}&:=\mathcal{d}_{\rm H}\Bigg(\bigcap_{s\in[0,t\wedge\mathcal{t}]}I_{2,s},\bigcap^{(K-1)\wedge\hat{\mathcal{l}}}_{l=0}I_{2,t_{l+1|K}}\Bigg)\\ &\leq\mathcal{d}_{\rm H}\big(I_{2,0},\hat{I}^{(K)}_{2,0}\big)+\mathcal{d}_{\rm H}\Bigg(\bigcap^{(K-1)\wedge\hat{\mathcal{l}}}_{l=0}I_{2,t_{l+1|K}}, \bigcap_{s\in[0,t\wedge\mathcal{t}]\setminus\{t_{l+1|K}:l\in\mathds{N}\cap[0,K-1]\}}I_{2,s}\Bigg),
\end{align*}
where the second distance in the last inequality also goes to zero as $K\rightarrow\infty$, $\PP$-a.s., because of path continuity and the fact that $\lim_{K\rightarrow\infty}\mathrm{Leb}_{\mathds{R}}([ t_{\hat{\mathcal{l}}|K},\mathcal{t}])\rightarrow0$, $\PP$-a.s.

Putting things together and consulting (\ref{2.2.1}) and (\ref{6.2}) we have
\begin{align*}
  \E\big[\mathcal{d}_{\rm H}\big(\mathcal{I}_{t},\hat{\mathcal{I}}^{(K)}_{t_{K|K}}\big)\big]&\leq\E\big[\mathcal{d}_{\rm H}\big(I_{1,t},\hat{I}^{(K)}_{1,t_{K|K}}\big)+D_{2}+D_{3}\big]\\
  &\rightarrow\lim_{K\rightarrow\infty}\Bigg(\E\big[\mathcal{d}_{\rm H}\big(I_{1,t},\hat{I}^{(K)}_{t_{K|K}}\big)\big]+\E\big[\mathcal{d}_{\rm H}\big(I_{2,0},\hat{I}^{(K)}_{2,0}\big)\big]\\
  &\qquad+\sum^{K-1}_{l=0}\E\big[\mathcal{d}_{\rm H}\big(I_{3,t_{l+1|K}},\hat{I}^{(K)}_{3,t_{l+1|K}}\big)\big]\Bigg)\\
  &=0.
\end{align*}
The last equality follows straightway from proven approximation theorems for Aumann stochastic integrals and set-valued It\^{o} integrals; see, respectively, \cite[Kisielewicz, 2020, \text{Thm.} 4.4.2]{K3} and \cite[Kisielewicz, 2020, \text{Thm.} 5.7.4]{K3}.
\end{proof}

\vspace{0.2in}

\section{A brief review of set-valued stochastic processes}\label{B}

\renewcommand{\theequation}{B.\arabic{equation}}
\newcommand{\dec}{\mathrm{dec}}

We consider the complete filtered probability space $(\Omega,\mathcal{F},\PP;\mathbb{F}\equiv\{\mathscr{F}_{t}\}_{t\in[0,T]})$ with $\mathbb{F}$ satisfying the usual conditions and a Euclidean space $E$ for the value space. By
\begin{equation*}
  \Sigma_{\mathbb{F}}:=\{A\in\mathcal{B}([0,T])\otimes\mathcal{F}:\;\{\omega:(t,\omega)\in A\}\in\mathscr{F}_{t},\forall t\in[0,T]\}
\end{equation*}
we denote the sub-sigma-field of $\mathcal{B}([0,T])\otimes\mathcal{F}$ with respect to which a stochastic process is measurable if and only if it is $\mathbb{F}$-non-anticipating.

We start with some basic notions in the general setting. For a complete finite measure space $(D,\mathcal{A},\mathbb{M})$, $\mathds{L}^{p}_{\mathcal{A}}(D;E)\equiv\mathds{L}^{p}(D,\mathcal{A},\mathbb{M};E)$, $p\geq1$, denotes the space of all equivalent (under $\mathbb{M}$-a.e.) classes of $p$-integrable functions $X:D\mapsto E$. $\mathrm{Cl}(E)$ denotes the space of nonempty closed subsets of $E$ and $\mathcal{P}\big(\mathds{L}^{p}_{\mathcal{A}}(D;E)\big)$ denotes the space of all nonempty subsets of $\mathds{L}^{p}_{\mathcal{A}}(D;E)$.

A closed set-valued mapping defined on $(D,\mathcal{A},\mathbb{M})$ is a multifunction $\mathcal{X}:D\mapsto\mathrm{Cl}(E)$ that is $\mathcal{A}$-measurable, which means that the pre-image $\mathcal{X}^{-1}(A):=\{a\in D:\mathcal{X}(a)\cap A\neq\emptyset\}\in\mathcal{A}$, for any open subset $A\subseteq E$. For the closure and convex hull operators $\cl$ and $\co$, respectively, it holds that the closed convex hull operator $\overline{\co}\equiv\cl\co$ for normed spaces, as is the case of $\mathds{L}^{p}_{\mathcal{A}}(D;E)$. If the set-valued mapping $\mathcal{X}$ is measurable, $\cl_{\mathds{L}^{p}}\mathcal{X}$ and $\overline{\co}_{\mathds{L}^{p}}\mathcal{X}$ are both measurable mappings from $D$ to $\mathrm{Cl}(E)$. A selector $X$ of the (closed) set-valued mapping $\mathcal{X}$ is such that $X(a)\in\mathcal{X}(a)$ for all $a\in D$, whose existence follows from the Zermelo axiom of choice. By the Kuratowski--Ryll--Nardzewski theorem, $\mathcal{X}$ admits a measurable selector $X$ (namely a single-valued mapping from $D$ to $E$) such that $X\in\mathcal{X}$ for all $a\in D$.

With $D=\Omega$, $\mathcal{A}=\mathcal{F}$, and $\mathbb{M}=\PP$, the set-valued mapping $\mathcal{X}$ is called a set-valued random variable. With $D=[0,T]\times\Omega$, $\mathcal{A}=\Sigma_{\mathbb{F}}$, and $\mathbb{M}=\mathrm{Leb}_{[0,T]}\times\PP$, it becomes a set-valued $\mathbb{F}$-non-anticipating stochastic process.

The following lemma is a result of \cite[Kisielewicz, 2020b, \text{Thm.} 2.2.3]{K3} and the separability of $E$, which highlights a useful correspondence between set-valued mappings as multifunctions and their output values.

\begin{lemma}\label{lem:1}
For a measurable set-valued mapping $\mathcal{X}:D\mapsto\mathrm{Cl}(E)$, there exists a sequence $\{X_{k}:k\in\mathds{N}_{++}\}\subseteq\mathcal{X}$ of single-valued mappings (referred to as the measurable selectors of $\mathcal{X}$) such that $\mathcal{X}=\cl_{E}\{X_{k}:k\in\mathds{N}_{++}\}$, $\mathbb{M}$-a.e.
\end{lemma}

In particular, if $D=\Omega$, $\mathcal{A}=\mathcal{F}$, and $\mathbb{M}=\PP$, Lemma \ref{lem:1} implies that the value of a set-valued random variable can be well approximated by those of a sequence of single-valued random variables and their $E$-closure. It provides an intuitive way to consider set-valued random variables as being spanned by single-valued ones and has been used in the proof of Proposition \ref{pro:1} and Theorem \ref{thm:4} shown in \ref{A}.

The Hausdorff distance is defined for arbitrary $A,B\in\mathrm{Cl}(E)$ as
\begin{equation*}
  \mathcal{d}_{\rm H}(A,B):=\max\Big\{\sup_{a\in A}\inf_{b\in B}\|a-b\|_{2},\sup_{b\in B}\inf_{a\in A}\|a-b\|_{2}\Big\}.
\end{equation*}
$(\mathrm{Cl}(E),\mathcal{d}_{\rm H})$ is known to be a complete metric space.

Let $\mathcal{G}$ be a sub-sigma-field of $\mathcal{A}$. To construct set-valued integrals, for a set-valued mapping $\mathcal{X}$, the set of $\mathcal{G}$-measurable $p$-integrable selectors, \text{a.k.a.} the set of subtrajectory integrals, of $\mathcal{X}$ is written as $\mathscr{S}^{p}_{\mathcal{G}}(\mathcal{X}):=\{X\in\mathds{L}^{p}_{\mathcal{G}}(D;E):X\in\mathcal{X},\;\mathbb{M}\text{-a.e.}\}$. $\mathcal{X}$ is said to be $p$-integrable if $\mathscr{S}^{p}(\mathcal{X})\equiv\mathscr{S}^{p}_{\mathcal{A}}(\mathcal{X})\neq\emptyset$ (recall $\mathcal{A}\supseteq\mathcal{G}$). It is called $p$-integrably bounded if there exists $Y\in\mathds{L}^{p}_{\mathcal{A}}(D;E)$ such that $\mathcal{d}_{\rm H}(\mathcal{X},\{\0\})\leq Y$, $\mathbb{M}$-a.s., or equivalently, $\E\big[\mathcal{d}^{p}_{\rm H}(\mathcal{X},\{\0\})\big]<\infty$ (see \cite[Malinowski, 2013]{M1}). With $D=\Omega$, $\mathcal{A}=\mathcal{F}$, and $\mathbb{M}=\PP$, it is then understood that the integrable bounded-ness of a set-valued random variable is the same as the integrability of its Hausdorff distance. $\mathds{A}^{p}_{\mathcal{G}}(D;E)\equiv\mathds{A}^{p}(D,\mathcal{G},\mathbb{M};E)$ denotes the family of $\mathcal{G}$-measurable $p$-integrable set-valued mappings from $D$ to $\mathrm{Cl}(E)$.

We proceed to the notion of decomposability of subsets of function spaces. Decomposability is an important property of set-valued mappings signifying that decomposable, namely state-dependent, combinations of known measurable selectors are also valid measurable selectors; indeed, such combinations are by construction guaranteed to belong to the set value for all states (input variables).

For a sub-sigma-field $\mathcal{G}\subseteq\mathcal{A}$, a subset $Z$ of the space $\mathds{L}^{p}_{\mathcal{A}}(D;E)$ is said to be $\mathcal{G}$-decomposable if for any $X,Y\in Z$ and $A\in\mathscr{F}_{t}$ it holds that $\mathds{1}_{A}X+\mathds{1}_{A^{\complement}}Y\in Z$. For any subset $Z$ of $\mathds{L}^{p}_{\mathcal{A}}(D;E)$, $\dec_{\mathcal{G}}Z$ is written for the ($\mathcal{G}$-)decomposable hull of $Z$ and $\overline{\dec}_{\mathcal{G}}Z$ its closure (in $\mathds{L}^{p}$). In other words,
\begin{align}\label{B.1}
  \dec_{\mathcal{G}}Z&=\Bigg\{\sum^{N}_{k=1}X_{k}\mathds{1}_{A_{k}}:\{X_{k}:k\in\mathds{N}\cap[1,N]\}\subseteq Z\text{ and }\{A_{k}:k\in\mathds{N}\cap[1,N]\}\in\mathfrak{P}(D,\mathcal{G}), \nonumber\\
  &\qquad\forall N\in\mathds{N}_{++}\Bigg\},
\end{align}
where $\mathfrak{P}(D,\mathcal{G})$ denotes the collection of all $\mathcal{G}$-measurable partitions of $D$, i.e., $\mathfrak{P}(D,\mathcal{G})$ contains all sequences of nonempty, pairwise disjoint subsets $\{A_{k}:k\in\mathds{N}\cap[1,N]\}\subseteq\mathcal{G}$ such that $\bigcup^{N}_{k=1}A_{k}=D$.

If $D=\Omega$ and $\mathcal{G}=\mathscr{F}_{t}\subseteq\mathcal{F}=\mathcal{A}$ for some $t\in[0,T]$, then immediately $\max\{X_{k}:k\in\mathds{N}\cap[1,N]\}\in\dec_{\mathscr{F}_{t}}Z$ for any $N\in\mathds{N}_{++}$. Then, for any subset $Z\subseteq\mathds{L}^{p}_{\mathscr{F}_{t}}(D;E)$ with $\Int Z\neq\emptyset$, it can be claimed (\cite[Fryszkowski, 2004, \text{Prop.} 51]{F2} that $\dec_{\mathscr{F}_{t}}Z=\mathds{L}^{p}_{\mathscr{F}_{t}}(\Omega;E)$. Put differently, any decomposable set smaller than the whole space must have empty interior.

The following lemma (see \cite[Michta, 2015, \text{Thm.} 2.2]{M5}) clarifies the relationship between integrability and decomposability.

\begin{lemma}\label{lem:2}
Let $Z\subseteq\mathds{L}^{p}_{\mathcal{G}}(D;E)$ be a nonempty subset. Then the following three assertions are equivalent: \medskip\\
(i) $Z$ is $p$-integrably bounded;\\
(ii) $\overline{\dec}_{\mathcal{G}}Z$ is a bounded subset of $\mathds{L}^{p}_{\mathcal{G}}(D;E)$;\\
(iii) $\{\max\{X_{k}:k\in\mathds{N}\cap[1,N]\}\subseteq Z:N\in\mathds{N}_{++}\}$ is $p$-integrably bounded.
\end{lemma}

The main takeaway from Lemma \ref{lem:2} is that the decomposability of a set-valued random variable cannot be a stronger condition than its integrable bounded-ness -- they are in fact equivalent. The third assertion is heavily tied to the definition of the decomposable hull, (\ref{B.1}).

The following important theorem explains the one-to-one correspondence between a set-valued mapping and the decomposability of the collection of its integrable selectors (see \cite[Hiai and Umegaki, 1977, \text{Thm.} 3.1]{HU}).

\begin{theorem}\label{thm:5}
Let $Z\subseteq\mathds{L}^{p}_{\mathcal{G}}(D;E)$, for fixed $t\in[0,T]$, be a nonempty closed set. Then there exists a $\mathcal{G}$-measurable mapping $\mathcal{X}:D\mapsto\mathrm{Cl}(E)$ such that $\mathscr{S}^{p}_{\mathcal{G}}(\mathcal{X})=Z$ if and only if $Z$ is decomposable.
\end{theorem}

Since Theorem \ref{thm:5} gives an ``if and only if'' statement, the most suitable way of defining (truly) set-valued stochastic integrals, which must be used to construct set-valued stochastic processes, is by decomposable subsets of $\mathds{L}^{p}$-function spaces.

Now we are ready to lay out the formal definitions of the Aumann stochastic integral (\cite[Kisielewicz, 2020, \text{Chap.} IV \text{Sect.} 4.2]{K3}) and the set-valued It\^{o} integral (\cite[Kisielewicz, 2020, \text{Chap.} V \text{Sect.} 5.1 and 5.2]{K3}). From now on, we shall specify the measure space $(D,\mathcal{A},\mathbb{M})$ and the Euclidean space $E$ and write $\mathds{L}^{p}_{\mathbb{F}}([0,T]\times\Omega;\mathds{R}^{d})\equiv\mathds{L}^{p}([0,T]\times\Omega,\Sigma_{\mathbb{F}}, \mathrm{Leb}_{[0,T]}\times\PP;\mathds{R}^{d})$, as has appeared already in Subsection \ref{ss:2.2}; similarly, we write $\mathds{A}^{p}_{\mathbb{F}}([0,T]\times\Omega;\mathds{R}^{d})\equiv\mathds{A}^{p}([0,T]\times\Omega,\Sigma_{\mathbb{F}}, \mathrm{Leb}_{[0,T]}\times\PP;\mathds{R}^{d})$.

\begin{definition}\label{def:7}
Let $d\in\mathds{N}_{++}$. For a $\Sigma_{\mathbb{F}}$-measurable $p$-integrably bounded set-valued stochastic process $\mathcal{X}$ with values in $\mathrm{Cl}(\mathds{R}^{d})$, the Aumann stochastic integral functional is defined for fixed $t\in[0,T]$ to be the multifunction
\begin{equation}\label{B.2}
  \mathds{A}^{p}_{\mathbb{F}}([0,T]\times\Omega;\mathds{R}^{d})\ni\mathcal{X}\mapsto \mathfrak{I}_{t}\big(\mathscr{S}^{p}_{\Sigma_{\mathbb{F}}}(\mathcal{X})\big):=\bigg\{\int^{t}_{0}X_{s}\dd s:X\in \mathscr{S}^{p}_{\Sigma_{\mathbb{F}}}(\mathcal{X})\bigg\}\in\mathcal{P}\big(\mathds{L}^{p}_{\mathscr{F}_{t}}(\Omega;\mathds{R}^{d})\big).
\end{equation}
Then, the Aumann stochastic integral of $\mathcal{X}$ can be defined for fixed $t\in[0,T]$ to be the multifunction
\begin{equation}\label{B.3}
  \Omega\ni\omega\mapsto\int^{t}_{0}\mathcal{X}_{s}(\omega)\dd s:=\mathfrak{I}_{t}\big(\mathscr{S}^{p}_{\Sigma_{\mathbb{F}}}(\mathcal{X})\big)(\omega)\in\mathrm{Cl}(\mathds{R}^{d}).
\end{equation}
\end{definition}

Note that (\ref{B.2}) is for $\mathbb{F}$-non-anticipating integrands. If the Aumann stochastic integral is defined for the larger space $\mathds{L}^{p}_{\mathscr{B}([0,T])\otimes\mathcal{F}}([0,T]\times\Omega;\mathds{R}^{d})$, then the decomposability of $\mathfrak{I}_{t}(\mathscr{S}^{p}(\mathcal{X}))$ will follow from that of $\mathscr{S}^{p}(\mathcal{X})$. Unfortunately, this is not true for the case of $\mathds{L}^{p}_{\mathbb{F}}([0,T]\times\Omega;\mathds{R}^{d})$; therefore the decomposable hull $\overline{\dec}_{\mathscr{F}_{t}}$ should be taken on the integral functional to ensure that it is a (closed) set-valued random variable.

Importantly, if $\mathcal{X}$ is convex-valued, then (\ref{B.3}) coincides (in the $\PP$-\text{a.s.} sense) with defining a set-valued random variable whose subtrajectory integrals satisfy $\mathscr{S}^{p}_{\mathscr{F}_{t}}\big(\int^{t}_{0}\mathcal{X}_{s}\dd s\big)=\overline{\dec}_{\mathscr{F}_{t}}\mathfrak{I}_{t}\big(\mathscr{S}^{p}_{\Sigma_{\mathbb{F}}}(\mathcal{X})\big)$.

\begin{definition}\label{def:8}
Let $W$ be an $m$-dimensional Brownian motion defined on $(\Omega,\mathcal{F},\PP;\mathbb{F})$, for a $\Sigma_{\mathbb{F}}$-measurable square-integrably bounded set-valued stochastic process $\mathcal{X}$ with values in $\mathrm{Cl}(\mathds{R}^{d\otimes m})$, the set-valued It\^{o} integral functional is defined for fixed $t\in[0,T]$ to be the multifunction
\begin{align}\label{B.4}
  \mathds{A}^{2}_{\mathbb{F}}([0,T]\times\Omega;\mathds{R}^{d\otimes m})\ni\mathcal{X}\mapsto \mathfrak{J}_{t}\big(\mathscr{S}^{2}_{\Sigma_{\mathbb{F}}}(\mathcal{X})\big)&:=\bigg\{\int^{t}_{0}X_{s}\dd W_{s}:X\in \mathscr{S}^{2}_{\Sigma_{\mathbb{F}}}(\mathcal{X})\bigg\} \nonumber\\
  &\in\mathcal{P}\big(\mathds{L}^{2}_{\mathscr{F}_{t}}(\Omega;\mathds{R}^{d})\big).
\end{align}
Then, the set-valued It\^{o} integral of $\mathcal{X}$, $\int^{t}_{0}\mathcal{X}_{s}\dd W_{s}$, is defined to be a set-valued random variable whose subtrajectory integrals satisfy $\mathscr{S}^{2}_{\mathscr{F}_{t}}\big(\int^{t}_{0}\mathcal{X}_{s}\dd W_{s}\big)=\overline{\dec}_{\mathscr{F}_{t}}\mathfrak{J}_{t}\big(\mathscr{S}^{2}_{\Sigma_{\mathbb{F}}}(\mathcal{X})\big)$.
\end{definition}

Similar to the defined Aumann stochastic integral functionals, set-valued It\^{o} integral functionals defined by (\ref{B.4}) are not guaranteed to form decomposable subsets of $\mathds{L}^{2}_{\mathscr{F}_{t}}(\Omega;\mathds{R}^{d})$, and so $\overline{\dec}_{\mathscr{F}_{t}}$ should subsequently be taken to construct the It\^{o} integrals.

As a remark, the Aumann stochastic integral and the set-valued It\^{o} integral are natural set-valued analogs of the Lebesgue stochastic integral and the (usual) It\^{o} integral, respectively. For generic time, they are both constructed from decomposable subsets of corresponding $\mathds{L}^{p}$-function spaces. The basic intuition of Definition \ref{def:7} and Definition \ref{def:8} is that for any subset of $p$-integrable $\mathbb{F}$-non-anticipating processes the set-valued integral functionals are collections of integral functionals defined in the usual sense for every (non-anticipating) selector of this subset.

Theorem \ref{thm:6} below summarizes some basic properties of the two types of set-valued stochastic integrals (see \cite[Kisielewicz, 2020b, \text{Chap.} IV and V]{K3} for references).

\begin{theorem}\label{thm:6}
Let $\mathcal{X}$ and $\mathcal{Y}$ be any two $\Sigma_{\mathbb{F}}$-measurable $p$-integrably bounded set-valued stochastic processes with values in $\mathrm{Cl}(E)$, where, for Aumann stochastic integrals and set-valued It\^{o} integrals, respectively, $p\geq1$ and $p=2$ and $E=\mathds{R}^{d}$ $E=\mathds{R}^{d\otimes m}$. Then, the following assertions hold for any fixed $t\in[0,T]$: \medskip\\
(i) $\int^{t}_{0}\cl_{\mathds{L}^{p}}\mathcal{X}_{s}\dd s=\int^{t}_{0}\mathcal{X}_{s}\dd s$, $\PP$-a.s.; \\
(ii) $\int^{t}_{0}\cl_{\mathds{L}^{2}}\mathcal{X}_{s}\dd W_{s}=\int^{t}_{0}\mathcal{X}_{s}\dd W_{s}$, $\PP$-a.s.; \\
(iii) $\int^{t}_{0}\overline{\co}_{\mathds{L}^{p}}\mathcal{X}_{s}\dd s=\overline{\co}_{\mathds{R}^{d}}\int^{t}_{0}\mathcal{X}_{s}\dd s$, $\PP$-a.s.; \\
(iv) $\int^{t}_{0}\overline{\co}_{\mathds{L}^{2}}\mathcal{X}_{s}\dd W_{s}=\overline{\co}_{\mathds{R}^{d}}\int^{t}_{0}\mathcal{X}_{s}\dd W_{s}$, $\PP$-a.s.; \\
(v) if $\mathfrak{I}_{t}\big(\mathscr{S}^{p}_{\Sigma_{\mathbb{F}}}(\mathcal{X})\big)$ and $\mathfrak{I}_{t}\big(\mathscr{S}^{p}_{\Sigma_{\mathbb{F}}}(\mathcal{Y})\big)$ are bounded subsets of $\mathds{L}^{p}_{\mathscr{F}_{t}}(\Omega;\mathds{R}^{d})$, $\int^{t}_{0}(\mathcal{X}_{s}+\mathcal{Y}_{s})\dd s=\int^{t}_{0}\mathcal{X}_{s}\dd s+\int^{t}_{0}\mathcal{Y}_{s}\dd s$, $\PP$-a.s., where addition is in the Minkowski sense in $\mathds{R}^{d}$; \\
(vi) if $\mathfrak{J}_{t}\big(\mathscr{S}^{2}_{\Sigma_{\mathbb{F}}}(\mathcal{X})\big)$ and $\mathfrak{J}_{t}\big(\mathscr{S}^{2}_{\Sigma_{\mathbb{F}}}(\mathcal{Y})\big)$ are bounded subsets of $\mathds{L}^{2}_{\mathscr{F}_{t}}(\Omega;\mathds{R}^{d})$, $\int^{t}_{0}(\mathcal{X}_{s}+\mathcal{Y}_{s})\dd W_{s}=\int^{t}_{0}\mathcal{X}_{s}\dd W_{s}+\int^{t}_{0}\mathcal{Y}_{s}\dd W_{s}$, $\PP$-a.s.
\end{theorem}

A somewhat unfortunate result which hinders most applications of (truly) set-valued It\^{o} integrals is the following (see \cite[Kisielewicz, 2020b, \text{Thm.} 5.3.1]{K3}).

\begin{theorem}\label{thm:7}
For fixed $t\in[0,T]$, the set-valued It\^{o} integral $\int^{t}_{0}\mathcal{X}_{s}\dd W_{s}$ as defined in Definition \ref{def:8} is square-integrably bounded if and only if it admits a $\Sigma_{\mathbb{F}}$-measurable representation Castaing $\{X_{k}:k\in\mathds{N}_{++}\}$ such that $\E\big[\int^{t}_{0}\|X_{k,s}-X_{k',s}\|^{2}_{2}\dd s\big]=0$, $\forall k,k'\geq1$.
\end{theorem}

An immediate consequence from Theorem \ref{thm:7} is that if $\mathscr{S}^{2}_{\Sigma_{\mathbb{F}}}(\mathcal{X})$, which must be decomposable due to Theorem \ref{thm:5}, is not a singleton, then the set-valued It\^{o} integral $\int^{t}_{0}\mathcal{X}_{s}\dd W_{s}$ will not be square-integrably bounded. Conversely, in order for such an integral to be square-integrably bounded, it must be single-valued, with $\mathcal{X}$ being a single-valued process.

A possible remedy is to consider indecomposable subsets of subtrajectory integrals and extend the definition of set-valued It\^{o} integral functionals in (\ref{B.4}) to any nonempty, indecomposable subsets of $\mathds{L}^{2}_{\mathbb{F}}([0,T]\times\Omega;\mathds{R}^{d\otimes m})$ in the obvious way. A sufficient condition is that the functional is operated on a finite set, i.e., with $t$ fixed, $\mathfrak{J}_{t}(Z)$, where $Z\subseteq\mathds{L}^{2}_{\mathbb{F}}([0,T]\times\Omega;\mathds{R}^{d\otimes m})$, and then the integral $\int^{t}_{0}Z(s)\dd W_{s}$, i.e., the random variable with $\mathscr{S}^{2}_{\mathscr{F}_{t}}(\Omega;\mathds{R}^{d})=\overline{\dec}_{\mathscr{F}_{t}}\mathfrak{J}_{t}(Z)$, will be square-integrably bounded. A more relaxed condition, as adopted in the set-valued It\^{o} process (\ref{2.3.3}), is for $Z$ to be the $\mathds{L}^{p}$-closed convex hull of a sequence of absolutely summable square-integrable processes, as shown in (\ref{2.2.4}). Even so, these conditions are not necessary to guarantee the square-integrable bounded-ness of resultant set-valued It\^{o} integrals.

Lastly, based on Definition \ref{def:7} and the above remark, for indefinite time, the Aumann stochastic integrals and the set-valued It\^{o} integral are defined, respectively, to be the set-valued stochastic processes $\int^{\imath}_{0}\mathcal{X}_{s}\dd s\equiv\big(\int^{t}_{0}\mathcal{X}_{s}\dd s\big)_{t\in[0,T]}$ and $\int^{\imath}_{0}Z(s)\dd W_{s}\equiv\big(\int^{t}_{0}Z(s)\dd W_{s}\big)_{t\in[0,T]}$ as mappings from $[0,T]\times\Omega$ to $\mathrm{Cl}(\mathds{R}^{d})$. For any $\mathcal{X}\in\mathds{A}^{p}_{\mathbb{F}}([0,T]\times\Omega;\mathds{R}^{d})$ and any $Z=\overline{\co}_{\mathds{L}^{2}}\{g_{k}:[0,T]\times\Omega\mapsto\mathds{R}^{d\otimes m}\}$ with (\ref{2.2.4}) satisfied by the elements $g_{k}$'s, the processes $\int^{\imath}_{0}\mathcal{X}_{s}\dd s$ and $\int^{\imath}_{0}Z(s)\dd W_{s}$ have ($\PP$-a.s.) continuous sample paths. The fundamental reason behind these continuity properties is the continuity of the integrators $\imath$ and $W$ as in the single-valued case (see \cite[Kisielewicz, 2020b, \text{Corol.} 4.2.1 and \text{Thm.} 5.5.2]{K3}). The sum of these two types of processes leads to what is called a ``set-valued It\^{o} process,'' or (\ref{2.2.3}) in its most general form.

\clearpage

\section{List of symbols}\label{C}

The following symbols are presented roughly in the order of appearance in the paper.

\begin{table}[H]\small
  \centering
  \begin{tabular}{ll|ll}
    \hline
    \multicolumn{4}{c}{\textbf{Mathematical meanings}} \\ \hline
    $\intercal$ & matrix transpose & $\|\;\|_{2}$ & Euclidean norm\\
    $\mathbb{F}$ & \makecell[l]{filtration of the probability space \\ $(\Omega,\mathcal{F},\PP)$} & $\mathrm{diag}$ & vector-to-diagonal matrix operator \\
    $\mathds{N}_{++}$ & set of positive integers & $\mathds{Q}$ & set of rational numbers \\
    $\mathrm{cl}$ & closure & $\overline{\mathrm{co}}$ & closed convex hull \\
    $\mathrm{card}$ & cardinality of a set & $\|\;\|_{\mathrm{F}}$ & Frobenius norm \\
    $\mathrm{Cl}(\;)$ & space of nonempty closed subsets & $\prod$ & Cartesian product \\
    $\mathcal{C}_{\rm b}(\; ;\;)$ & space of bounded continuous functions & $\mathds{L}^{p}_{\mathbb{F}}(\; ;\;)$ & \makecell[l]{space of $p\geq1$-integrable \\ $\mathbb{F}$-non-anticipating functions} \\
    $\mathcal{P}(\;)$ & space of nonempty subsets & $\mathcal{d}_{\rm H}$ & Hausdorff distance \\
    $\0$ & zero element & $\1$ & element of all ones \\
    $\mathrm{epi}$ & epigraph & $\mathrm{hyp}$ & hypograph \\
    $\Int$ & interior & $\emptyset$ & empty set \\
    $\complement$ & set complement & $\langle\;,\;\rangle$ & inner product or duality pairing \\
    $\dag$ & dual space or conjugate & $\|\;\|_{1}$ & \makecell[l]{total variation norm or \\ Taxicab norm} \\
    $\mathcal{B}(\;)$ & standard Borel space & $\mathcal{C}_{0}(\; ;\;)$ & classical Wiener space \\
    $\delta$ & dirac measure & $\imath$ & identity map (time variable) \\
    $\ddag$ & bi-conjugate & $\mathds{1}$ & indicator function \\
    $\mathbf{v}$ & vector field & $\lrcorner$ & interior product \\
    $\pd$ & boundary of a compact set & $\mathcal{D}$ & Malliavin derivative operator \\
    $\mathrm{erfc}$ & complementary error function & $\mathrm{dist}$ & point-to-set distance \\
    $\upharpoonright$ & restriction of measure & $\mathbb{D}^{1,2}$ & \makecell[l]{space of Malliavin-differentiable \\ functionals in $\mathds{L}^{2}_{\mathcal{F}}(\Omega)$} \\
    $\nu$ & unit normal & $\mathrm{Leb}$ & Lebesgue measure \\
    $\mathcal{X}$ & generic set-valued mapping & $\mathds{A}^{p}(\; ;\;)$ & \makecell[l]{space of measurable \\ $p$-integrable set-valued mappings} \\
    $\overline{\dec}$ & decomposable closure & $\mathscr{S}^{p}(\; ;\;)$ & set of subtrajectory integrals \\
    $\mathfrak{I}(\;)$ & Aumann integral functional & $\mathfrak{J}(\;)$ & set-valued It\^{o} integral functional \\
    \hline
    \multicolumn{4}{c}{\textbf{Financial and economic meanings}} \\ \hline
    $r$ & risk-free rate & $m$ & number of risky assets\\
    $S$ & risky asset price process & $D$ & dividend process \\
    $n$ & number of commodities & $P$ & commodity price process \\
    $\succeq$ & preference relation & $c$ & consumption level \\
    $d$ & number of preferential parameters & $\mathcal{I}$ & multi-utility index set process \\
    $u$ & consumption multi-utility function & $U$ & bequest utility function \\
    $\circleddash$ & incomplete part of $\succeq$ & $X$ & inter-temporal wealth process \\
    $C$ & total consumption expenditure & $\Pi$ & portfolio process \\
    $\mathfrak{A}(X_{0})$ & admissibility set & $\mathfrak{B}(X_{0})$ & budget set \\
    $\xi$ & state price density & $\theta$ & market price of risk \\
    $w$ & totaling rule or weighting factor & $J$ & reduced index set process \\
    $\chi$ & attention degree & $p$ & risk aversion degree \\
    \hline
  \end{tabular}
\end{table}


\vspace{0.2in}

\begin{thebibliography}{99}\footnotesize
\bibitem{AO} Agranov, M. and Ortoleva, P. (2017). Stochastic choice and preferences for randomization. {\sl Journal of Political Economy}, 125: 40--68.
\bibitem{AF} Ararat, \c{C}. and Feinstein, Z. (2021). Set-valued risk measures as backward stochastic difference inclusion and equations. {\sl Finance and Stochastics}, 25: 43--76.
\bibitem{AMW} Ararat, \c{C}., Ma, J., and Wu, W. (2023). Set-valued backward stochastic differential equations. {\sl Annals of Applied Probability}, 33: 3418--3448.
\bibitem{A1} Aumann, R.J. (1962). Utility theory without the completeness axiom. {\sl Econometrica}, 30: 445--462.
\bibitem{A2} Aumann, R.J. (1965). Integrals of set-valued functions. {\sl Journal of Mathematical Analysis and Applications}, 12: 1--12.
\bibitem{BEX} Bekaert, G., Engstrom, E., and Xing, Y. (2009). Risk, uncertainty, and asset prices. {\sl Journal of Financial Economics}, 91: 59--82.
\bibitem{BC} Benedetti, G. and Campi, L. (2012). Multivariate utility maximization with proportional transaction costs and random endowment. {\sl SIAM Journal on Control and Optimization}, 50: 1283--1308.
\bibitem{B1} Bewley, T.F. (2002). Knightian decision theory: part I. {\sl Decisions in Economics and Finance}, 25: 79--110.
\bibitem{BP} Biagini, S. and P{\i}nar, M.\c{C}. (2017). The robust Merton problem of an ambiguity averse investor. {\sl Mathematics and Financial Economics}, 11: 1--24.
\bibitem{BGZ} Bollerslev, T., Gibson, M., and Zhou, H. (2011). Dynamic estimation of volatility risk premia and investor risk aversion from option-implied and realized volatilities. {\sl Journal of Econometrics}, 160: 235--245.
\bibitem{B2} Bornstein, R.F. (1989). Exposure and affect: Overview and meta-analysis of research, 1968--1987. {\sl Psychological Bulletin}, 106: 265--289.
\bibitem{BPPS} Boutilier, C., Patrascu R., Poupart P., Schuurmans, D. (2006). Constraint-based optimization and utility elicitation using the minimax decision criterion. {\sl Artificial Intelligence}, 170: 686--713.
\bibitem{B3} Bucklin, L.P. (1963). Retail strategy and the classification of consumer goods. {\sl Journal of Marketing}, 27: 50--55.
\bibitem{CO1} Campi, L. and Owen, M.P. (2011). Multivariate utility maximization with proportional transaction costs. {\sl Finance and Stochastics}, 15: 461--499.
\bibitem{CO2} \c{C}anako\u{g}lu, E. and \"{O}zekici, S. (2012). HARA frontiers of optimal portfolios in stochastic markets. {\sl European Journal of Operational Research}, 221: 129--137.
\bibitem{CR1} Cettolin, E. and Riedl, A. (2019). Revealed preferences under uncertainty: Incomplete preferences and preferences for randomization. {\sl Journal of Economic Theory}, 181: 547--585.
\bibitem{CR2} Coeurdacier, N. and Rey, H\'{e}l\`{e}ne. (2012). Home bias in open economy financial macroeconomics. {\sl Journal of Economic Literature}, 51: 63--115.
\bibitem{DP} Danan, E., Ziegelmeyer, A., (2006). Are preferences complete? An experimental measurement of indecisiveness under risk. Mimeo. Universit\'{e} de Cergy-Pontoise.
\bibitem{DMOPH} Deparis, S., Mousseau, V., \"{O}zt\"{u}rk, M., Pallier, C., and Huron, C. (2012). When conflict induces the expression of incomplete preferences. {\sl European Journal of Operational Research}, 221: 593--602.
\bibitem{DGR} Detemple, J., Garcia, R., and Rindisbacher, M. (2003). A Monte Carlo method for optimal portfolios. {\sl The Journal of Finance}, 58: 401--446.
\bibitem{DGW} Doidge, C., Griffin, J., and Williamson, R. (2006). Measuring the economic importance of exchange rate exposure. {\sl Journal of Empirical Finance}, 13: 550--576.
\bibitem{D} Dubra, J. (2009). A theory of time preferences over risky outcomes. {\sl Journal of Mathematical Economics}, 45: 576--588.
\bibitem{DMO} Dubra, J., Maccheroni, F., and Ok, E.A. (2004). Expected utility theory without the completeness axiom. {\sl Journal of Economic Theory}, 115: 118--133.
\bibitem{E1} Ehrgott, M. (2005). {\sl Multicriteria Optimization}, 2nd Edn, Springer-Verlag, Berlin--Heidelberg.
\bibitem{E2} Eichfelder, G. (2008). {\sl Adaptive Scalarization Methods in Multiobjective Optimization}, Springer-Verlag, Berlin--Heidelberg.
\bibitem{EV} Eisenhauer, J.G. and Ventura, L. (2003). Survey measures of risk aversion and prudence. {\sl Applied Economics}, 35: 1477--1484.
\bibitem{E3} Elias, N. (1982). {\sl The civilizing process: Sociogenetic and Psychogenetic Investigations}, Translated by Jephcott, E. Rev. Ed. Blackwell, Oxford.
\bibitem{EJ} Epstein, L.G. and Ji, S. (2013). Ambiguous volatility and asset pricing in continuous time. {\sl Review of Financial Studies}, 26: 1740--1786.
\bibitem{E4} Evren, \"{O}. (2014). Scalarization methods and expected multi-utility representations. {\sl Journal of Economic Theory}, 151: 30--63.
\bibitem{EO} Evren, \"{O}. and Ok, E.A. (2011). On the multi-utility representation of preference relations. {\sl Journal of Mathematical Economics}, 47: 554--563.
\bibitem{F1} Flanders, H. (1973). Differentiation under the integral sign. {\sl The American Mathematical Monthly}, 80: 615--627.
\bibitem{FPS} Fouque, J-P., Papanicolaou, G., and Sircar, K.R. (2000). Mean-reverting stochastic volatility. {\sl International Journal of Theoretical and Applied Finance}, 3: 101--142.
\bibitem{FPW} Fouque, J-P., Pun, C.S., and Wong, H.Y. (2016). Portfolio optimization with ambiguous correlation and stochastic volatilities. {\sl SIAM Journal on Control and Optimization}. 54: 2309--2338.
\bibitem{FS} Frost, R.O. and Shows, D.L. (1993). The nature and measurement of compulsive indecisiveness. {\sl Behaviour Research and Therapy}, 31: 683--692.
\bibitem{F2} Fryszkowski, A. {\sl Fixed Point Theory for Decomposable Sets}, Kluwer Academic Publishers, New York, 2004.
\bibitem{GSS} Gaar, E., Scherer, D., and Schiereck, D. (2022). The home bias and the local bias: A survey. {\sl Management Review Quarterly}, 72: 21--57.
\bibitem{GK} Galaabaatar, T. and Karni, E. (2012). Expected multi-utility representations. {\sl Mathematical Social Sciences}, 64: 242--246.
\bibitem{GS} Gass, S. and Saaty, T. (1955). The computational algorithm for the parametric objective function. {\sl Naval Research Logistics Quarterly}, 2: 39--45.
\bibitem{GSZ} Guiso, L., Sapienza, P., and Zingales, L. (2018). Time varying risk aversion. {\sl Journal of Financial Economics}, 128: 403--421.
\bibitem{H} Hamel, A.H. (2009). A duality theory for set-valued functions I: Fenchel conjugation theory. {\sl Set-Valued Analysis}, 17: 153--182.
\bibitem{HHLRS} Hamel, A.H., Heyde, F., L\"{o}hne, A., Rudloff, B., and Schrage, C. (2015). Set Optimization -- A Rather Short Introduction. In: {\sl Set Optimization and Applications -- The State of the Art}, Springer Proceedings in Mathematics and Statistics, vol 151. Springer, Berlin, Heidelberg.
\bibitem{HL} Hamel, A.H. and L\"{o}hne, A. (2014). Lagrange duality in set optimization. {\sl Journal of Optimization Theory and Applications}, 161: 368--397.
\bibitem{HW} Hamel, A.H. and Wang, S.Q. (2017). A set optimization approach to utility maximization under transaction costs. {\sl Decisions in Economics and Finance}, 40: 257--275.
\bibitem{HH} He, T-S. and Hong, F. (2022). Risk breeds risk aversion. {\sl Experimental Economics}, 21: 815--835.
\bibitem{HU} Hiai, F. and Umegaki, H. (1977). Integrals, conditional expectations and martingales of multivalued functions. {\sl Journal of Multivariate Analysis}, 7: 149--182.
\bibitem{JJ} Janssen, M.A. and Jager, W. (2001). Fashions, habits, and changing preferences: Simulation of psychological factors affecting market dynamics. {\sl Journal of Economic Psychology}, 22: 745--772.
\bibitem{KLS} Karatzas, I., Lehoczky, J. and Shreve, S. (1987). Optimal portfolio and consumption decisions for a ``small investor'' on a finite horizon. {\sl SIAM Journal on Control and Optimization}, 25: 1557--1586.
\bibitem{KR} Keeney, R.L. and Raiffa, H. (1993). {\sl Decisions with Multiple Objectives}. Cambridge University Press, Cambridge.
\bibitem{KY} Kelsey, D. and Yalcin, E. (2007). The arbitrage pricing theorem with incomplete preferences. {\sl Mathematical Social Sciences}, 54: 90--105.
\bibitem{K1} Kisielewicz, M. (1991). {\sl Differential Inclusions and Optimal Control}. Kluwer Academic Publishers, New York.
\bibitem{K2} Kisielewicz, M. (2012). Some properties of set-valued stochastic integrals. {\sl Journal of Mathematical Analysis and Applications}, 388: 984--995.
\bibitem{K3} Kisielewicz, M. (2020). {\sl Set-Valued Stochastic Integrals and Applications}. Springer Nature Switzerland, Switzerland.
\bibitem{LL1} Levy, H. and Levy, M. (2014). The home bias is here to stay. {\sl Journal of Banking and Finance}, 47: 29--40.
\bibitem{LWZ} Li, H., Wu, C., and Zhou, C. (2022). Time-varying risk aversion and dynamic portfolio allocation. {\sl Operations Research}, 70: 23--37.
\bibitem{LL2} Li, J. and Li, S. (2009). It\^{o} type set-valued stochastic differential equation. {\sl Journal of Uncertain Systems}, 3: 52--63.
\bibitem{LLL} Li, S., Li, J., and Li, X. (2010). Stochastic integral with respect to set-valued square integrable martingales. {\sl Journal of Mathematical Analysis and Applications}, 370: 659--671.
\bibitem{LM1} Liang, Z. and Ma, M. (2020). Robust consumption--investment problem under CRRA and CARA utilities with time-varying confidence sets. {\sl Mathematical Finance}, 30: 1035--1072.
\bibitem{LP} Loomes, G., Pogrebna, G. (2014). Measuring individual risk attitudes when preferences are imprecise. {\sl The Economic Journal}, 124: 569--593.
\bibitem{LSS} Loomes, G., Starmer, C., and Sugden, R. (1991). Observing violations of transitivity by experimental methods. {\sl Econometrica}, 59: 425--439.
\bibitem{LS} Loomes, G. and Sugden, R. (1982). An alternative theory of rational choice under uncertainty. {\sl The Economic Journal}, 92: 805--824.
\bibitem{L1} Liu, L. (2004). A new foundation for the mean--variance analysis. {\sl European Journal of Operational Research}, 158: 229--242.
\bibitem{LM2} Luttmer, E.G.J. and Mariotti, T. (2003). Subjective discounting in an exchange economy. {\sl Journal of Political Economy}, 111: 959--989.
\bibitem{L2} Lyasoff, A. (2017). {\sl Stochastic Methods in Asset Pricing}, MIT Press, Cambridge, MA.
\bibitem{MRW} Ma, J., Rokhlin, V., and Wandzura, S. (1996). Generalized Gaussian quadrature rules for systems of arbitrary functions. {\sl SIAM Journal on Numerical Analysis}, 33: 971--996.
\bibitem{MMR} Maccheroni, F., Marinacci, M., and Ruffino, D. (2013). Alpha as ambiguity: Robust mean--variance portfolio analysis. {\sl Econometrica}, 81: 1075--1113.
\bibitem{M1} Malinowski, M.T. (2013). On a new set-valued stochastic integral with respect to semimartingales and its applications. {\sl Journal of Mathematical Analysis and Applications}, 408: 669--680.
\bibitem{M2} Mandler, M. (2005). Incomplete preferences and rational intransitivity of choice. {\sl Games and Economic Behavior}, 50: 255--277.
\bibitem{M3} Merton, R.C. (1969). Lifetime portfolio selection under uncertainty: the continuous time case. {\sl Review of Economics and Statistics}, 51: 247--257.
\bibitem{M4} Merton, R.C. (1971). Optimum consumption and portfolio rules in a continuous-time model. {\sl Journal of Economic Theory}, 3: 373--413.
\bibitem{ML} Michelbrink, D. and Le, H. (2012). A martingale approach to optimal portfolios with jump--diffusions. {\sl SIAM Journal on Control and Optimization}, 50: 583--599.
\bibitem{M5} Michta, M. (2015). Remarks on unboundedness of set-valued It\^{o} stochastic integrals. {\sl Journal of Mathematical Analysis and Applications}, 424: 651--663.
\bibitem{MMCE} Mrad, M., Majdalani, J., Cui, C.C., and El Khansa, Z. (2020). Brand addiction in the contexts of luxury and fast-fashion brands. {\sl Journal of Retailing and Consumer Services}, 55: 102089.
\bibitem{N1} Nau, R. (2006). The shape of incomplete preferences. {\sl Annals of Statistics}, 34: 2430--2448.
\bibitem{NO} Nishimura, H. and Ok, E.A. (2016). Utility representation of an incomplete and nontransitive preference relation. {\sl Journal of Economic Theory}, 166: 164--185.
\bibitem{N2} Nualart, D. (1995). {\sl The Malliavin Calculus and Related Topics}, Springer-Verlag, Berlin--Heindelberg.
\bibitem{O} Ok, E.A. (2002). Utility representation of an incomplete preference relation. {\sl Journal of Economic Theory}, 104: 429--449.
\bibitem{OM} Ok, E.A. and Masatlioglu, Y. (2007). A theory of (relative) discounting. {\sl Journal of Economic Theory}, 137: 214--245.
\bibitem{OOR} Ok, E.A., Ortoleva, P., and Riella, G. (2012). Incomplete preferences under uncertainty: Indecisiveness in beliefs versus tastes. {\sl Econometrica}, 80: 1791--1808.
\bibitem{PW1} Patalano, A.L. and Wengrovitz, S.M. (2007). Indecisiveness and response to risk in deciding when to decise. {\sl Journal of Behavioral Decision Making}, 20: 405--424.
\bibitem{PW2} P\'{e}zier, J. and White, A. (2008). The relative merits of investable hedge fund indices and of funds of hedge funds in optimal passive portfolios. {\sl The Journal of Alternative Investments}, 10: 37--49.
\bibitem{RR1} Read, D. and Roelofsma, P.H.M.P. (2003). Subadditive versus hyperbolic discounting: A comparison of choice and matching. {\sl Organizational Behavior and Human Decision Processes}, 91: 140--153.
\bibitem{RDD-S} Regenwetter, M., Dana, J., and Davis-Stober, C.P. (2011). Transitivity of preferences. {\sl Psychological Review}, 118: 42--56.
\bibitem{RR2} Renaud, J-F. and R\'{e}millard, B. (2007). Explicit martingale representations for Brownian functionals and applications to option hedging. {\sl Stochastic Analysis and Applications}, 25: 801--820.
\bibitem{RS} Rigotti, L. and Shannon, C. (2005). Uncertainty and risk in financial markets. {\sl Econometrica}, 73: 203--243.
\bibitem{RR3} Roelofsma, P.H.M.P. and Read, D. (2000). Intransitive intertemporal choice. {\sl Journal of Behavioral Decision Making}, 13: 161--177.
\bibitem{RU} Rudloff, B. and Ulus, F. (2021). Certainty equivalent and utility indifference pricing for incomplete preferences via convex vector optimization. {\sl Mathematics and Financial Economics}, 15: 397--430.
\bibitem{S1} Sautua, S. (2017). Does uncertainty cause inertia in decision making? An experimental study of the role of regret aversion and indecisiveness. {\sl Journal of Economic Behavior and Organization}, 136: 1--14.
\bibitem{S2} Schoutens, W. (2003). {\sl L\'{e}vy Processes in Finance: Pricing Financial Derivatives}. John Wiley \& Sons Ltd, The Atrium, Southern Gate, Chichestor, England.
\bibitem{S3} Starr, R.M. (1969). Quasi-equilibria in markets with non-convex preferences. {\sl Econometrica}, 37: 25--38.
\bibitem{T} Tu, L.W. (2011). {\sl An Introduction to Manifolds}, 2nd Edn. Universitext, Springer, New York.
\bibitem{WWZ} Wu, H., Weng, C., and Zeng, Y. (2018). Equilibrium consumption and portfolio decisions with stochastic discount rate and time-varying utility functions. {\sl OR Spectrum}, 40: 541--582.
\bibitem{X} Xia, W. (2025). Set-valued stochastic integrals for convoluted L\'{e}vy processes. {\sl Journal of Mathematical Analysis and Applications}, 545: 129076.
\bibitem{Z} Zadeh, L. (1963). Optimality and non-scalar-valued performance criteria. {\sl IEEE Transactions on Automatic Control}, 8: 59--60.
\bibitem{ZLMO} Zhang, J., Li, S., Mitoma, I., and Okazaki, Y. (2009). On the solutions of set-valued stochastic differential equations in M-type 2 Banach spaces. {\sl Tohoku Mathematical Journal}, 61: 417--440.
\end{thebibliography}
\end{document}